\let\orienddocument\enddocument
\documentclass{aa}
\let\enddocument\orienddocument

\usepackage[hyperindex=true,colorlinks=true,citecolor=blue,linkcolor=blue,breaklinks=true]{hyperref}
\usepackage{graphicx}
\usepackage{txfonts}
\usepackage{color}
\usepackage{lscape}
\usepackage{natbib,twoopt}
\usepackage{lastpage}
\usepackage{multirow}
\usepackage{placeins}

\definecolor{red}{rgb}{1.,0.0,0.}

\definecolor{orange}{rgb}{1.,.65,0.}

\definecolor{vert}{rgb}{.0,.65,0.}

\newcommand{\kms}{$\mathrm{km\,s^{-1}}\,$}
\newcommand{\ms}{$\mathrm{m\,s^{-1}}\,$}
\newcommand{\parsec}{\texttt{PARSEC}}
\newcommand{\basti}{\texttt{BaSTI}}
\newcommand{\mist}{\texttt{MIST}}
\newcommand{\dsep}{\texttt{DSEP}}

\bibpunct{(}{)}{;}{a}{}{,} 
\newcommandtwoopt{\citeads}[3][][]{\href{http://adsabs.harvard.edu/abs/#3}%
	{\citealp[#1][#2]{#3}}} 
\newcommandtwoopt{\citepads}[3][][]{\href{http://adsabs.harvard.edu/abs/#3}%
	{\citep[#1][#2]{#3}}} 
\newcommandtwoopt{\citetads}[3][][]{\href{http://adsabs.harvard.edu/abs/#3}%
	{\citet[#1][#2]{#3}}}
\newcommandtwoopt{\citeyearads}[3][][]%
{\href{http://adsabs.harvard.edu/abs/#3}{\citeyear[#1][#2]{#3}}}

\begin{document}

	\title{The Araucaria project: High-precision orbital parallaxes and masses of binary stars}
	
	\subtitle{I. VLTI/GRAVITY observations of ten double-lined spectroscopic binaries\thanks{Based on observations made with ESO telescopes at Paranal and La Silla observatory under programme IDs 0104.D-0191(A), 105.207C.001, 106.210E.001, 106.210E.002, 108.221X.001, 108.221X.003, 109.22XY.001, 109.22XY.002, 109.22XY.003, and 110.23QY.001.}}
	
	\author{A.~Gallenne\inst{1,2},
		A.~M\'erand\inst{3},
		P.~Kervella\inst{4},
		D.~Graczyk\inst{5},
		G.~Pietrzy\'nski\inst{6},
		W.~Gieren\inst{1}
		\and
		B.~Pilecki\inst{6}
	}
	
	\authorrunning{A. Gallenne et al.}
	\titlerunning{GRAVITY observations of 10 double-lined spectroscopic binaries}
	
	\institute{Universidad de Concepci\'on, Departamento de Astronom\'ia, Casilla 160-C, Concepci\'on, Chile
		\and Unidad Mixta Internacional Franco-Chilena de Astronom\'ia (CNRS UMI 3386)
		\and European Southern Observatory, Karl-Schwarzschild-Str. 2, 85748 Garching, Germany
		\and LESIA, Observatoire de Paris, CNRS UMR 8109, UPMC, Universit\'e Paris Diderot, 5 Place Jules Janssen, F-92195 Meudon, France
		\and Centrum Astronomiczne im. Miko\l{}aja Kopernika, Polish Academy of Sciences, Rabia\'nska 8, 87-100 Toru\'n, Poland
		\and Centrum Astronomiczne im. Miko\l{}aja Kopernika, Polish Academy of Sciences, Bartycka 18, 00-716 Warsaw, Poland
	}
	
	
	   \date{Received December 19, 2022; accepted February 8, 2023}
	
	
	\abstract
	{}
	{We aim to measure very precise and accurate model-independent masses and distances of detached binary stars. Precise masses at the $< 1$\,\% level are necessary to test and calibrate stellar interior and evolution models, while precise and independent orbital parallaxes are essential to check for the next Gaia data releases.}
	{We combined RV measurements with interferometric observations to determine orbital and physical parameters of ten double-lined spectroscopic systems. We report new relative astrometry from VLTI/GRAVITY and, for some systems, new VLT/UVES spectra to determine the radial velocities of each component.}
	{We measured the distance of ten binary systems and the mass of their components with a precision as high as 0.03\,\% (average level 0.2\,\%). They are combined with other stellar parameters (effective temperatures, radii, flux ratios, etc.) to fit stellar isochrones and determine their evolution stage and age. We also compared our orbital parallaxes with Gaia and showed that half of the stars are beyond $1\sigma$ with our orbital parallaxes; although, their RUWE is below the frequently used cutoff of 1.4 for reliable Gaia astrometry. By fitting the telluric features in the GRAVITY spectra, we also estimated the accuracy of the wavelength calibration to be $\sim 0.02$\,\% in high and medium spectral resolution modes.}
	{We demonstrate that combining spectroscopic and interferometric observations of binary stars provides extremely precise and accurate dynamical masses and orbital parallaxes. As they are detached binaries, they can be used as benchmark stars to calibrate stellar evolution models and test the Gaia parallaxes.}
	
	
	\maketitle
	
	
	\section{Introduction}
	\label{section__introduction}
	
	Stars in detached binary systems are the only tool enabling direct and precise distance and mass measurements. When the spectral lines of both components can be detected, the radial velocity (RV) of the orbital motion around the centre of mass of the two stars can be measured, providing the spectroscopic orbit of the system. However, this only provides the spectroscopic mass function, which is a combination of the stellar masses. With the apparent orbit from astrometry, we can measure the orbital inclination and obtain the individual masses of the system. In addition, spectroscopy provides the projected linear semi-major axis while astrometry gives its angular size, which directly provides the distance to the system. This is the only geometric and model-independent way of measuring masses and distances of stars \citep[see e.g.][]{McAlister_1976_12_4,Pourbaix_2004_09_1,Torres_2004_12_1,Docobo_2013_01_4,Torres_2015_07_0,Gallenne_2016_02_0}. 
	
	The main source providing precise stellar parameters is eclipsing binary (EB) systems, which combines RVs with photometric measurements during the eclipses \citep[see e.g.][]{Andersen_1984_01_0,Milone_1992_03_0,Pietrzynski_2009_05_0,Pietrzynski_2013_03_0,Kirkby-Kent_2016_06_0,Pietrzynski_2019_03_0,Graczyk_2020_11_0}. Although a precision level of $1-3$\,\% is routinely achieved, it is still subject to some modelling of the light curves, such as for the limb darkening, oblateness, or stellar spots. To determine the distance, a surface brightness colour relation is usually used to estimate the angular diameter of the stars, which is then combined with the linear value measured from the eclipses. EBs are a powerful tool and already provide precise masses and distances; however, this still depends on some modelling and relations that need to be well calibrated. 
	
	Combining astrometry with RV is the best way to measure the basic stellar properties using a minimum of theoretical assumptions. An astrometric orbit can be measured using different observing techniques. The most commonly used one is direct imaging which allows one to resolve the components to then monitor their relative position with time to cover the orbit. Direct imaging is sensitive to wide binaries, and therefore to systems with very long orbital periods making observations impractical. If the faintest (secondary) component is not seen as a separate star, we can sometimes infer its presence from the gravitational influence on the main visible (primary) component. The detection of the primary wobble requires long-term observations and a good precision on the measured positions and proper motions, in addition to needing the secondary companion to be massive enough to produce a detectable effect. Thousands of binaries were detected this way by the Hipparcos telescope \citep{Perryman_1997_07_0,Lindegren_1997_07_5}  and in the Gaia data release 3 (DR3) \citep{Halbwachs_2022_06_7,Gaia-Collaboration_2016_11_0}. Many more will be discovered with the improved precision of Gaia \citep[see e.g.][]{Kervella_2019_03_1,Kervella_2019_03_0}. Speckle imaging is also a well-established technique for obtaining diffraction-limited images and to monitor orbits of binary stars, providing an overlap with long baseline interferometry \citep[see e.g.][]{Horch_2020_05_3,Tokovinin_2016_11_0}. Interferometric observing techniques allow for astrometric measurements of binary stars by probing a different spatial scale, well below the diffraction limit of a single-dish telescope. Very high spatial resolution can be explored with optical long-baseline interferometry \citep[LBI, see e.g.][]{Lawson_2000_08_1}, down to a few milli-arcsecond (mas). LBI enables the astrometric detection of close-in binaries \citep[$< 20$\,mas, see e.g.][]{Hummel_1994_05_2,Hummel_2001_03_0,Zhao_2007_04_0,Konacki_2010_08_0}, with such a small angular resolution so that more complex systems can be observed, such as interacting binaries \citep{Zhao_2008_09_0} or an eclipsing system with one component enshrouded in an accretion disk \citep{Kloppenborg_2010_04_0}. LBI has now proven its efficiency in terms of angular resolution and accuracy for close-in binary stars \citep[see e.g.][]{Baron_2012_06_0,Le-Bouquin_2013_03_0,Gallenne_2013_04_0,Gallenne_2014_01_0,Gallenne_2015_07_0,Gallenne_2016_02_0,Gallenne_2018_11_0,Pribulla_2018_08_0,Gardner_2018_03_0,Gallenne_2019_10_0,Lester_2020_08_8,Gardner_2021_01_5}, and it can now reach a few $\mu$as in astrometric precision.
	
	In this series of papers, we use LBI to observe simple double-lined spectroscopic binary (SB2) systems, that is non-interacting binary systems. This way, the systems are free of any modelling assumptions, and our main objective is to provide very precise and accurate mass and distance measurements. We have shown in our previous studies \citep{Gallenne_2016_02_0,Gallenne_2019_10_0} that mass accuracies as high as 0.05\,\% can be achieved combining RVs and LBI. The mass is a fundamental parameter in order to understand the structure and evolution of stars, and very precise measurements are necessary to check the consistency with theoretical models and to tighten the constraints. For now, stellar parameters (e.g. the effective temperature and radius) predicted from different stellar evolution codes can lead to discrepancies with the empirical values, and therefore provide a large range of possible ages for a given system \citep[see e.g.][]{Torres_2010_02_0,Gallenne_2016_02_0}. Stellar interior models differ in various ways, for instance in input physics, initial chemical compositions, their treatment of convective-core overshooting, and using  rotational mixing or the mixing length parameter \citep{Marigo_2017_01_0,Bressan_2012_11_0,Dotter_2008_09_8,Pietrinferni_2004_09_0}. With high-precision measurements, evolutionary models can be tightly constrained and provide a better understanding of stellar interior physics, enabling the calibration of the physics within evolutionary models. Recent work has shown that a significant improvement in precision on the stellar mass $<< 1$\,\% is necessary to obtain reliable determinations of the stellar interior model parameters \citep[overshooting, initial helium abundance, etc., see e.g.][]{Valle_2017_04_0,Claret_2018_06_0,Higl_2018_09_0}.
	
	Our previous works combining SB2 systems with LBI \citep{Gallenne_2016_02_0,Gallenne_2019_10_0} also demonstrate that very precise and accurate distances can be measured, down to 0.35\,\%. The knowledge of distances is important in many fields of astrophysics. Gaia is revolutionising the field with precise parallax measurements, but it still suffers from some calibration or systematic biases \citep{Lindegren_2021_05_7}. Independent, precise, and accurate geometric distance measurements from binary systems provide a unique benchmark on which to test Gaia parallaxes \citep[see e.g.][]{Southworth_2022_07_0,Graczyk_2021_05_9,Gallenne_2019_10_0}. 
	
	In this paper, we report new spectroscopic and interferometric observations of ten binary systems, including three eclipsing ones. In Sect.~\ref{section__observations_and_data_reduction}, we describe the observations and the data reduction methods. The main improvement since our previous works is the use of the  GRAVITY beam combiner of the Very Large Telescope Interferometer (VLTI) to perform relative astrometry. We previously used the VLTI/PIONIER instrument \citep[Precision Integrated-Optics Near-infrared Imaging ExpeRiment,][]{Le-Bouquin_2011_11_0}, which limited the accuracy of any dimensional measurement to 0.35\,\% \citep{Gallenne_2018_08_0} due to the internal calibration of the instrument wavelength scale. Thanks to a dedicated internal reference laser source, the GRAVITY instrument can provide a much better accuracy, down to 0.02\,\% with high spectral resolution and 0.05\,\% in medium resolution (Gravity collaboration, priv. comm.), which we verified in Sect.~\ref{subsection__astrometric_accuracy_of_interferometric_data} by fitting telluric lines. We present in Sect.~\ref{section__orbit_fitting} our fitting formalism and the results obtained for each system in Sect.~\ref{section__results_for_individual_systems}. We conclude in Sect.~\ref{section__conclusions}.
	
	
	\section{Observations and data reduction}
	\label{section__observations_and_data_reduction}
	
	\subsection{Spectroscopic data \& radial velocities}
	
	We report new observations with the UV Echelle Spectrograph \citep[UVES,][]{Dekker_2000_08_8} mounted on the ESO Very Large Telescope (VLT). We simultaneously used the UVES blue and red arms of the instrument, centred on 390\,nm and 584\,nm with a slit width of 0.4\arcsec\ and 0.3\arcsec, respectively. This provides a resolving power of $\sim 80~000$ and $\sim 110~000$ in the blue and red arm, respectively, covering the spectral range $0.3-0.7\,\mu$m. For each observation, a thorium-argon lamp exposure was taken immediately after each science spectrum in order to achieve the best calibration of a few $\mathrm{m~s^{-1}}$ in precision.
	
	As our stars are bright, each observation has only one exposure of a few seconds. This programme was executed using the 'filler'-type observations, and the stars were observed in non-optimal conditions in terms of seeing, cloud coverage, etc. Despite these limitations, we have spectra of really good quality with a high signal-to-noise ratio (S/N), ranging from $\sim 30$ to $\sim 500$. Raw data were reduced using the UVES pipeline v6.1.6 within the EsoReflex environment \citep{Freudling_2013_11_0}.
	
	To extract the RVs, we used the broadening function (BF) formalism \citep{Rucinski_1999__0} implemented in the \texttt{RaveSpan} software\footnote{\url{https://users.camk.edu.pl/pilecki/ravespan/}} \citep[][see also e.g. \citealt{Pilecki_2018_07_0, Gallenne_2016_02_0, Graczyk_2015_09_0,Pilecki_2013_12_0}]{Pilecki_2017_06_0}. We used templates from the synthetic library of LTE spectra of \citet{Coelho_2005_11_0} in the range $3760-4520\AA, 4625-5600\AA$, and $5675-6645\AA$ (corresponding to the BLUE, REDL, and REDU UVES spectra). For some stars, the S/N was low in the blue part, so we only kept the red part. The templates were chosen to match the atmospheric properties of the hottest star in the systems. They were taken from the literature and are listed in Table~\ref{table__stellar_properties}. The errors of RVs were measured from the broadening function profiles. Our measured UVES RVs are listed in Table~\ref{table__rv}.
	
	We also gathered precise RVs from the literature. In some cases, fitting both the literature's and our RVs degraded the final mass and distance accuracy, either because those from the literature were not precise enough or they had a large scatter, so we only used our RVs whenever necessary. In other cases, we did not have UVES observations and only RVs from the literature were used. Additional observations for some targets were also executed as backup targets by our team with the High-Accuracy Radial-velocity Planet Searcher (HARPS) spectrograph mounted on the 3.6 m telescope at La Silla observatory \citep[][]{Pepe_2002_12_0}. The HARPS spectrograph provides a high spectral resolution of $R \sim 115~000$ in the wavelength range $3800- 6900$ and allows for precise RV measurements. The data were reduced by the instrument data reduction software. Our spectroscopic dataset was supplemented with public HARPS-reduced spectra downloaded from the ESO archive\footnote{\url{http://archive.eso.org/wdb/wdb/adp/phase3_spectral/form}} and public SOPHIE-reduced spectra downloaded from the dedicated database\footnote{\url{http://atlas.obs-hp.fr/sophie/}}. SOPHIE \citep[Spectrographe pour l'Observation des Ph\'enom\`enes des Int\'erieurs stellaires et des Exoplan\`etes,][]{Bouchy_2006_02_0} is also a visible echelle spectrograph providing a resolution $R \sim 40~000$ in high-efficiency mode. All RVs were estimated the same way as described previously. A summary of the RVs used is listed in Table~\ref{table__ref_RVs}.

	\begin{table}[!h]
		\centering
		\caption{Stellar atmospheric parameters of the hottest component used for the spectral templates to estimate the radial velocities.}
		\begin{tabular}{ccccc}
			\hline
			\hline
			Star    &       $T_\mathrm{eff}$        & $\log g$ & [Fe/H]  & Ref. \\
			&       (K)                                                     &                                &                                       &   \\
			\hline
			HD9312          &       5367 & 4.30 & 0.03 & Ki18  \\
			HD41255    &    6150 &  4.10  & $-0.12$  & Ca11 \\
			HD70937    &   6555 & 4.10 & 0.09  & Bo16 \\  
			HD210763  & 6388 & 3.62 & 0.21 & Ca11 \\
			HD224974  &   6232  &  3.93  &  0.17   &  Ca11  \\
			HD188088  &   4868   & 3.50   &  -0.32  & Bo16  \\ 
			$o$~Leo   &  6172       &       3.06    &       -0.06   &  Ad14 \\
			\hline
		\end{tabular}
		\label{table__stellar_properties}
		\tablefoot{Ki18: \citet{Kiefer_2018_02_6}; 
			Bo16: \citet{Boeche_2016_03_1}; 
			Ca11: \citet{Casagrande_2011_06_0}; and
			Ad14: \citet{Adamczak_2014_08_1}.                }
	\end{table}
	
	\begin{table}[!h]
		\centering
		\caption{References for the radial velocity we used in this work.}
		\begin{tabular}{cc}
			\hline
			\hline
			Star    &       Ref. \\
			&                                   \\
			\hline
			AK~For        &  He14, HARPS \\
			HD9312          &       Ki18, UVES  \\
			HD41255    &   UVES, HARPS  \\
			HD70937    &  UVES, HARPS \\  
			HD210763  &  Fe11, UVES \\
			HD224974  & UVES, HARPS \\
			HD188088   &  UVES, HARPS \\  
			LL~Aqr         &  Gr16 \\
			$o$~Leo      &  UVES, SOPHIE \\
			V963~Cen    & Gr22 \\
			\hline
		\end{tabular}
		\label{table__ref_RVs}
		\tablefoot{Fe11:  \citet{Fekel_2011_09_0};
			HARPS: RVs estimated from reduced spectra downloaded from the ESO archive; 
			He14: \citet{Heminiak_2014_07_7}; 
			Ki18: \citet{Kiefer_2018_02_6};
			UVES: RVs estimated from our UVES observations;
			Gr16: \citet{Graczyk_2016_10_0};
			and                        Gr22: \citet{Graczyk_2022_10_0}.
		}
	\end{table}
	
	\subsection{Interferometric data}
	
	Observations were performed with the four-telescope combiner GRAVITY \citep{Eisenhauer_2011_03_0}, the second-generation instrument of the VLTI \citep{Haguenauer_2010_07_0}. As our stars are bright, we used the auxiliary telescopes (ATs) of the VLTI and the single-field mode of the instrument, meaning that the incoming light is split between the fringe tracker (FT) and the science combiner (SC) with a 50\,\%-50\,\% beam splitter. GRAVITY simultaneously combines the light coming from four telescopes in the $K$ band and delivers spectrally dispersed interference fringes for three spectral resolution for the science channel ($R \sim 22, 500,$ and 4000), while the FT only operates at low spectral resolution ($R \sim 22$). The FT allows for a longer integration time on the SC thanks to a real-time analysis of the fringe position to correct for the atmospheric and instrumental piston. The final main observables are the visibilities ($|V|$), squared visibilities ($V^2$), and the closure phases ($CPs$) (the bispectrum amplitude $T3_\mathrm{amp}$ has not been validated in commissioning yet). In some cases, when a spectral feature is expected (e.g. H$\alpha$ emission line for a Be star), differential visibilities and phases are also useful in the spectral model, but not particularly needed to study astrometric orbits.
	
	We observed ten double-lined spectroscopic binaries from 2019 to 2021 in medium and high spectral resolution mode with the medium and large quadruplets, providing baselines up to 140\,m. To monitor the instrumental and atmospheric contributions, we used the standard observational procedure, which consists of observing a reference star before or after the science target. The calibrators, listed in Table~\ref{table__astrometry} and detailed in Table~\ref{table__calibrators}, were selected using the SearchCal\footnote{Available at \url{http://www.jmmc.fr/searchcal}} software \citep{Bonneau_2006_09_0,Bonneau_2011_11_0,Chelli_2016_05_0} provided by the Jean-Marie Mariotti Center (JMMC).
	
\begin{figure*}[!h]
		\centering
		\resizebox{\hsize}{!}{\includegraphics{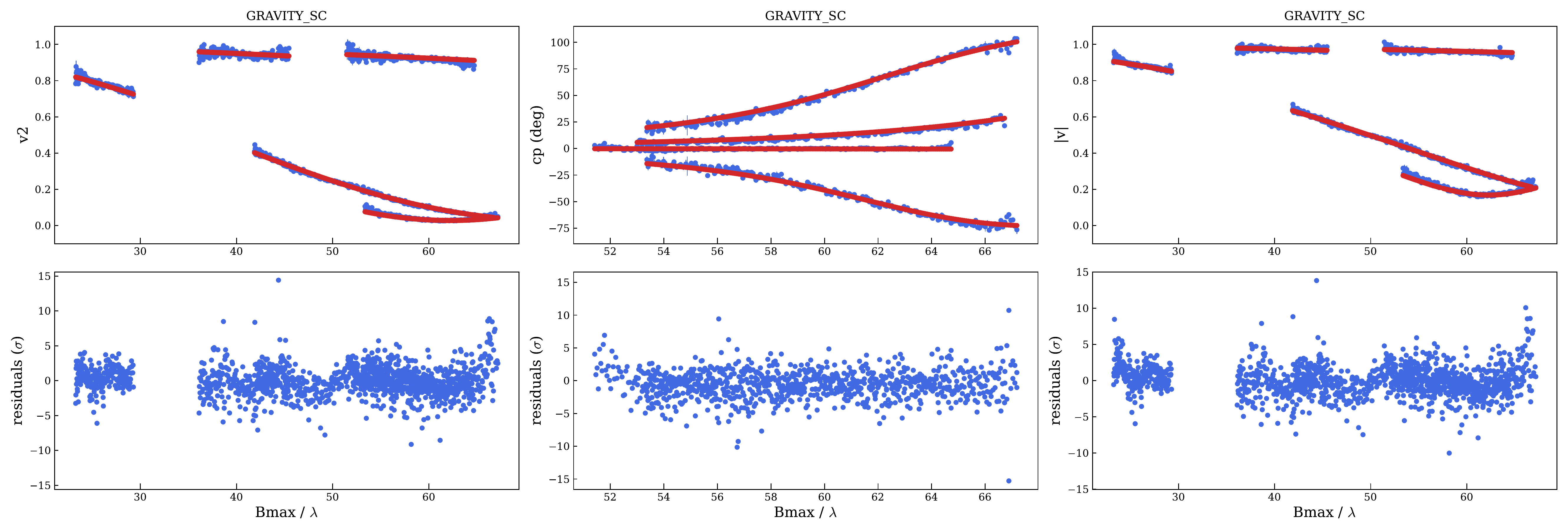}}
		\caption{Squared visibility, closure phase, and visibility measurements from the science combiner for AK~For observed on 2021 November 08. The data are in blue, while the red dots represent the fitted binary model for this epoch. The residuals (in number of sigma) are also shown in the bottom panels.}
		\label{figure__visibility_akfor}
	\end{figure*}

	The data were reduced with the GRAVITY data reduction pipeline described in \citet{Lapeyrere_2014_07_0}. The main procedure is to compute squared visibilities and triple products for each baseline and spectral channel, and to correct for photon and readout noises. In Fig.~\ref{figure__visibility_akfor}, we present an example of the observables for one observation of AK~For. The binary nature of the system is clearly detected.
	
	For each epoch, we proceeded to a grid search to find the global minimum and the location of the companions. For this, we used the interferometric tool \texttt{CANDID}\footnote{Available at \url{https://github.com/amerand/CANDID}} \citep{Gallenne_2015_07_0} to search for the companion using all available observables. \texttt{CANDID} allows for a systematic search for point-source companions performing an $N \times N$ grid of fit, whose minimum required grid resolution is estimated a posteriori. The tool delivers the binary parameters, that is, the flux ratio $f$ and the relative astrometric separation ($\Delta \alpha, \Delta \delta$). \texttt{CANDID} can also fit the angular diameter of both components; however, in our case, we kept them fixed for most systems (nine out of ten) during the fitting process as the VLTI baselines do not allow for reliable measurements of small diameters. For each epoch, \texttt{CANDID} found the global best-fit separation vector. The final astrometric positions for all epochs of all systems are listed in Table~\ref{table__astrometry}. We estimated the uncertainties from the bootstrapping technique (with replacement) and 10~000 bootstrap samples (also included in the \texttt{CANDID} tool). For the flux ratio, we considered the median value and the maximum value between the 16th and 84th percentiles from the
	distributions as uncertainty. For the astrometry, the $1\sigma$ error region of each position $(\Delta \alpha, \Delta \delta)$ is defined with an error ellipse parametrised with the semi-major axis $\sigma_\mathrm{maj}$, the semi-minor axis $\sigma_\mathrm{min}$, and the position angle $\sigma_\mathrm{PA}$ measured from north to east. 
	
	To reach the highest astrometric precision, we only used data from the science combiner because of the systematic uncertainty from the precision of the GRAVITY wavelength calibration. The SC provides a calibration at a sub-percent level, but the FT has a lower precision due to its low spectral resolution and it would limit our final precision on the distance of the binary systems.
	
	As mentioned previously, the primary angular diameters of nine systems are too small to be spatially resolved by the VLTI. We therefore kept them fixed during the grid search with the values given in Table~\ref{table__diam_values}. We note that angular diameters do not significantly affect the measured astrometry (the flux ratio is the most affected). In some cases where there is no binary signature in the CPs, the flux ratio is poorly constrained; we therefore fixed it to the average value estimated from the epoch with strong binary signature  in all observables. \citet{Gallenne_2019_10_0} investigated the effect of fitting or fixing the flux ratio in deriving astrometric positions for very nearby components (i.e. $< \lambda/2B$) and show that the agreement in fixing the flux ratio or not stays within $1\sigma$. The astrometric measurements are reported in Table~\ref{table__astrometry}. 
	
	
	\begin{table}[!ht]
		\centering
		\caption{Angular diameters taken from the literature. Fitted angular diameters are reported in the corresponding sections.}
		\begin{tabular}{cccc}
			\hline
			\hline
			Star    &       $\theta_\mathrm{LD_1}$  & $\theta_\mathrm{LD_2}$ & Ref. \\
			&       (mas) & (mas) &  \\
			\hline
			AK~For & $0.103\pm0.005$ & $0.091\pm0.004$ &1 \\
			HD9312 & $0.525\pm0.005$ & -- & 2  \\
			HD41255 & $0.245\pm0.002$ & -- & 2 \\
			HD70937 & $0.413\pm0.005$ & -- &  2 \\
			HD210763  & $0.367\pm0.009$ & -- & 2 \\
			HD224974  & $0.308\pm0.003$ & -- & 2 \\
			HD188088  & $0.746\pm0.007$  &  -- &  2 \\
			LL~Aqr         & $0.092\pm0.003$ & $0.070\pm0.002$ &3 \\
			$o$~Leo         &  fitted                         & $0.49\pm0.14$ & 4 \\
			V963~Cen  & $0.114\pm0.011$ & $0.111\pm0.010$ &5 \\
			\hline
		\end{tabular}
		\label{table__diam_values}
		\tablefoot{1 - \citet{Heminiak_2014_07_7}; 
			2 -  from the surface brightness relation for dwarf stars of \citet{Kervella_2004_07_0}; 
			3 -  \citet{Graczyk_2016_10_0}; 
			4 - \citet{Hummel_2001_03_0}; and 
			5 - \citet{Sybilski_2018_07_0}.
		}
	\end{table}
	
	\subsection{Astrometric accuracy of interferometric data}
	\label{subsection__astrometric_accuracy_of_interferometric_data}
	
	The accuracy of the interferometric astrometry is dominated by the uncertainties in the knowledge of the spatial frequency $B/\lambda$, where B is the projected baseline and $\lambda$ the wavelength of observations. The VLTI baselines are determined by observing stars all over the sky and recording their absolute fringe position using the delay line laser metrology. This process leads to a baseline scaled to the calibrated metrology laser, which is typically known to 0.02 parts per million. 
	
	The wavelength of the instrument is a much larger source of potential inaccuracy of the spatial frequency scaling. GRAVITY is equipped with a spectrograph calibrated using a dedicated Argon lamp. Because GRAVITY is not a conventional spectrograph, and since spectral data are transformed and resampled \citep{Lapeyrere_2014_07_0}, the resulting spectral accuracy is expected to be of the order of the spectral resolution: in high resolution mode (R$\sim$4000), the spectral accuracy is expected to be $\sim 0.55\,\mu$m (0.025)\,\%, whereas in medium resolution (R$\sim$500), it is expected to be $\sim 4.4\,\mu$m (0.2\,\%). One can verify the final accuracy of the spectral calibration by fitting the telluric features in the GRAVITY spectra. We did this using \texttt{PMOIRED}\footnote{\url{https://github.com/amerand/PMOIRED}} \citep{Merand_2022_08_0} using a synthetic interpolated grid of telluric spectra computed with \texttt{MOLECFIT} \citep{Smette_2015_04_0}. For the AK~For dataset, we find an average wavelength bias per epoch ranging from $\sim-0.006$\,\% to $\sim+0.019$\,\% in both medium and high spectral resolution mode. The spectral calibration is displayed in Fig.~\ref{appendix__specCal}. This confirms that taking the spectral resolution for the accuracy of the spectral calibration is a reasonable choice in high resolution, whereas it is better than expected in medium resolution. We chose to add a 0.02\,\% systematic error to each astrometric measurement for observations for both spectral resolutions.
	
	We also investigated the effect of static optical aberrations on the visibility measurements across the field of view (FoV). This was first assessed with GRAVITY in low resolution mode for the Galactic Centre by \citet{GRAVITY-Collaboration_2021_03_0}, who developed a full analytical model describing the effect of these aberrations on the measurement of complex visibilities. Their analysis has shown that small optical imperfections induce field-dependent phase errors, which can affect the measured binary separation. In addition, misalignments of the injection fibres with respect to the centre of the FoV can also introduce phase errors. They used the GRAVITY calibration units \citep{Blind_2014_07_0} to measure the static aberrations of the science channel and to construct a complex aberration map, which is then used in fitting a binary model to complex visibilities. They present a binary test case with a 200\,mas separation observed with the ATs, and show that accurate astrometry mostly depends on a consistent treatment of the pupil-plane distortions rather than a precise fibre alignment. They also demonstrate that in the specific case of the S2 orbit around the Galactic Centre, phase aberrations introduced a shift up to 0.5\,mas on the separation, which is not negligible. In this paper, our binaries have angular separations substantially smaller than the binary cases reported by \citet{GRAVITY-Collaboration_2021_03_0}. To quantify the effect of static aberrations on smaller binary separations, we performed our own test using the public script provided by the GRAVITY team to create aberration maps\footnote{\url{https://github.com/widmannf/GRAVITY-Phasemaps}} and extract the amplitude, phase error, and intensity at the given position in the fibre. Our tests consisted in comparing different binary star models (i.e. various separations and flux ratios) with and without static aberrations in order to assess an additional systematic uncertainty. We created binary models of two unresolved stars with flux ratios of 5\,\%, 10\,\%, 30\,\%, 50\,\%, and 80\,\%; separations of 2, 5, 10, 15, 20, 30, and 50\,mas; and projection angles of 0, 45, 90, 135, 180, 225, 270, and 315$^\circ$ (we set the typical error of 2\,\% on the visibilities and 0.5$^\circ$ for the closure phases). In addition, for each position $(\Delta \alpha, \Delta \delta),$ we added a random offset in the range $[-0.3,0.3]$\,mas at each telescope (from a uniform distribution) to take possible fibre misalignments into account. We chose this range because it corresponds to about twice the mean value of all offsets of our entire dataset\footnote{The offset of the fibres can be extracted from the header of the fits files with the keyword \texttt{ESO QC MET SOBJ DRAX} and \texttt{ESO QC MET SOBJ DDECX} where X = 1, 2, 3, or 4.}. We then fitted all models around the expected astrometric positions with the flux ratios kept fixed and compared the fitted positions with the expected ones. We report the standard deviation of the residual in $\Delta \alpha$ and $\Delta \delta$ of all fits in Table~\ref{table__aberation_map_results} (i.e. the standard deviation of the fitted minus expected values). We see that fibre misalignments slightly contribute to the error and optical aberrations have more of an impact on low flux ratios, and they decrease with increasing flux ratios. For our binary systems, we have flux ratios $> 8$\,\%, and therefore the additional errors due to optical aberrations and fibre misalignments would be $< 50\,\mu$as. We quadratically added this error to our astrometric measurements according to the flux ratio of a given system, that is 50, 20, 12, and 8\,$\mu$as for $f \sim$ 10, 30, 50, and 80\,\%, respectively.

	\begin{table}[!ht]
		\centering
		\caption{Standard deviation of the residuals between the fitted and expected positions of our synthetic companions.}
		\begin{tabular}{ccc}
			\hline
			\hline
			Flux ratio &    No fibre offsets        & With fibre offsets \\
			\hline
			\multirow{2}{*}{5\,\%} & $\sigma_{\Delta \alpha} = 94\,\mu$as &  $\sigma_{\Delta \alpha} = 103\,\mu$as   \\
			&  $\sigma_{\Delta \delta} = 84\,\mu$as  &  $\sigma_{\Delta \delta} = 94\,\mu$as  \\
			\hline
			\multirow{2}{*}{10\,\%} & $\sigma_{\Delta \alpha} = 48\,\mu$as &  $\sigma_{\Delta \alpha} = 53\,\mu$as   \\
			&  $\sigma_{\Delta \delta} = 42\,\mu$as  &  $\sigma_{\Delta \delta} = 46\,\mu$as  \\
			\hline
			\multirow{2}{*}{30\,\%} & $\sigma_{\Delta \alpha} = 18\,\mu$as &  $\sigma_{\Delta \alpha} = 19\,\mu$as  \\
			&  $\sigma_{\Delta \delta} = 16\,\mu$as  &  $\sigma_{\Delta \delta} = 17\,\mu$as  \\
			\hline
			\multirow{2}{*}{50\,\%} & $\sigma_{\Delta \alpha} = 12\,\mu$as &  $\sigma_{\Delta \alpha} = 12\,\mu$as  \\
			&  $\sigma_{\Delta \delta} = 10\,\mu$as  &  $\sigma_{\Delta \delta} = 10\,\mu$as  \\
			\hline
			\multirow{2}{*}{80\,\%} & $\sigma_{\Delta \alpha} = 8\,\mu$as &  $\sigma_{\Delta \alpha} = 8\,\mu$as  \\
			&  $\sigma_{\Delta \delta} = 7\,\mu$as  &  $\sigma_{\Delta \delta} = 7\,\mu$as  \\
			\hline
		\end{tabular}
		\label{table__aberation_map_results}
	\end{table}
	
	%
	%
	%
	%
	%

	\section{Orbit fitting}
	\label{section__orbit_fitting}
	
	We used the same formalism as in \citet{Gallenne_2019_10_0}. We simultaneously fitted the RVs and astrometric positions using a Markov chain Monte Carlo (MCMC) routine\footnote{With the Python package emcee developed by \citet{Foreman-Mackey_2013_03_0}.}. The log-likelihood function is defined with the following:
	\begin{displaymath}
		\log(\mathcal{L}) = - \dfrac{1}{2}\,\chi^2,\, \mathrm{with}\,\chi^2 = \chi^2_\mathrm{RV} + \chi^2_\mathrm{ast}.
	\end{displaymath}
	
	Radial-velocity measurements are related to the orbital elements with
	\begin{displaymath}
		\chi^2_\mathrm{RV} =  \sum \dfrac{(V_1 - V_\mathrm{1m})^2}{\sigma_\mathrm{V_1}^2} +  \sum \dfrac{(V_2 - V_\mathrm{2m})^2}{\sigma_\mathrm{V_2}^2}, 
	\end{displaymath}
	in which $V_\mathrm{i}$ and $\sigma_\mathrm{V_i}$ denote the measured RVs and uncertainties for the component $i$.\ In addition, ($V_\mathrm{1m}, V_\mathrm{2m})$ are the Keplerian velocity models of both components, defined by \citep{Heintz_1978__0}
	\begin{eqnarray*}
		V_\mathrm{1m} &=& \gamma + K_1\,[\cos(\omega + \nu) + e\cos{\omega}], \\
		V_\mathrm{2m} &=& \gamma - K_2\,[\cos(\omega + \nu) + e\cos{\omega}], \\
		\tan \dfrac{\nu}{2} &= & \sqrt{\dfrac{1 + e}{1 - e}} \tan \dfrac{E}{2}, \\
		E - e \sin E &=& \dfrac{2\pi (t - T_\mathrm{p})}{P_\mathrm{orb}},
	\end{eqnarray*}
	where $\gamma$ is the systemic velocity, $e$ the eccentricity, $\omega$ the argument of periastron, $\nu$ the true anomaly,  $E$ the eccentric anomaly, $t$ the observing date, $P_\mathrm{orb}$ the orbital period, and $T_\mathrm{p}$ the time of periastron passage. The parameters $K_1$ and  $K_2$ are the RV amplitude of both stars. 
	
	The astrometric measurements are fitted as
	\begin{eqnarray*}
		&\chi^2_\mathrm{ast} =& \chi^2_\mathrm{a} + \chi^2_\mathrm{b}, \\
		&\chi^2_\mathrm{a} =& \sum \frac{[ (\Delta \alpha - \Delta \alpha_\mathrm{m}) \sin \sigma_\mathrm{PA} + (\Delta \delta - \Delta \delta_\mathrm{m}) \cos \sigma_\mathrm{PA} ]^2}{\sigma^2_\mathrm{maj}}, \\
		&\chi^2_\mathrm{b} =& \sum \frac{[ -(\Delta \alpha - \Delta \alpha_\mathrm{m}) \cos \sigma_\mathrm{PA} + (\Delta \delta - \Delta \delta_\mathrm{m}) \sin \sigma_\mathrm{PA} ]^2}{\sigma^2_\mathrm{min}}, \\
	\end{eqnarray*}
	in which $(\Delta \alpha, \Delta \delta, \sigma_\mathrm{PA}, \sigma_\mathrm{maj}, \sigma_\mathrm{min})$ denote the relative astrometric measurements with the corresponding error ellipses, and $(\Delta \alpha_\mathrm{m}, \Delta \delta_\mathrm{m})$ the astrometric model can be defined with
	\begin{eqnarray*}
		\Delta \alpha_\mathrm{m} &=& r \,[ \sin \Omega \cos(\omega + \nu) + \cos i \cos \Omega \sin(\omega + \nu) ], \\
		\Delta \delta_\mathrm{m} &=& r \,[ \cos \Omega \cos(\omega + \nu) - \cos i \sin \Omega \sin(\omega + \nu) ], \\
		r &=& \dfrac{a (1 - e^2)}{1 + e\cos \nu},\\
	\end{eqnarray*}
	where $\Omega$ is the longitude of ascending node, $i$ the orbital inclination, and $a$ the angular semi-major axis.
	
	As a starting point for our 100 MCMC walkers, we performed a least squares fit using orbital values from the literature as first guesses. We then ran 100 initialisation steps to thoroughly explore the parameter space and get settled into a stationary distribution. The prior distributions used are uniform for all parameters and are listed in Table~\ref{table__priors}. For all cases, the chain converged before 50 steps. Finally, we used the last position of the walkers to generate our full production run of 1000 steps, discarding the initial 50 steps. All the orbital elements, that is $P_\mathrm{orb}, T_\mathrm{p}, e, \omega, \Omega, K_1, K_2, \gamma, a$, and $i$, were estimated from the distribution considering the median value and the maximum value between the 16th and 84th percentiles as uncertainty (although the distributions were roughly symmetrical).
	
	\begin{figure*}[!h]
		\centering
		\resizebox{0.8\hsize}{!}{\includegraphics{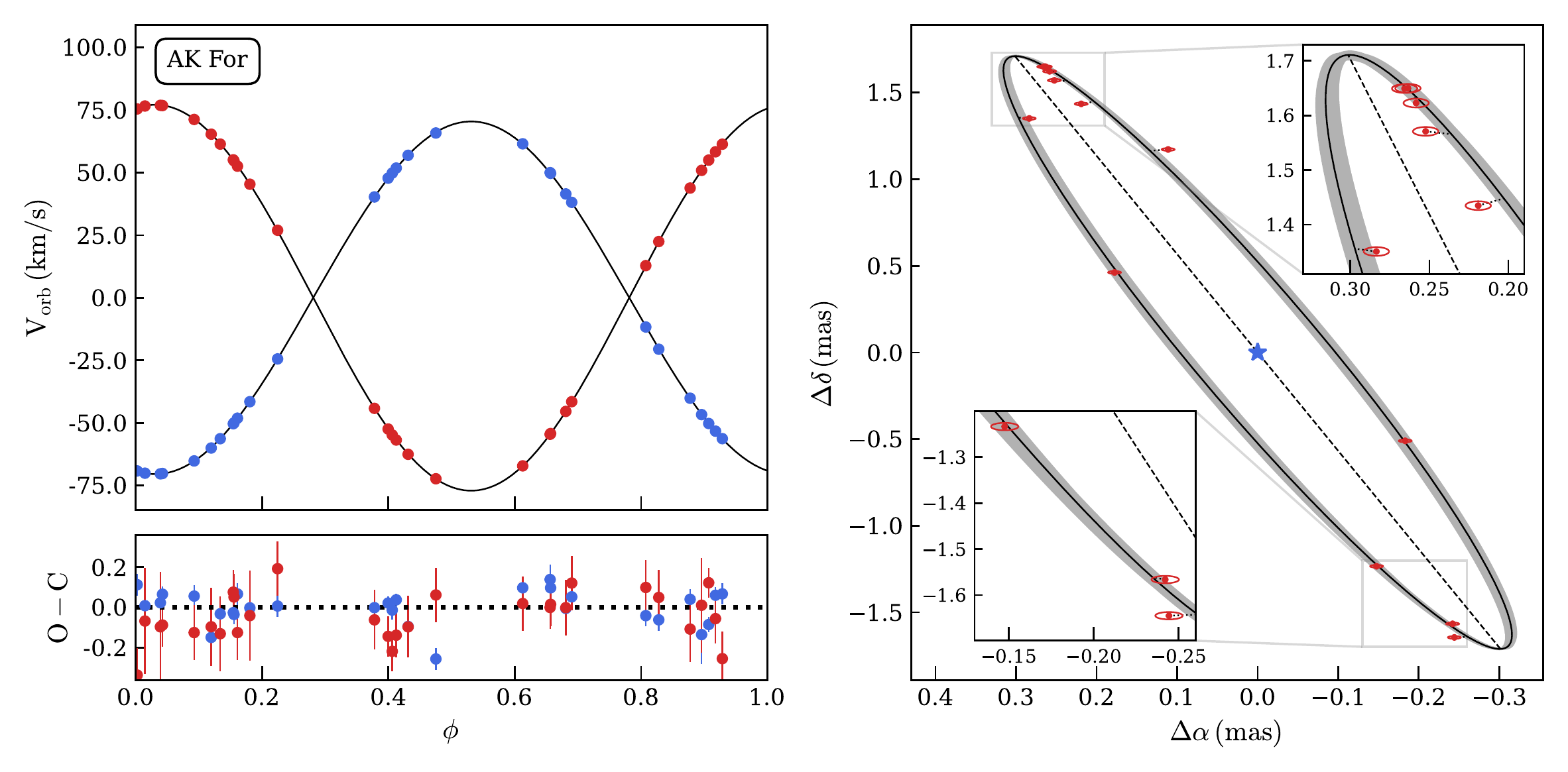}}
		\resizebox{0.8\hsize}{!}{\includegraphics{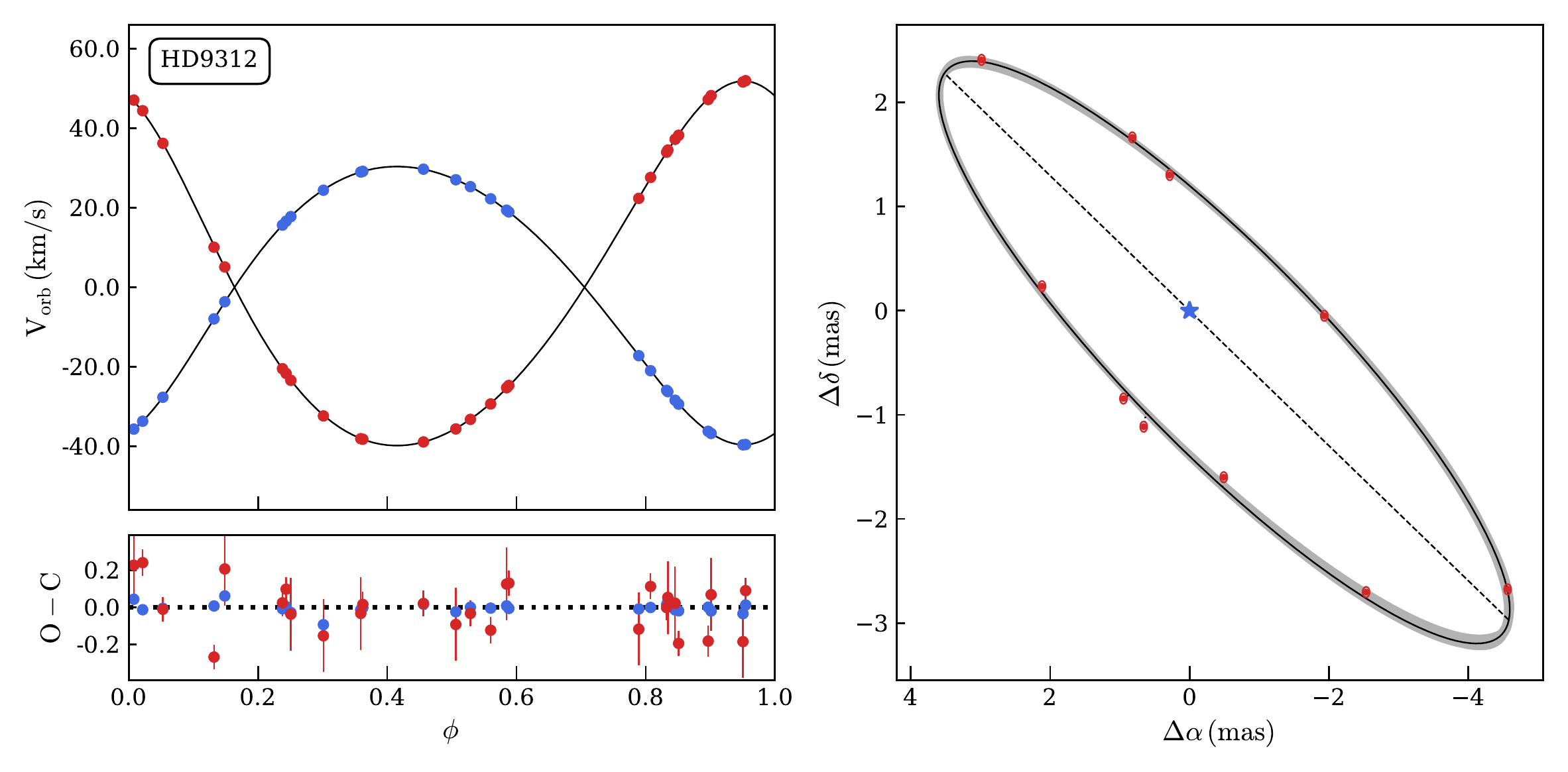}}
		\resizebox{0.8\hsize}{!}{\includegraphics{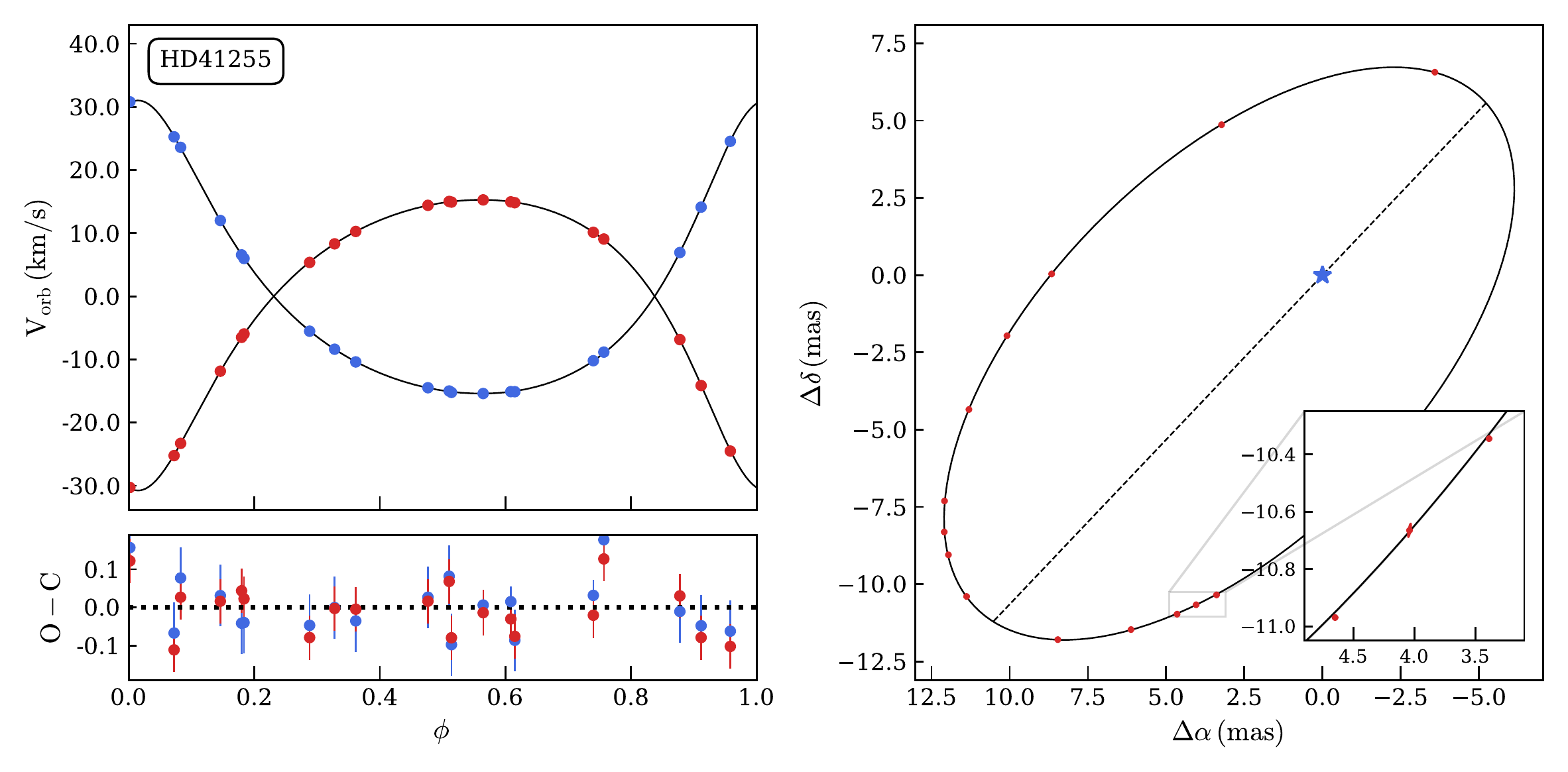}}
		\caption{From top to bottom: Combined fit of AK~For, HD9312, and HD41255. \textit{Left}: Radial velocities of the primary (blue) and the secondary (red) star. \textit{Right}: GRAVITY astrometric orbit. The shaded grey area represents the $1\sigma$ orbit.}
		\label{figure__orbit_akfor_hd9312_hd41255}
	\end{figure*}
	
	\begin{table}[!ht]
		\centering
		\caption{Uniform prior distributions used.}
		\begin{tabular}{cc}
			\hline
			\hline
			Parameters  &  Distributions \\
			\hline
			$P_\mathrm{orb}$ (days)                                                 & [$P_\mathrm{orb,0} - 10$, $P_\mathrm{orb,0} + 10$]      \\
			$T_\mathrm{p}$ (days)                                                     & [$T_\mathrm{p,0} - 10$, $T_\mathrm{p,0} + 10$]  \\
			$e$                                                                                                                                & [0, 1]      \\
			$K_1$ (\kms)                                                                                             & [$K_{1,0} - 10$, $K_{1,0} + 10$] \\
			$K_2$ (\kms)                                                                                        & [$K_{2,0} - 10$, $K_{2,0} + 10$]                          \\
			$\gamma_1$      (\kms)                                                               & [$\gamma_{1,0} - 10$, $\gamma_{1,0} + 10$]         \\
			$\gamma_2$      (\kms)                                                              & [$\gamma_{1,0} - 10$, $\gamma_{1,0} + 10$]  \\
			$\omega$        ($\degr$)                                                                 & [0, 360]     \\
			$\Omega$        ($\degr$)                                                                & [0, 360]       \\
			$a$ (mas)                                                                                                         & [0,300]        \\
			$i$ ($\degr$)                                                                                            &  [0, 180]      \\
			\hline
		\end{tabular}
		\label{table__priors}
		\tablefoot{Parameters with index 0 are the results from the least squares fit using orbital values from the literature.}
	\end{table}
	
	The distributions of the mass of both components and the distance were derived from the MCMC distributions with \citep{Torres_2010_02_0}
	\begin{eqnarray*}
		M_1 &=& \dfrac{1.036149\times 10^{-7} (K_1 + K_2)^2 K_2\,P\,(1 - e^2)^{3/2} }{\sin ^3 i},\\
		M_2 &=& \dfrac{1.036149\times 10^{-7} (K_1 + K_2)^2 K_1\,P\,(1 - e^2)^{3/2} }{\sin ^3 i},\\
		a_\mathrm{AU} &=& \dfrac{9.191940\times 10^{-5} (K_1 + K_2)\,P \sqrt{1 - e^2} }{\sin i}, \\
		d &=& \dfrac{a_\mathrm{AU}}{a},
	\end{eqnarray*}
	where the masses are expressed in solar units, the distance in parsec, $K_1$ and $K_2$ in \kms, $P$ in days, and $a$ in arcsecond. The parameter $a_\mathrm{AU}$ is the linear semi-major axis expressed in astronomical units (the constant value of \citet{Torres_2010_02_0} is expressed in solar radii, and was converted using the astronomical constants $R_\odot = 695.658 \pm 0.140 \times 10^6$\,m from \citealt{Haberreiter_2008_03_0} and $AU = 149~597~870~700 \pm 3$\,m from \citealt{Pitjeva_2009_04_0}). As was previously done, we then took the median value and the maximum value between the 16th and 84th percentiles as uncertainty. The fitting results are presented in the next section for all systems. 
	
	\section{Results for individual systems}
	\label{section__results_for_individual_systems}
	
	
	\subsection{AK~Fornacis}
	
	This is a low-mass eclipsing binary system composed of two similar main-sequence stars in a 3.98\,d orbit. \citet{Heminiak_2014_07_7} studied this system using both RVs and photometry during eclipses and derived very precise stellar parameters. We used the RVs they reported (from the CORALIE and FEROS spectrographs) and we complemented the dataset with HARPS data. A zero-point offset between our RVs and those from \citet{Heminiak_2014_07_7} were determined by fitting each dataset independently, and taking the difference in the systemic velocities. We found an offset of $+261\,$\ms which we added to the HARPS RVs. Our final parameters measured from our combined astrometric and RVs' orbital fit are listed in Table~\ref{table__orbit_fit}, and the orbit is displayed in Fig.~\ref{figure__orbit_akfor_hd9312_hd41255}. We reached a r.m.s. of the astrometric orbit of $11\,\mu$as. 
	
	Our measured masses have a precision $\leqslant 0.12$\,\%. They are in excellent agreement with the values derived by \citet{Heminiak_2014_07_7} at $<1\sigma$, demonstrating that our measurements are both precise and accurate. We measured a distance to the system with an accuracy level of 0.17\,\%, and in agreement at $0.7\sigma$ with the last Gaia data release (here and in the following, we applied
	a zero-point offset following the correction from \citealt{Lindegren_2021_05_7}). We also measured an average flux ratio $f_\mathrm{K} = 71.7\pm1.6$\,\% between the two components.
	
	\subsection{HD9312}
	
	This SB2 system has two similar stars orbiting each other in a $36.5$\,d orbit. It was mainly studied using spectroscopy \citep[see e.g.][]{Katoh_2021_02_7,Kiefer_2018_02_6,Halbwachs_2014_12_2}, and no individual mass has been measured, but a mass ratio was derived to be $q = 0.7624\pm0.0015$ from the cross-correlation function (CCF) of several spectra \citep[][]{Halbwachs_2014_12_2}. The SB2 orbit was only published recently by \citet{Kiefer_2018_02_6}. \citet{Wang_2015_10_9} used an iterative method to determine self-consistent orbital solutions via a combined fit of the SB1 orbit and the Hipparcos Intermediate Astrometric Data, but their estimated mass ratio of 0.88 is larger than the one estimated by \citet{Halbwachs_2014_12_2}, although with a large uncertainty of $\sim 45$\,\%. Combining their fitted stellar parameters with evolutionary tracks, they show that the primary is on the subgiant giant branch while the secondary is on the main sequence.
	
	Our combined fit with astrometry provides measured masses with a precision level of 0.3\,\%. It is displayed in Fig.~\ref{figure__orbit_akfor_hd9312_hd41255}. The r.m.s. of the astrometric orbit is $\sim 50\,\mu$as. We corrected for a RV offset between our UVES RVs and \citet{Kiefer_2018_02_6} of $+0.304$\,\kms. Our mass ratio is in very good agreement with the value derived by \citet{Halbwachs_2014_12_2} at the $< 0.1\sigma$ level. The spectroscopic orbital parameters estimated by \citet{Wang_2015_10_9} are in agreement with our values, but their inclination is not consistent and it is $\sim3\sigma$ smaller. The other astrometric orbital parameters, $a$ and $\Omega$, are also quite different, with an $\sim 1.5$\,mas difference for the semi-major axis. Their derived masses have a relative uncertainty of $\sim 30$\,\%, meaning that they are in agreement within $1\sigma$ with our measurements. The method applied by \citet{Wang_2015_10_9} is not a direct measurement; it needs evolutionary models to estimate the primary mass and a mass luminosity relation for the secondary mass. This may explain the low precision on the masses.
	
	We measured the distance to the system with a precision of 0.6\,\% and this is in agreement with the Gaia measurement at $\sim 1.8\sigma$. Our interferometric observations also provide a $K$-band flux ratio of $7.84\pm0.49$\,\%.
	
	\subsection{HD41255}
	
	The spectroscopic binary nature was only recently discovered by the Geneva-Copenhagen survey of the Solar neighbourhood \citep[GCS,][]{Holmberg_2009_07_3}, but no orbital parameters were determined. The double-line nature was detected by \citet{Gorynya_2014_07_5} and they estimate a preliminary orbital period of 163\,d and a mass ratio $q = 0.98$. The system is composed of two similar stars whose primary is a F8/G0V star. The spectroscopic orbital parameters were later determined by \citet{Gorynya_2018_03_9} with an updated period of 148.3\,d. They estimate a semi-major axis of 11\,mas and predict an orbital inclination around $53^\circ$, which are very close to our measured values listed in Table~\ref{table__orbit_fit}. 
	
	We obtained a very precise orbit, as we can see in Fig.~\ref{figure__orbit_akfor_hd9312_hd41255}, which enabled us to measure the mass of both components at a 0.21\,\% precision level. The stars have an equal mass with a ratio $q = 1.008\pm0.003$, and they are solar-type stars. This is also consistent with the average $K$-band flux ratio $f_\mathrm{H} = 95.9\pm1.0$\,\% we determined from GRAVITY. We also measured the distance to the system at a 0.08\,\% precision that is at $2.1\sigma$ to the Gaia estimate. We reached an unprecedented astrometric orbit r.m.s. of $\leq 11\,\mu$as.
	
	\subsection{HD70937}
	
	\begin{figure*}[!h]
		\centering
		\resizebox{0.8\hsize}{!}{\includegraphics{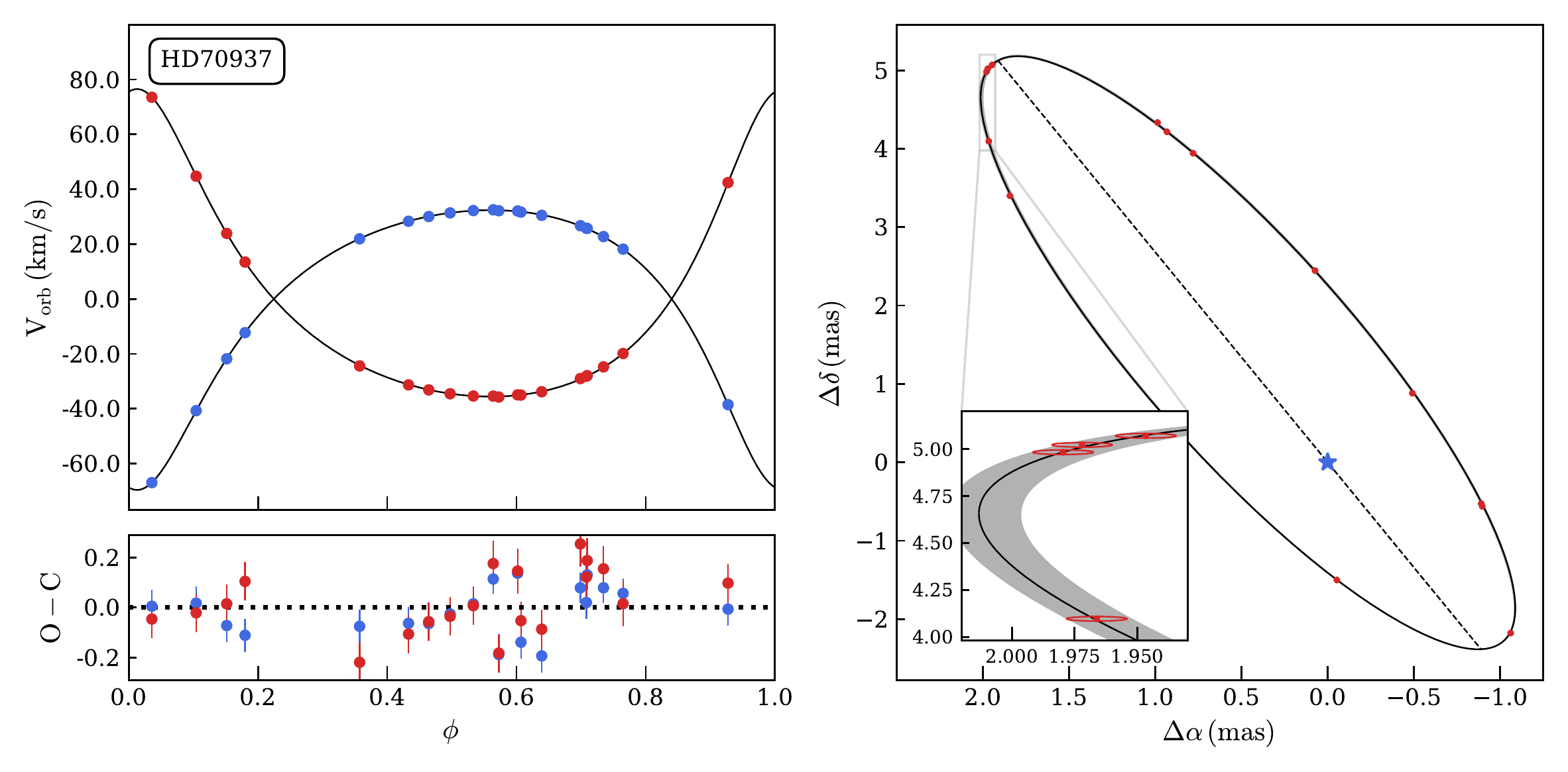}}
		\resizebox{0.8\hsize}{!}{\includegraphics{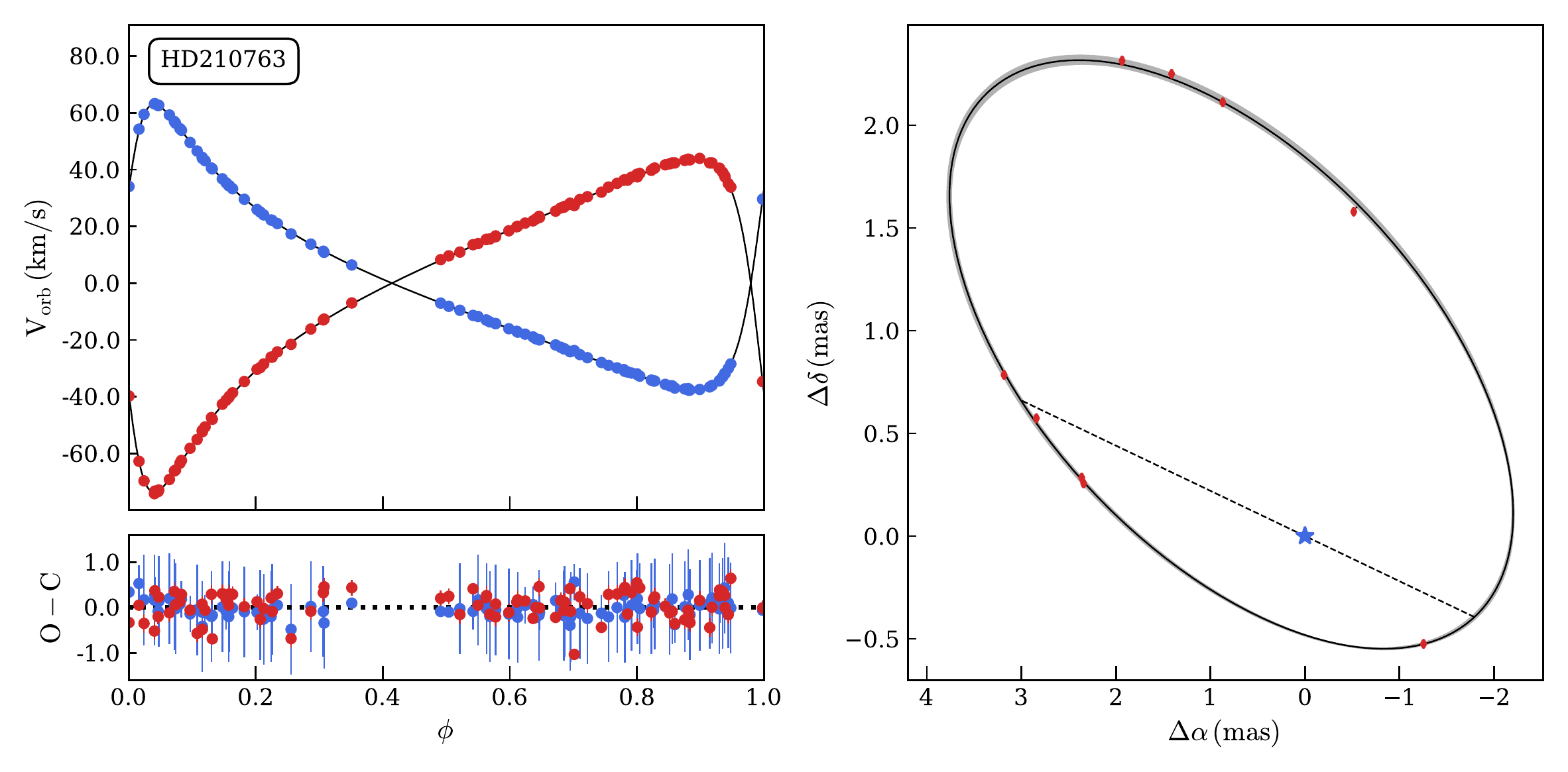}}
		\resizebox{0.8\hsize}{!}{\includegraphics{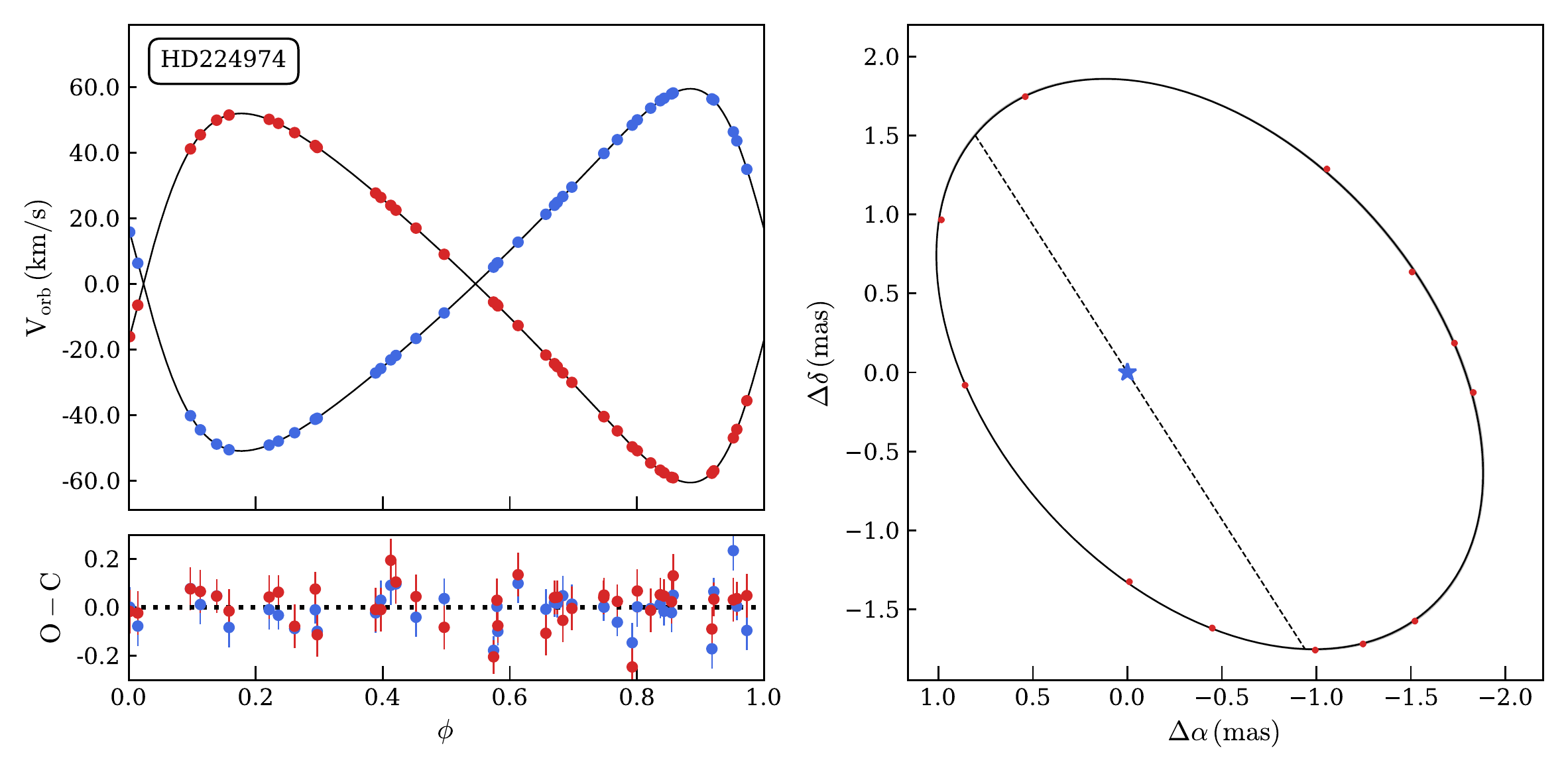}}
		\caption{Same as Fig. \ref{figure__orbit_akfor_hd9312_hd41255}, but for the system HD70937, HD210763, and HD224974 from top to bottom, respectively.}
		\label{figure__orbit_hd70937_hd210763_hd224974}
	\end{figure*}
	
	This is a 28\,d SB2 binary whose full spectroscopic orbital elements have been recently determined by \citet{Gorynya_2018_03_9}. They measured a mass ratio of $0.77\pm0.12$ from the RV semi-amplitudes, while \citet{Nordstrom_2004_05_9} estimated $q = 0.862$. The primary star was classified as a F5 main-sequence star by \citet{1999MSS...C05....0H} and no additional information is known about this system.
	
	Our precise orbit, displayed in Fig.~\ref{figure__orbit_hd70937_hd210763_hd224974}, allows for the orbital parameters to be fully determined (listed in Table~\ref{table__orbit_fit}), and provides a mass ratio of $0.9099\pm0.0018$ which is more in agreement with the \citet{Nordstrom_2004_05_9} value. We measured a semi-major axis of $\sim 4$\,mas and this explains the undetected companion from speckle interferometry \citep{Horch_2020_05_3,Horch_2015_11_1,Hartkopf_2012_02_8}. We found that the stars have similar masses around $1.5\,\mathrm{M_\odot}$, with an average flux ratio in the $H$ band of $59.7\pm1.0$\,\%. We measured the mass with an accuracy of 0.14\,\% and the distance at 0.11\,\%, which is in good agreement with Gaia at $0.7\sigma$. We reached an astrometric orbit r.m.s. $\lesssim 10\,\mu$as.
	
	\subsection{HD210763}
	
	The first RV observations of this system were performed at Mount Wilson Observatory and published by \citet{Wilson_1950_03_1} who already noticed the variability of the velocities. The double-line signature was later detected by \citet{Nadal_1983_05_9} who determined the first spectroscopic orbit. With new observations from the Observatoire de Haute Provence, they determined a period of 42.4\,d and an eccentricity of 0.616. The spectral type of the primary is between F8 IV and F6 V. \citet{Fekel_2011_09_0} revised the spectroscopic orbit with extensive new and more precise RVs. They also classified the spectral type of the two components to be F6 V and F6 IV.  They measured a mass ratio $q = 0.855\pm0.003$. We used their RVs together with ours derived from the UVES spectra. We corrected our RVs by $-0.1$\kms due to a RV offset.
	
	We displayed our combined fit in Fig.~\ref{figure__orbit_hd70937_hd210763_hd224974}. We have a very precise astrometric orbit with a r.m.s. of $\sim15\,\mu$as. The masses were measured with an accuracy of 0.26\,\%, with the same mass ratio as \citet{Fekel_2011_09_0}. Our measured orbital parameters are listed in Table~\ref{table__orbit_fit}. We measured a $K$-band flux ratio of $37.1\pm0.8$\,\%, and an orbital parallax of $10.691\pm0.037$\,mas. We reached a similar precision as Gaia, which is in disagreement by $2.3\sigma$ with our measurements.
	
	\subsection{HD224974}
	
	
	This system is composed of twin main-sequence stars orbiting each other with a 10.7\,d period. The full spectroscopic orbital solutions were determined by \citet{Gorynya_2014_07_5}, who derived a mass ratio of $q = 0.982\pm0.003$. This is consistent with the previous estimate of $1.000\pm0.011$ from the GCS. 
	
	We present in Fig.~\ref{figure__orbit_hd70937_hd210763_hd224974} our combined fit using only our new RVs determined from UVES and HARPS spectra. Our astrometric orbit is precise at a $11\,\mu$as level. We measured the masses with an accuracy of 0.5\,\%, providing a mass ratio $q = 0.981\pm0.007$, which is in very good agreement with \citet{Gorynya_2014_07_5}. Our measured distance is accurate at 0.3\,\% and at $2.1\sigma$ with the Gaia value. The fitted parameters are listed in Table~\ref{table__orbit_fit}. We also measured an average $K$-band flux ratio of $90.0\pm0.9$\,\%.
	
	\subsection{HD188088}
	
	\begin{figure*}[!h]
		\centering
		\resizebox{0.8\hsize}{!}{\includegraphics{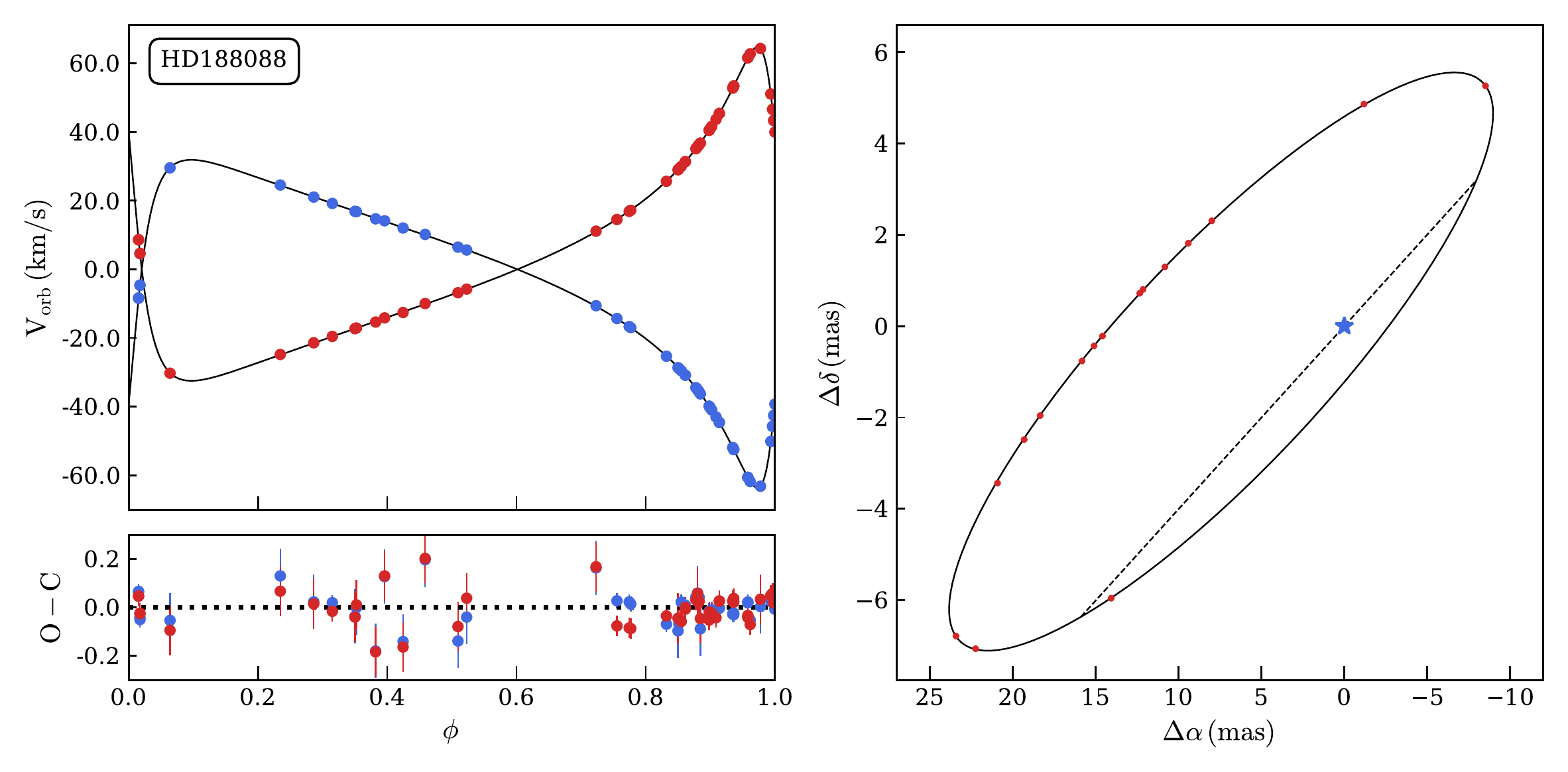}}
		\resizebox{0.8\hsize}{!}{\includegraphics{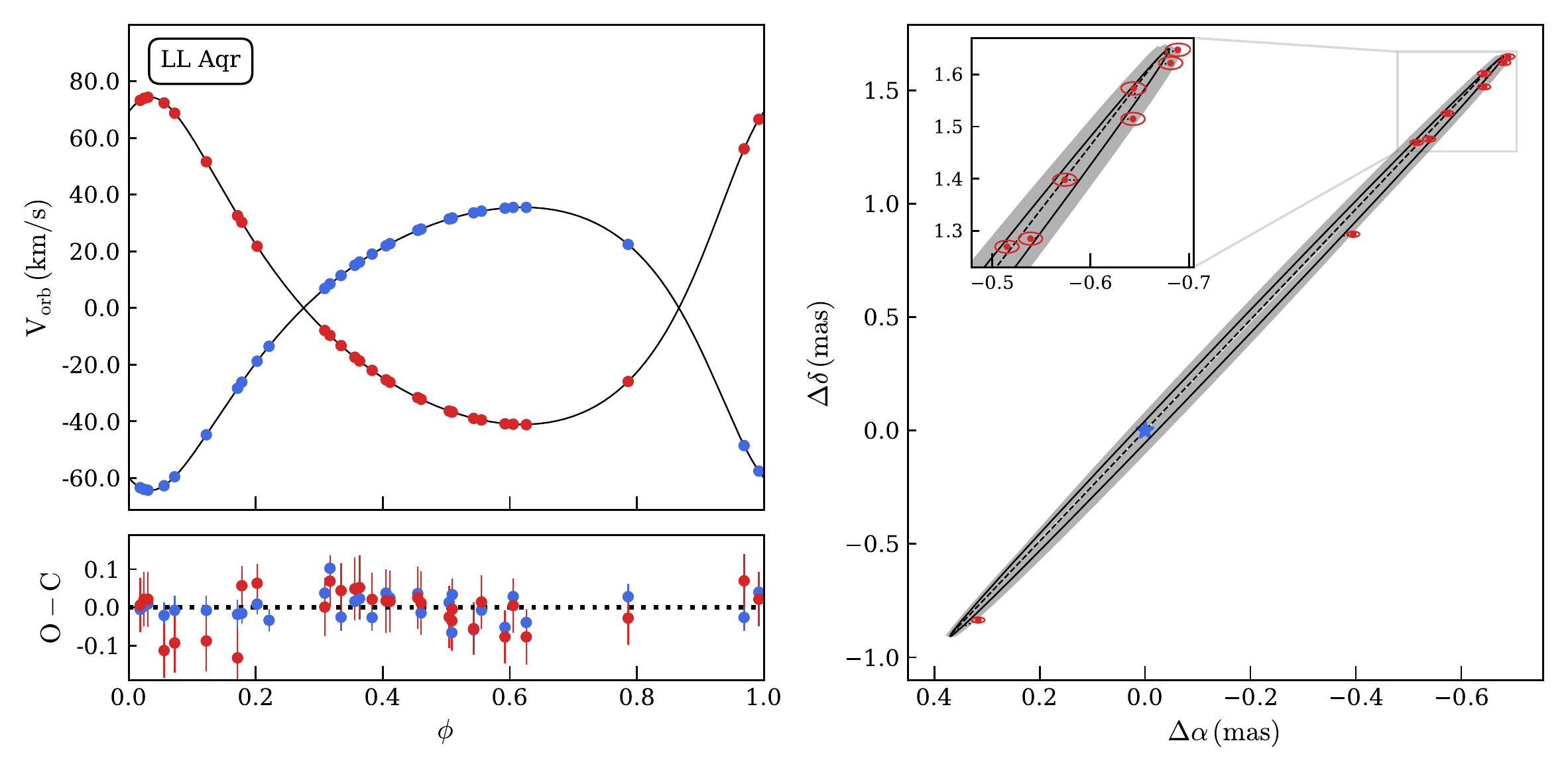}}           
		\resizebox{0.8\hsize}{!}{\includegraphics{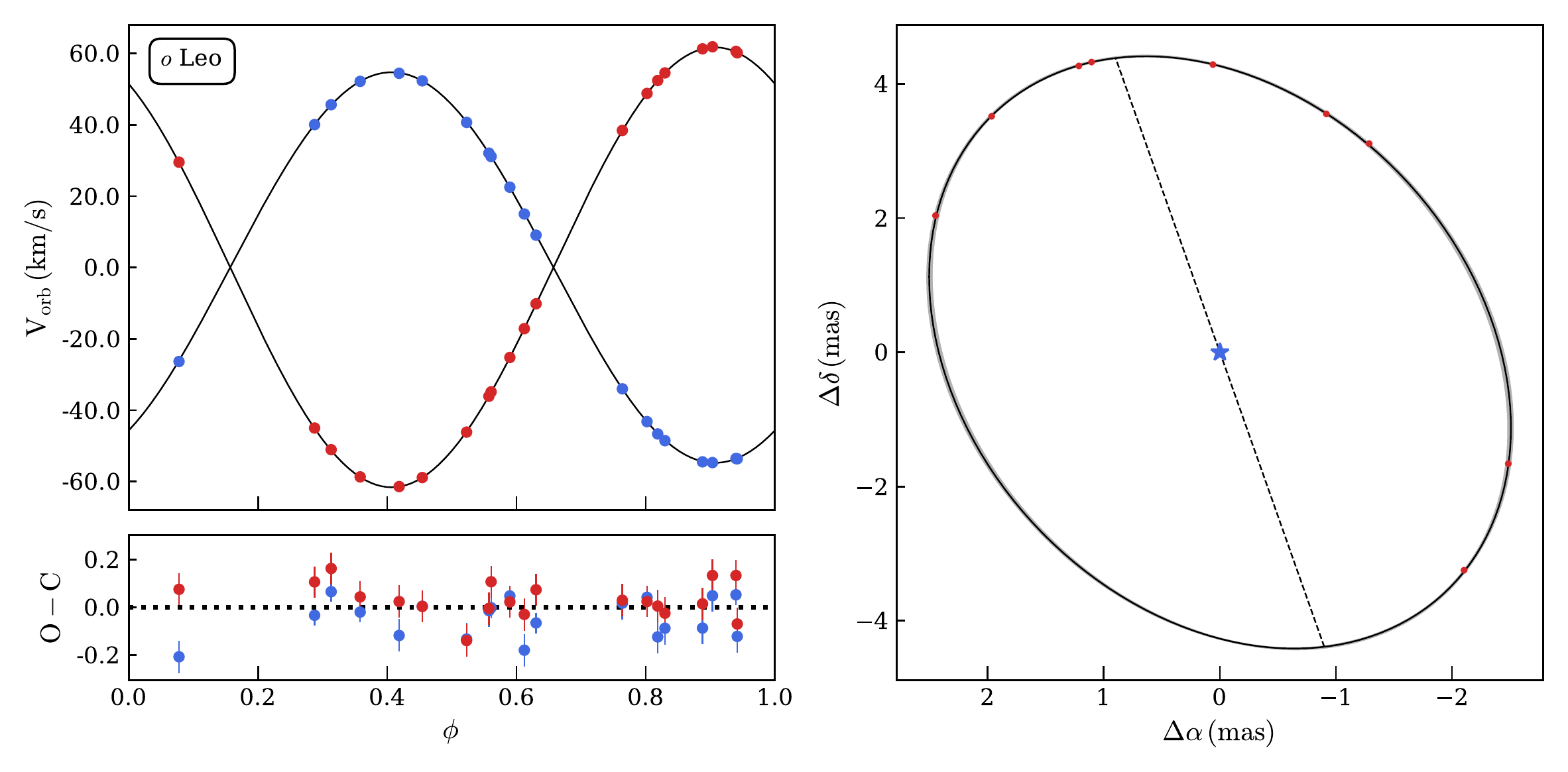}}
		\caption{Same as Fig. \ref{figure__orbit_akfor_hd9312_hd41255}, but for the system HD188088, LL~Aqr, and $o$~Leo from top to bottom, respectively.}
		\label{figure__orbit_hr7578_llaqr_omileo}
	\end{figure*}
	
	
	This star is a BY Dra variable and a member of a triple system containing an inner spectroscopic binary and an outer M5 companion. The third component is, however, located at about 41\arcsec \citep{Allen_2012_08_7,Chini_2014_01_2} and therefore  has a negligible impact on the inner system we are studying here. The spectroscopic components have a similar brightness with spectral type K3V and orbit each other in 46.8\,d. The first spectroscopic observations were acquired by \citet{Evans_1968_12_8} who suggested the binary nature of the system. This was later confirmed by \citet{Fekel_1983_04_5} who detected the lines of both components and determined the first spectroscopic orbital solutions. They were then refined by \citet{Fekel_2017_09_0} with new and more precise RVs. They estimated a minimum mass of $0.86\,M_\odot$ for both stars.
	
	Our orbital fit is displayed in Fig.~\ref{figure__orbit_hr7578_llaqr_omileo} and the solutions are listed in Table~\ref{table__orbit_fit}. We only used our more precise and uniform HARPS and UVES observations to avoid additional offsets (there is no improvement in precision by adding the RVs from \citet{Fekel_2017_09_0}). A zero-point offset of $+0.039$\,\kms was added to the HARPS measurements. We reached a final r.m.s. of the orbit of $\sim 18\,\mu$as. We measured the masses and the distance with a precision of $\sim0.03$\,\%. Our mass ratio is in agreement at $0.05\sigma$ with \citet{Fekel_2017_09_0}. The Gaia parallax is at $3.0\sigma$ with our value and we reached a better accuracy. We also measured an average $K$-band flux ratio of $90.1\pm1.6$\,\%.
	
	\subsection{LL~Aqr}
	
	This is a well-studied detached eclipsing system composed of main-sequence stars \citep[F9V + G3V,][]{Graczyk_2016_10_0} orbiting each other in 20.2\,d. First discovered as being variable by the Hipparcos mission, the first combined photometric and RV solution was obtained by \citet{Ibanoglu_2008_11_0}. The orbital parameters and absolute dimensions were later refined with new and more precise measurements. The latest data set was provided by \citet{Graczyk_2016_10_0}, who determined the masses with a precision of 0.06\,\% and a distance precise at 3\,\%. We used their RVs in our combined fit.
	
	Our orbital solutions are displayed in Fig.~\ref{figure__orbit_hr7578_llaqr_omileo} and the fitted parameters are listed in Table~\ref{table__orbit_fit}. We reached the same level of precision as \citet{Graczyk_2016_10_0} for the masses. In addition, our values are in very good agreement with theirs at the $< 0.3\sigma$ level, demonstrating the accuracy of the measurements as well. Our distance has a better accuracy with 0.27\,\%, but it is still in agreement with their value. Our distance uncertainty is similar to Gaia and is in agreement at $0.9\sigma$. We also measured an average $K$-band flux ratio of $53.3\pm0.7$\,\%.

	\subsection{$o$~Leo}
	
	
	This system is composed of a F8-G0 giant star and a hotter A7m III-IV companion, as identified by \citet{Ginestet_2002_12_9}. \citet{Hummel_2001_03_0} presented the first three-dimensional solution by combining photoelectric RVs and astrometry from interferometry. They measured $M_1 = 2.12\pm0.01\,M_\odot$, $M_2 = 1.87\pm0.01\,M_\odot$, and $d = 41.4\pm0.1$\,pc. The composite spectra were more deeply studied by \citet{Griffin_2002_02_8} who determined effective temperatures of 7600\,K and 6100\,K for the giant and dwarf component, respectively. \citet{Piccotti_2020_02_0} updated the orbit with new RVs from \citet{Massarotti_2008_01_7} and measured $M_1 = 2.093\pm0.068\,M_\odot$ and $M_2 = 1.857\pm0.058\,M_\odot$.
	
	We obtained new UVES spectroscopic and interferometric data to improve the solutions and to obtain more precise masses and distances. We also complemented the RVs with archived SOPHIE spectra. We found a zero-point offset of $-0.098$\kms for SOPHIE . As we can see in Fig.~\ref{figure__orbit_hr7578_llaqr_omileo}, we obtained a very precise orbit, with an average r.m.s. of $\sim 10\,\mu$as, demonstrating that the 20\,$\mu$as systematic error we added (see Table~\ref{table__aberation_map_results}) is probably too conservative. We measured the masses with a precision of 0.6\,\%, which are in agreement at $< 0.3\sigma$ with \citet{Piccotti_2020_02_0}. However, they differ from \citet{Hummel_2001_03_0} by $\sim2-3\sigma$. We fitted their RVs with our astrometry and we also found the masses to be in disagreement by about the same amount. We performed the same check with the RVs from \citet{Massarotti_2008_01_7} and found the masses to be in agreement within $1\sigma$ with our measurements. We therefore suspect that the disagreement with the \citet{Hummel_2001_03_0} masses comes from their photoelectric RVs. 
	
	We measured a distance of $40.964\pm0.135$\,pc, which is in agreement within $1\sigma$ with \citet{Piccotti_2020_02_0} but at $3\sigma$ with \citet{Hummel_2001_03_0}. It is also consistent at $0.7\sigma$ with Gaia.
	
	For each epoch, we were able to measure the uniform-disk (UD) angular diameter of the primary component as it is large enough to be spatially resolved. We measured an average value of $\theta_\mathrm{UD_1} = 1.285\pm0.075$\,mas. We converted this value to a limb-darkened (LD) angular diameter using a linear-law parametrisation $I_\lambda(\mu) = 1 - u_\lambda(1 - \mu)$. The LD coefficient $u_\lambda = u_\mathrm{K} = 0.1956\pm0.0265$ \citep[][we took the mean and standard deviation of the coefficient given by the least-square and flux conservation methods]{Claret_2011_05_0} was chosen taking the stellar parameters as close as possible to the values listed in Table~\ref{table__atmospheric_parameter}. The conversion is then given by the approximate formula of \citet{Hanbury-Brown_1974_06_0}:
	\begin{displaymath}
		\theta_\mathrm{LD}(\lambda) = \theta_\mathrm{UD}(\lambda) \sqrt{\frac{1-u_\lambda/3}{1-7u_\lambda/15}},
	\end{displaymath}
	which gives $\theta_\mathrm{LD,1} =  1.303\pm0.076$\,mas. It is in agreement with the $1.31\pm0.23$\,mas determined by \citet{Hummel_2001_03_0}. We also estimated an average flux ratio in $K$ of $25.03\pm0.20$\,\%, which is also in excellent agreement with $25.4\pm1.2$ measured by \citet{Hummel_2001_03_0}.
	
	\subsection{V963~Cen}
	
	This is a thoroughly studied high-eccentricity solar-type eclipsing system composed of two G2V-IV stars \citep{Graczyk_2022_10_0}. First discovered as likely being variable by \citet{Olsen_1993_11_0}, the first photometric measurements covering the eclipses were only obtained later by \citet{Clausen_1999_01_5,Clausen_2001_08_7} who refined the orbital period. The first orbital solutions combining both photometry and spectroscopy were obtained by \citet{Sybilski_2018_07_0} who measured masses precisely at a 0.08\,\% level. In \citet{Graczyk_2022_10_0}, we refined the solutions by using more precise photometry from the Transiting Exoplanet Survey Satellite \citep[TESS,][]{Ricker_2014_08_0} and new HARPS observations. We also noticed an apsidal motion of about 55~000\,yr, but in our case this is negligible as the time span by our data set is small.
	
\begin{figure*}[!h]
		\centering
		\resizebox{\hsize}{!}{\includegraphics{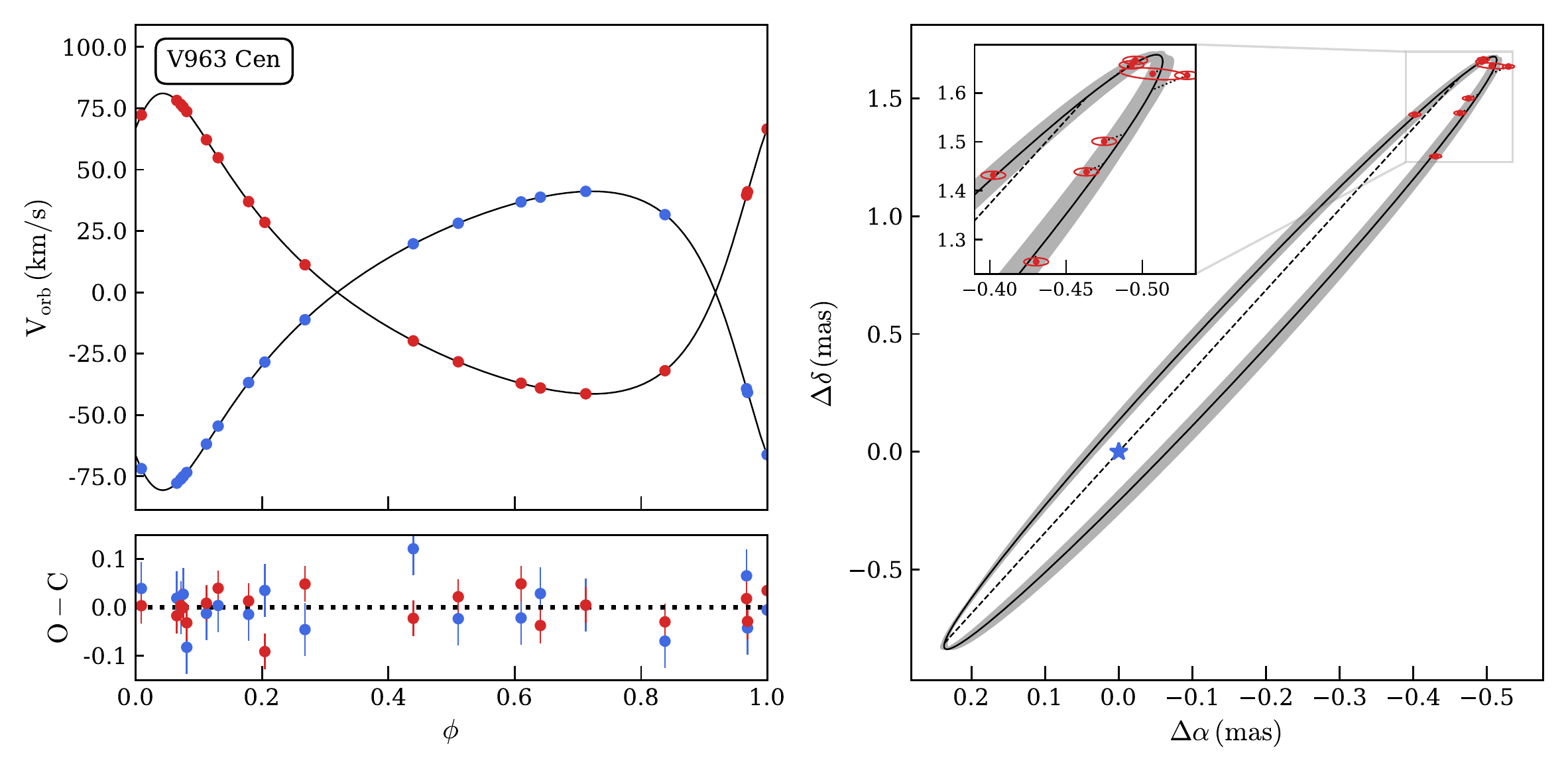}}
		\caption{Same as Fig. \ref{figure__orbit_akfor_hd9312_hd41255}, but for the system V963~Cen.}
		\label{figure__orbit_v963Cen}
	\end{figure*}

	In our combined fit displayed in Fig.~\ref{figure__orbit_v963Cen}, we used the RVs of \citet{Graczyk_2022_10_0}, which are the most precise and they were also determined with the broadening function method from the \texttt{RaveSpan} software. We measured the masses at a 0.08\,\% precision and the distance at 0.2\,\%. Our masses are in good agreement with those measured by \citet{Graczyk_2022_10_0} at a $\sim0.5\sigma$ level, but they are in slight agreement ($\sim2\sigma$) with \citet{Sybilski_2018_07_0}. However, they assumed a fixed eccentricity in their fitting process which likely results in underestimated uncertainties because the masses are directly linked to the eccentricity such that $M \propto (1 - e^2)^{3/2}$. As a check, we combined our astrometry with their RVs and found an agreement at $1.1\sigma$ with our values. In addition, RVs from \citet{Sybilski_2018_07_0} are less precise and accurate than ours, providing a reduced $\chi^2$ of the secondary RVs larger ($\chi^2_r = 15.3$) than ours ($\chi^2_r = 2.4$).
	
	Our distance is also in agreement within $1\sigma$ with the photometric distance estimated by \citet{Graczyk_2022_10_0}, and at $0.8\sigma$ from Gaia. We measured a flux ratio in $K$ of $96.0\pm1.1$\,\%, consistent with the extrapolated value of \citet{Graczyk_2022_10_0}.
	
	\begin{table*}[!ht]
		\centering
		\caption{Best-fit orbital elements and parameters for our binary systems.}
		\begin{tabular}{cccccc}
			\hline
			\hline
			Parameter       & AK~For & HD9312 & HD41255 & HD70937 & HD210763 \\
			\hline
			$P_\mathrm{orb}$ (days)         &  $3.9809943(4)$  &  $36.51920(2)$     &  $148.329(5)$      & $27.8858(4)$      &  $42.38113(2)$      \\
			$T_\mathrm{p}$ (days)       &  $2451904(1)$  &  $2456614.654(4)$  &  $2459569.95(2)$  & $2459577.754(4)$  &  $2454276.430(3)$      \\
			$e$                         &  $0.0002(2)$     &  $0.1433(1)$       &  $0.3432(2)$      & $0.3716(4)$       &  $0.6228(3)$      \\
			$K_1$ (\kms)                    &  $70.42(1)$      &  $34.971(3)$       &  $23.22(2)$       & $50.99(3)$        &  $50.28(3)$      \\
			$K_2$ (\kms)                    &  $77.06(4)$      &  $45.83(2)$        &  $23.03(2)$       & $56.05(3)$        &  $58.76(3)$      \\
			$\gamma_1$      (\kms)              &  $2.63(1)$       &  $0.772(2)$        &  $-2.099(9)$       & $-30.29(1)$       &  $14.94(1)$      \\
			$\gamma_2$      (\kms)              &  --              &  $1.27(2)$         &  --               & --                &  --     \\
			$\omega$        ($\degr$)               &  $170(107)$       &  $203.386(41)$     &  $348.404(60)$    & $169.192(63)$     &  $293.965(37)$      \\
			$\Omega$        ($\degr$)       &  $190.0(1)$     &  $237.0(3)$        &  $316.79(2)$      & $20.46(6)$        &  $257.6(2)$      \\
			$a$ (mas)                           &  $1.738(3)$      &  $4.85(3)$         &  $11.573(4)$      & $4.031(4)$        &  $3.76(1)$      \\
			$a$ (AU)                                    &  $0.0541(2)$    &  $0.276(2)$        &  $0.7081(6)$      & $0.2598(4)$       &  $0.351(2)$      \\
			$i$ ($\degr$)                           &  $86.9(1)$      &  $103.4(2)$        &  $56.76(3)$       & $78.60(7)$        &  $71.0(1)$      \\
			$M_1$ ($M_\odot$)               &  $0.69460(72)$   &  $1.1917(36)$      &  $1.0719(23)$     & $1.5761(22)$      &  $1.7377(45)$      \\
			$M_2$ ($M_\odot$)               &  $0.63475(50)$   &  $0.9093(26)$      &  $1.0808(24)$     & $1.4341(20)$      &  $1.4871(39)$      \\
			$d$ (pc)                            &  $31.107(53)$    &  $56.96(33)$       &  $61.185(51)$     & $64.462(74)$      &  $93.53(32)$      \\
			$\varpi$ (mas)                      &  $32.147(55)$    &  $17.56(10)$      &  $16.344(14)$     & $15.513(18)$     &  $10.692(37)$      \\
			\hline
			& HD224974 & HD188088 & LL~Aqr & $o$~Leo & V963~Cen \\
			\hline
			$P_\mathrm{orb}$ (days)         &  $10.658451(5)$   &  $46.81614(3)$      &  $20.17845(4)$    & $14.498068(6)$   &  $15.269309(6)$    \\
			$T_\mathrm{p}$ (days)       &  $2459388.250(1)$ &  $2455441.0406(3)$  &  $2455100.568(2)$ & $2450623.9(9)$   &  $2456807.2172(7)$ \\
			$e$                         &  $0.3391(2)$      &  $0.68664(6)$       &  $0.3165(2)$      & $0.0007(4)$      &  $0.4220(2)$       \\
			$K_1$ (\kms)                    &  $55.26(2)$       &  $47.79(1)$        &  $49.937(9)$      & $54.75(2)$       &  $60.90(2)$        \\
			$K_2$ (\kms)                    &  $56.31(2)$       &  $48.63(1)$         &  $57.73(2)$       & $61.66(2)$       &  $61.24(1)$        \\
			$\gamma_1$      (\kms)              &  $-21.203(8)$     &  $-5.019(7)$        &  $-9.830(9)$      & $26.24(1)$       &  $-30.450(8)$      \\
			$\gamma_2$      (\kms)              &  --               &  --                 &  $-9.58(2)$       & --               &  --          \\
			$\omega$        ($\degr$)               &  $76.861(39)$     &  $241.056(11)$      &  $155.695(37)$    & $214(22)$        &  $140.144(31)$     \\
			$\Omega$        ($\degr$)       &  $28.21(8)$       &  $111.83(1)$        &  $337.7(2)$      & $191.6(1)$      &  $343.8(1)$       \\
			$a$ (mas)                           &  $2.073(4)$       &  $21.643(5)$        &  $1.394(4)$       & $4.477(9)$       &  $1.363(3)$        \\
			$a$ (AU)                                    &  $0.1282(4)$      &  $0.3054(1)$        &  $0.1895(7)$      & $0.1834(7)$      &  $0.1555(4)$       \\
			$i$ ($\degr$)                           &  $53.4(1)$        &  $99.048(7)$        &  $89.2(4)$        & $57.8(2)$       &  $87.6(3)$         \\
			$M_1$ ($M_\odot$)               &  $1.2479(61)$     &  $0.87492(32)$      &  $1.19476(81)$    & $2.074(13)$     &  $1.07972(80)$     \\
			$M_2$ ($M_\odot$)               &  $1.2247(59)$     &  $0.85978(29)$      &  $1.03350(54)$    & $1.841(11)$     &  $1.07377(87)$     \\
			$d$ (pc)                            &  $61.82(18)$      &  $14.1129(38)$      &  $135.90(36)$     & $40.96(14)$     &  $114.12(21)$      \\
			$\varpi$ (mas)                      &  $16.176(47)$     &  $70.857(19)$       &  $7.358(20)$     & $24.412(81)$     &  $8.763(16)$      \\
			\hline
		\end{tabular}
		\tablefoot{Values in parentheses are uncertainties as to the final digits. $P_\mathrm{orb}$: orbital period. $T_\mathrm{p}$: time passage through periastron. $e$: eccentricity. $K_1, K_2$: radial velocity semi-amplitude of the primary and secondary. $\gamma$: systemic velocity. $\omega$: argument of periastron. $\Omega$: position angle of the ascending node. $a$: semi-major axis. $i$: orbital inclination. $M_1, M_2$: mass of the primary and secondary. $d, \varpi$: distance and parallax. 
		}
		\label{table__orbit_fit}
	\end{table*}
	
	\section{Evolutionary state}
	\label{section__evolutionary_states}
	
	We employed the same fitting method as in \citet{Gallenne_2019_10_0,Gallenne_2018_08_0,Gallenne_2016_02_0}, that is to say we used several stellar evolution models. We fitted the \parsec\ \citep[PAdova and TRieste Stellar Evolution Code,][]{Bressan_2012_11_0}, \basti\ \citep[Bag of Stellar Tracks and Isochrones,][]{Pietrinferni_2004_09_0}, \mist\ \citep[Mesa Isochrones and Stellar Tracks,][]{Choi_2016_06_0}, and \dsep\ \citep[Dartmouth Stellar Evolution Program,][]{Dotter_2008_09_8} isochrone models to estimate the stellar age of our systems. These models are well suited for our targets as they include the horizontal and asymptotic giant branch evolutionary phases, and contain a wide range of initial masses and metallicities. In addition, it enable us to test the uncertainty of the age induced by different models.
	
	\parsec\ models are computed for a scaled-solar composition with $Z_\odot=0.0152$, and they follow a helium initial content relation $Y = 0.2485 + 1.78Z$ with a mixing length parameter $\alpha_\mathrm{MLT} = 1.74$. They include convective core overshooting during the main sequence phase, parametrised with the strength of convective overshooting in units of the pressure scale height $l_\mathrm{ov} = \alpha_\mathrm{ov}H_\mathrm{p}$. The overshooting parameter $\alpha_\mathrm{ov}$ is set depending on the mass of the star, that is  $\alpha_\mathrm{ov} = 0$ for $M \lesssim 1.1\,M_\odot$, $\alpha_\mathrm{ov} \sim 0.25$ for $M \gtrsim 1.4\,M_\odot$, and linearly ramps with the mass in between. The \basti\ models are computed for a scaled-solar composition with $Z_\odot=0.0198$, following the relation $Y = 0.245 + 1.4Z$ with $\alpha_\mathrm{MLT} = 1.913$. They also include convective core overshooting with the same parametrisation, but with the conditions $\alpha_\mathrm{ov} = 0$ for $M \lesssim 1.1\,M_\odot$, $\alpha_\mathrm{ov} = 0.20$ for $M \gtrsim 1.7\,M_\odot$, and $(M - 0.9M_\odot)/4$ in between. The \mist\ models use a scaled-solar composition with $Z_\odot=0.0142$, with the relation $Y = 0.2703 + 1.5Z$ and $\alpha_\mathrm{MLT} = 1.82$. They use an alternate prescription of the core overshooting with a diffusion coefficient $D_\mathrm{ov} = D_0\exp{(-2z/H_\nu)}$, where $z$ is the distance from the edge of the convective zone, $D_0$ is the coefficient at $z = 0$, and $H_\nu$ is defined with the overshooting parameter $f_\mathrm{ov}$ such that $H_\nu = f_\mathrm{ov}H_\mathrm{p}$. \mist\ models adopt a fixed value $f_\mathrm{ov} = 0.016$ for all stellar masses, which would be approximatively converted to $\alpha_\mathrm{ov} \sim 0.18$ \citep{Claret_2017_11_0}. The \dsep\ models use a scaled-solar composition with $Z_\odot=0.0166$, with the relation $Y = 0.245 + 1.5Z$ and a solar-calibrated mixing length $\alpha_\mathrm{MLT} = 1.938$. The amount of core overshooting is also parametrised as a multiple of the pressure scale height such as, for solar metallicity, $\alpha_\mathrm{ov} = 0.05, 0.1$, and 0.2 for $M \lesssim 1.2\,M_\odot, 1.2\,M_\odot< M < 1.3$, and $M \gtrsim 1.3\,M_\odot$, respectively.
	
	We retrieved several isochrones from the \parsec\ database tool\footnote{\url{http://stev.oapd.inaf.it/cgi-bin/cmd}}, with ages ranging from $\log t = 6.6$ to 10.13 with a step of 0.05 (i.e. $\sim 0.005-13$\,Ga), and metallicities from $Z = 0.003$ to 0.06 (i.e. $-0.7 \leq \mathrm{[Fe/H]} \leq +0.6$, using $[\mathrm{Fe/H}] \sim \log{(Z/Z_\odot)}$), with a step of 0.001 (fine enough to avoid re-interpolation). The \basti\ isochrones are pre-computed in their database\footnote{\url{http://basti.oa-teramo.inaf.it/index.html}}, we downloaded models for $t = 0.1-9.5$\,Ga by a step of $\sim 0.2$\,Ma and $Z = 0.002, 0.004, 0.008, 0.01, 0.0198, 0.03$, and 0.04 (i.e. $-1.0 \leq \mathrm{Fe/H} \leq 0.3$). For fitting purposes, we created an interpolated grid of the \basti\ isochrones in $Z$, from 0.002 to 0.04 with a step of 0.001. We also computed \mist\ isochrones from their database tool\footnote{\url{http://waps.cfa.harvard.edu/MIST/interp_isos.html}} using the standard age grid from 0.1\,Ma to 20\,Ga with a step of $\sim 1$\,Ma, and for metallicities in the range $0.001 \leq Z \leq 0.045$ (i.e. $-1.15 \leq \mathrm{Fe/H} \leq 0.5$) with a step of 0.001. We downloaded \dsep\ isochrones from their website\footnote{\url{http://stellar.dartmouth.edu/models/isolf_new.html}} with a grid with an age from 1\,Ga to 10\,Ga with a step of 0.02\,Ga and metallicity from 0.001 to 0.058 with a step of 0.001 (i.e. $-1.2 \leq \mathrm{Fe/H} \leq 0.5$).
	
	When possible, we searched for the best-fit age in stellar effective temperature, radius, mass, and $K$-band absolute magnitude for both components simultaneously, assuming coeval stars and following $\chi^2$ statistics:
	\begin{displaymath}
		\chi^2 = \sum_{i=1}^2 \left[ \left(\dfrac{\Delta T_\mathrm{eff}}{\sigma_\mathrm{T_{eff}}}\right)_i^2 +
		\left(\dfrac{\Delta R}{\sigma_R}\right)_i^2 + \left(\dfrac{\Delta M}{\sigma_M}\right)_i^2 + \left(\dfrac{\Delta M_\mathrm{K}}{\sigma_{M_\mathrm{K}}}\right)_i^2 \right],  
	\end{displaymath}
	where the sum is over both components ($i = 1, 2$). The $\Delta$ symbol represents the difference between the predicted and observed quantities. The effective temperature and the radii are measured quantities and were taken from the literature. They are listed in Table~\ref{table__atmospheric_parameter}. The masses were also measured for this work and are reported in Table~\ref{table__orbit_fit}. When available, we took care of re-scaling the retrieved linear radii according to our own estimate of the linear semi-major axis. In our isochrone plots, we also display the stellar luminosity estimated from the Stefan-Boltzmann law, but this parameter was not included in the fit as this is not an independent measurement. Absolute magnitudes in the $K$ band,  $M_\mathrm{K}$, were also included in the fit using our measured flux ratio $f_\mathrm{K}$ following the relations
	\begin{eqnarray}
		M_\mathrm{K,1} &=& m_\mathrm{K} - A_\mathrm{K} + 2.5\log{(1 + f_\mathrm{K})}  - 5\log{d} + 5, \label{equation__magnitude1} \\
		M_\mathrm{K,2} &=& m_\mathrm{K} - A_\mathrm{K} + 2.5\log{(1 + 1/f_\mathrm{K})}  - 5\log{d} + 5, \label{equation__magnitude2}
	\end{eqnarray}
	where $m_\mathrm{K}$ is the combined magnitude as measured in the 2MASS catalogue \citep{Cutri_2003_03_0}, $d$ our measured distance, and $A_\mathrm{K}$ the extinction coefficient in the $K$ band, such as $A_\mathrm{K} = 0.119 A_\mathrm{V}$ \citep{Fouque_2007_12_0} and $A_\mathrm{V} = 3.1 E(B-V)$ \citep{Cardelli_1989_10_0}. The colour excess coefficient was estimated from the three-dimensional extinction map \texttt{STILISM}\footnote{\url{https://stilism.obspm.fr}} \citep{Lallement_2018_08_4}. It is worth mentioning that our systems are nearby and the effect of reddening is generally negligible. All errors have been propagated to the final magnitudes. 
	
	The stellar metallicity was kept fixed in the fitting process to a value from the literature (listed in Table~\ref{table__atmospheric_parameter}). Our fitting procedure was the following. For all isochrone models, we first chose the closest grid in $Z$ for a given metallicity. Then, we searched for the global $\chi^2$ minimum in age by fitting all isochrones for that given metallicity. A second fit was then performed around that global minimum value, and where the grid was interpolated in age at each iteration. To assess the uncertainties on the four isochrone models (i.e. \parsec, \basti, \mist, and \dsep), we repeated the process with $Z \pm \sigma$. Our final adopted age corresponds to the average and standard deviation between the four models.
	
\begin{figure}[!h]
	\centering
	\resizebox{\hsize}{!}{\includegraphics[width = \linewidth]{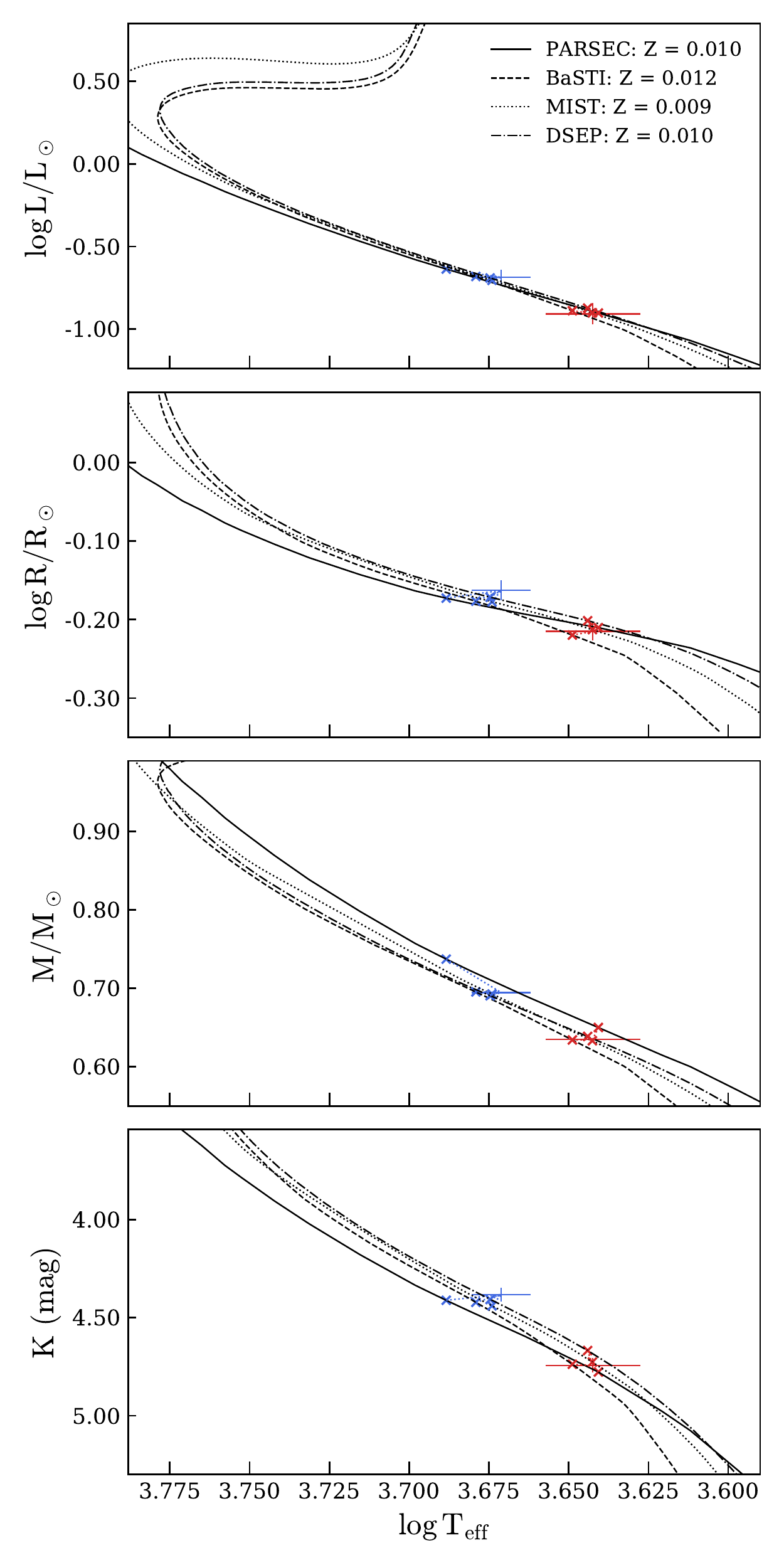}}
	\caption{Fitted \parsec, \basti, \mist, and \dsep\ isochrones for the AK~For system. We note that the luminosities were not fitted and were estimated from the Stefan-Boltzmann law.}
	\label{figure__isochrones_akfor}
\end{figure}

	\noindent \textsf{AK~For:} We adopted the metallicity from \citet{Heminiak_2014_07_7} and the stellar parameters (effective temperatures, radii, and flux ratio in $K$) listed in Table~\ref{table__atmospheric_parameter}. All models give a rather different estimate for the age, although all values are within $1\sigma$. \parsec\ gives a substantially younger system compared to the other models and the predicted masses are $> 30\sigma$ away from our measurements. The best model (i.e. the one with all parameters within $1-2\sigma$) is from \basti\ isochrones with all predicted stellar parameters within $1\sigma$, giving an age of $9.5\pm0.01$\,Ga. This is older than the 6\,Ga estimated by \citet{Heminiak_2014_07_7} using \dsep\ isochrones, although there is no error quoted by the authors.\ However, it would be in agreement within errors with our \dsep\ fit. Excluding the age estimated with \parsec, we adopted an average age of $7.9\pm1.3$\,Ga. As displayed in Fig.~\ref{figure__isochrones_akfor}, all models show that both stars still reside on the main sequence, which is in agreement with previous works, and that they are near the turnoff point. Using the stellar temperatures and adopting the spectral type-temperature calibration of \citet{Pecaut_2013_09_0}, we determined that the primary and the secondary component have spectral types of K4V and K5V, respectively, with a $\pm1$ index as uncertainty due to the temperature errors.

\noindent \textsf{HD9312:} The metallicity $0.03\pm0.1$\,dex from \citet{Kiefer_2018_02_6} was used for the isochrone fit. As observables,  we used their measured effective temperatures, our measured masses, and $K$-band flux ratio. Luminosities and radii can also be estimated using our measured masses, as well as the temperatures and surface gravities from \citet{Kiefer_2018_02_6}. However, they are not independent parameters, so they are displayed in Fig.~\ref{figure__isochrones_hd9312_hd41255} but not included in the isochrone fit.


We found a system that is slightly younger than \citet{Wang_2015_10_9}, that is the secondary still resides on the main sequence while the primary is exiting the turnoff point. This is because the work of \citet{Wang_2015_10_9} is based on the primary star properties only (SB1 at that time), a solar metallicity $Z = 0.019$, and a larger estimate ($\sim 15$\,\%) of the mass ratio compared to our measurement ($q = 0.763$) and the one from \citet[$q =0.762$][]{Halbwachs_2014_12_2}. We see in Fig.~\ref{figure__isochrones_hd9312_hd41255} that all models provide an acceptable fit of the observations, although they all provide a secondary effective temperature that is larger by $\sim 1.5\sigma$, while the primary temperature is within $1\sigma$ except for the \parsec\ model. To reconcile the observations with the models within $1-2\sigma$, we would need to increase the measured metallicity to 0.1\,dex. However, the \parsec\ models still predict masses $> 3\sigma$. The best agreement is given by \dsep\ and \mist, but all give a similar age for the system (see Table~\ref{table__stellar_properties}). We adopted an average age of $5.60\pm0.09$\,Ga. There are no significant changes if we choose the same metallicity [Fe/H] = 0.1\,dex as \citet{Wang_2015_10_9}, with all isochrones giving a similar age. As previously mentioned, using the calibration from \citet{Pecaut_2013_09_0}, we determined the spectral types to be G9V and K1V for the primary and secondary star, respectively.

\noindent \textsf{HD41255}: There are neither measurements for the effective temperature nor metallicity for this system. The temperature can be estimated from colour-temperature calibrations (using the $(V - K)$ colour for instance) \citep[see e.g.][]{Casagrande_2011_06_0}. However, measured magnitudes include the flux from both components. To estimate the individual magnitudes in $K$, we used our measured flux ratio and Equ.~\ref{equation__magnitude1} and \ref{equation__magnitude2}. We used the same equations for the $V$ band, assuming the same flux ratio as in $K$. We estimated an average effective temperature $T_\mathrm{eff,1} = 6107\pm39$\,K and $T_\mathrm{eff,2} = 5984\pm39$\,K combining the $(V - K)-T_\mathrm{eff,}$ relations from \citet{di-Benedetto_1998_11_8}, \citet{Houdashelt_2000_03_5}, \citet{Ramirez_2005_06_7}, \citet{Masana_2006_05_1}, \citet{Gonzalez-Hernandez_2009_04_3}, and \citet{Casagrande_2010_03_6}. As the flux ratio is close to 1, we found similar temperatures between the stars (error is the standard deviation between the relations). This is consistent with $5996\pm330$\,K estimated by Gaia (assuming a single star).

\begin{figure*}[!ht]
	\centering
	\resizebox{\hsize}{!}{\includegraphics[width = \linewidth]{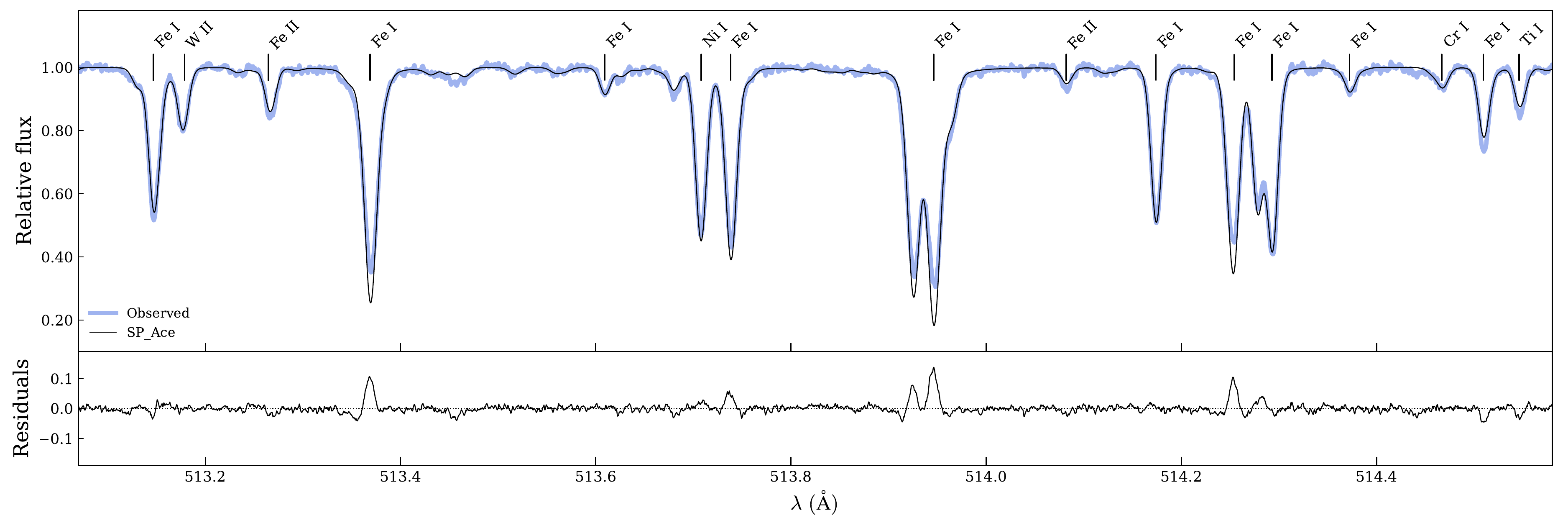}}
	\caption{Comparison of the decomposed primary spectrum (blue line) with the synthetic spectrum (red line) for the binary system HD41255.}
	\label{figure__spectrum_hd41255}
\end{figure*}

To go a step further, we disentangled our UVES spectra and estimated the effective temperatures. For disentangling, we used all spectra with the \texttt{RaveSpan} software which utilises the method presented in \citet{Gonzalez_2006_03_0}. We ran two iterations choosing a median value for the normalisation of the spectra. We then performed a spectral analysis using the Stellar Parameters And Chemical abundances Estimator code \texttt{SP\_Ace}\footnote{\url{https://dc.zah.uni-heidelberg.de/SP_ACE}} \citep{Boeche_2016_03_1}. This tool employs a new method based on a library of general curve-of-growth (GCOG) in the spectral range $4800--6860$\AA. Stellar parameters were derived from a $\chi^2$ minimisation between the observed and model spectra. However, \texttt{SP\_Ace} neither relies on a library of synthetic spectra, nor does it measure the equivalent width (EW) of absorption lines, but it constructs the models from a library of GCOG, which are coefficients of polynomial functions (one per absorption line) describing the EW of the lines as a function of the stellar parameters \citep[for more details, see][]{Boeche_2016_03_1}. This tool uses different elements such as Fe, C, N, O, Mg, Al, Si, Ca, and Ti (up to 21 elements) to estimate the stellar parameters $T_\mathrm{eff}, \log{g}$, [M/H], and elemental abundances. We found for the primary component $T_\mathrm{eff} = 6017\pm120$\,K, $\log{g} = 4.46\pm0.18$\,dex, and [Fe/H] = $-0.21\pm0.08$\,dex. The temperature is similar to the one predicted by colour-temperature calibrations. The uncertainties were estimated by repeated the process by fixing the temperature to values of $\pm200$\,K, and fixing $\log{g}$ to $\pm0.25$\,dex. A comparison of the observed average spectrum and the fitted \texttt{SP\_Ace} model is displayed in Fig.~\ref{figure__spectrum_hd41255}.

We performed the same spectral analysis for the secondary component. We found $T_\mathrm{eff} = 6064\pm98$\,K, $\log{g} = 4.57\pm0.21$\,dex, and [Fe/H] = $-0.09\pm0.07$\,dex. The metallicity of the two components is within $\sim1\sigma$ of each other, we therefore adopted an average and standard deviation of [Fe/H] = $-0.15\pm0.06$\,dex for the system. This is at $1\sigma$ with the $-0.17$\,dex derived from the colour-[Fe/H] calibration of the GCS, although assuming a single star. 


From the isochrone fit, we found the system to be composed of two main-sequence stars near the turnoff point. Isochrones are displayed in Fig.~\ref{figure__isochrones_hd9312_hd41255}. The \basti\ and \dsep\ models give a similar age, although \basti\ better fits all the observables ($<2.2\sigma$ difference). We adopted an average age between these two models of $t = 2.57\pm0.14$\,Ga. We also estimated the luminosity and radius of both components using $\log{g}$, Newton's law of gravitation, and the Stefan-Boltzmann law. They are also listed in Table~\ref{table__atmospheric_parameter} but they are not included as fitted parameters. The \parsec\ and \mist\ isochrones are not reliable as they provide a system that is too young, $<600$\,Ma. Our estimate is smaller than the value reported by \citet{Casagrande_2011_06_0} from a Bayesian analysis of isochrone matching based on temperature and metallicity relations. Assuming a single star, they derived $3.90^{-0.16}_{+1.37}$\,Ga using the \parsec\ models and $4.70^{-1.23}_{+0.41}$\,Ga with \basti. We note that their inferred mass is also not consistent with our measurement at $> 2\sigma$. The GCS survey also predicted a higher age of $3.6^{-0.3}_{+2.1}$\,Ma from fitting the \parsec\ isochrone with their effective temperature of 6081\,K determined from the IR flux method, a metallicity of $-0.17$\,dex, and the Hipparcos distance to estimate the absolute magnitudes. In general, we can see in Fig.~\ref{figure__isochrones_hd9312_hd41255} that the isochrone models predict a slightly higher temperature of about +200\,K. From the measured temperatures, we derived the spectral types to be F9V for both stars.

\noindent \textsf{HD70937}: No measurements of the effective temperature nor metallicity  is available in the literature. The average temperature given by colour-temperature relations cited previously is $6401.2+/-79.3$\,K for both components, using the same flux ratio in $V$ as in $K$. We also estimated the stellar parameters from our UVES disentangled spectra following the same analysis as previously discussed using both \texttt{SP\_Ace} codes.  For the primary star, we estimated $T_\mathrm{eff} = 6277\pm100$\,K, $\log{g} = 3.72\pm0.27$\,dex, and [Fe/H] = $-0.02\pm0.06$\,dex, and then $T_\mathrm{eff} = 6446\pm84$\,K, $\log{g} = 4.10\pm0.24$\,dex, and [Fe/H] = $0.05\pm0.06$\,dex  for the secondary. We adopted an average metallicity of $0.02\pm0.04$\,dex for the system. All parameters are reported in Table~\ref{table__atmospheric_parameter}.

We found that both stars are at the main-sequence turnoff, as shown in Fig.~\ref{figure__isochrones_hd70937_hd210763}. We also display the luminosity and radii as calculated previously, but they are not fitted. All models provide a similar age for the system, with \basti\ being the best fitted model. The most discrepant parameters are the temperature, which is higher by $2-5\sigma$. To reconcile the models within $1-3\sigma$, we would need to increase the measured metallicity to 0.1\,dex at $2.5\sigma$ with our measured value. We adopted the average value of $1.78\pm0.07$\,Ma as the age of the system. This is in agreement with $1.5\pm0.1$Ma derived by \citet{Holmberg_2009_07_3} from the GCS, although they used a temperature of 6412\,K and a [Fe/H] = 0.01\,dex. From the measured temperatures, we determined the spectral types to be F7V and F5V for the primary and secondary, respectively.

\noindent \textsf{HD210763}: There is no measurements of the temperatures or metallicity in the literature. The colour-temperature relations predict an average temperature around 6250\,K for both stars (also assuming the same flux ratio in $V$ as in $K$). Using the \texttt{SP\_Ace} code on the disentangled spectra, we measured$T_\mathrm{eff} = 6173\pm38$\,K, $\log{g} = 3.78\pm0.26$\,dex, and [Fe/H] = $-0.03\pm0.07$\,dex for the primary star, and  $T_\mathrm{eff} = 6523\pm87$\,K, $\log{g} = 4.08\pm0.29$\,dex, and [Fe/H] = $0.04\pm0.06$\,dex for the secondary component. We adopted an average metallicity of $0.01\pm0.04$\,dex, which is in agreement with the predicted value of 0.05 from \citet{Holmberg_2009_07_3}.

As for HD70937, both stars are located at the main-sequence turnoff, as shown in Fig.~\ref{figure__isochrones_hd70937_hd210763}. All models provide a similar age for the system, with \basti\ best fitting the data with the lowest $\chi^2_r$. We adopted an average age of $1.67\pm0.07$\,Ma (excluding \mist). This is similar to \citet{Holmberg_2009_07_3} predicting $1.3\pm0.2$\,Ma. We note that all models provide temperatures that are $>3\sigma$ with the measurements. Increasing the metallicity to 0.1\,dex would mitigate the discrepancy within $2\sigma$ for the \basti\ and \parsec\ models. Our final stellar parameters are listed in Table~\ref{table__atmospheric_parameter}. We derived their spectral type to be F8V and F5V.

\noindent \textsf{HD224974}: We performed the same analysis with \texttt{SP\_Ace} for this system as there are no measured temperatures or metallicities in the literature. The colour-temperature relations predict an average temperature around 6141\,K for both stars (assuming the same flux ratio in $V$ as in $K$). The \texttt{SP\_Ace} analysis provides $T_\mathrm{eff} = 6221\pm85$\,K, $\log{g} = 4.33\pm0.13$\,dex, and [Fe/H] = $0.16\pm0.05$\,dex for the primary star, and  $T_\mathrm{eff} = 6171\pm75$\,K, $\log{g} = 4.49\pm0.20$\,dex, and [Fe/H] = $0.07\pm0.07$\,dex for the secondary. We adopted an average metallicity of $0.04\pm0.03$\,dex, which is in agreement with the predicted value of 0.08 from \citet{Holmberg_2009_07_3}.

Our isochrone fit of Fig.\ref{figure__isochrones_hd224974_hr7578} shows that both stars exhausted the hydrogen fuel at their cores and are entering the turnoff point. We obtained the best fitted isochrone with the \basti\ models, with all parameters within $1.5\sigma$, while \mist\ and \parsec\ are not consistent, providing a system that is too young, $< 600$\,Ma. All parameters are reported in Table~\ref{table__atmospheric_parameter}. We adopted  $t = 2.98\pm0.07$\,Ga as a final
age for the system, corresponding to the average value between the \basti\ and \dsep\ models. This is consistent with the predicted value of $2.4\pm0.2$ from the GCS. We derived their spectral type to be F7V and F8V.

\noindent \textsf{HD188088}: \citet{Gray_2006_07_6} measured a metallicity of 0.29\,dex, $\log{g} = 4.24$, and a temperature of 4774\,K, assuming a single star. \citet{Luck_2017_01_5} measured a lower metallicity of 0.12\,dex (although within $1\sigma$), while their temperature of $4818\pm63$\,K and surface gravity og 4.20\,dex are similar to \citet{Gray_2006_07_6}. We did not succeed in converging with \texttt{SP\_Ace} on our average disentangled spectrum because it reached the upper limit metallicity range of 0.5\,dex. This may be due to the fact that the stars are chromospherically active, with light variability due to star spots, which can prevent a good spectral disentangling. \citet{Boeche_2016_03_1} succeeded in using \texttt{SP\_Ace} with a FEROS spectrum and determined $T_\mathrm{eff} = 4868\pm80$\,K, $\log{g} = 3.50\pm0.03$\,dex, and [Fe/H] = $-0.32\pm0.38$\,dex. The temperature is in good agreement with the previous authors, while the surface gravity is lower. The metallicity, however, is inconsistent with previous estimates. In our isochrone fit, we adopted the mean temperature and gravity of 4820\,K and 4.00\,dex for both stars, and considered 200\,K and 0.3\,dex as uncertainty, respectively.  For the metallicity, we tested different values in the range [$-0.35, 0.35]$\,dex that match the measurements best.

We found that the best fit of the isochrone models is given for a metallicity of 0.28\,dex. However, the masses are still $> 8\sigma$ for all models, except for \basti. The most acceptable fit is with \basti\ providing predicted values at $< 2\sigma$ from the measurements, except $M_2$ at $\sim 5\sigma$. However the given age of 910\,Ma is too young for such stars. The other models predict masses $> 8\sigma$, with rather different ages and large uncertainties. We adopted the average age $t = 1.6\pm1.1$Ga as it is not possible to constrain it better. The isochrones are displayed in Fig.~\ref{figure__isochrones_hd224974_hr7578} and the stellar parameters are with fitted ages in Table~\ref{table__stellar_properties}. This system is a good example of not being able to constrain the age even with very accurate mass measurements, demonstrating the importance of the other stellar parameters.

\noindent \textsf{LL~Aqr}: We used the metallicity, temperatures, radii, and the $V$-band flux ratio determined by \citet{Graczyk_2016_10_0}. The luminosity was also used, but not fitted as explained previously. Our isochrone fits of Fig.~\ref{figure__isochrones_llaqr_oleo} show that both stars are near the main-sequence turnoff. All models provide a consistent age with each other; however, no model satisfies all observables at less than $3\sigma$. All isochrones are too hot with respect to the measurements. This was also noticed by \citet{Graczyk_2016_10_0} who used the \parsec\ and MESA \citep[Modules for Experiments in Stellar Astrophysics,][]{Paxton_2015_09_5} isochrones. To mitigate the discrepancy, the metallicity would need to increase to $\sim 0.15$\,dex, which is at $3\sigma$ with the measured value. Instead, \citet{Graczyk_2016_10_0} reconciled the predicted and observed parameters by fine-tuning some internal stellar parameters of the stellar evolutionary tracks, more particularly the element diffusion and allowing for different mixing-length parameters for the two components. They estimated an age for this system ranging from 2.3 to 2.7\,Ga, which agrees with our average value of $2.76\pm0.20$\,Ga; although, the four models do not match the observables properly.

\noindent \textsf{$o$~Leo}: The metallicity and temperature of the giant star (primary) were estimated by \citet{Adamczak_2014_08_1} from new high-resolution spectra. They determined $T_\mathrm{eff} = 6173\pm59$\,K, $\log{g} = 3.06\pm0.25$\,dex, and [Fe/H] = $-0.06\pm0.09$\,dex. This is consistent with $6200\pm200$\,K reported by \citet{Griffin_2002_02_8}, who also determined the secondary effective temperature to be $7600\pm200$. We performed a \texttt{SP\_Ace} analysis with our disentangled spectra of the primary and estimated $T_\mathrm{eff} = 6107\pm93$\,K, $\log{g} = 2.91\pm0.24$\,dex, and [Fe/H] = $0.11\pm0.10$\,dex, in good agreement with the previous study. For the secondary, we could not use the \texttt{SP\_Ace} algorithm because the effective temperature is not in the stellar parameter ranges covered (3600\,K$ < T_\mathrm{eff} < 7400$\,K). We therefore adopted the temperature estimated by \citet{Griffin_2002_02_8}. As for HD9312, we can have an estimate for the secondary radius via Stefan's law and our measured primary radius. We assumed a bolometric flux ratio of $0.43\pm0.04$, similar to the measured flux ratio measured by \citet{Hummel_2001_03_0} at $0.55\,\mu$m.  

Our isochrone fit is displayed in Fig.~\ref{figure__isochrones_llaqr_oleo}. We note that $L_1, L_2$, and $R_2$ are displayed but not fitted as they are not independent parameters. All models provide a similar age, except for \mist, giving an average age of $1.06\pm0.03$\,Ga (excluding \mist). \basti\ and \parsec\ best fit the observables within $2.5\sigma$. The \mist\ model predicts a younger system than the other models. Our derived age is in good agreement with the 1.02\,Ga determined by \citet{Griffin_2002_02_8} assuming a metallicity $Z = 0.02$\,dex. In agreement with the previous works, we see that the primary is a giant and the secondary is a dwarf star located at the turnoff point. 

\noindent \textsf{V963~Cen}: We used the metallicity, temperatures, and radii determined by \citet{Graczyk_2022_10_0}. The luminosity was also used but not fitted as explained previously. Our isochrone fits of Fig.~\ref{figure__isochrones_v963cen} show that both stars are near the main-sequence turnoff. Although all models provide a similar age of $6.10\pm52$\,dex, none gives a consistent fit, with most of the parameters above $3\sigma$ with the predicted values. The models predict slightly hotter components with the given metallicity [Fe/H] = $-0.06\pm0.05$\,dex. To reconcile the isochrones within $1-2\sigma$ with the same stellar effective temperatures (which are easier to measure than the metallicity), we tested different metallicities. To reconcile the \basti\ and \mist\ isochrones, we would need a metallicity of $\sim 0.13$\,dex, while $\sim 0.1$\,dex would reconcile the \basti\ models. We did not find an acceptable metallicity to better match the \parsec\ models. A metallicity of $\sim 0.1$\,dex would provide slight agreement for some observables, but the masses and secondary radii are $> 5\sigma$ away.

In Fig.~\ref{figure__isochrones_v963cen}, we display the isochrones for two metallicities, -0.06\,dex as estimated in  \citet{Graczyk_2022_10_0} and 0.1\,dex, which better fit the observables. They give an age of $6.15\pm0.31$\,Ga and $7.31\pm0.06$, respectively. The stellar parameters are reported in Table~\ref{table__atmospheric_parameter}, together with the age given for each model for the second metallicity.

\begin{sidewaystable*}[!ht]
	\centering
	\caption{Stellar parameters used for the age determinations, together with our fitted and adopted age for the systems.}
	\begin{tabular}{ccccccccc|cccccc} 
		\hline
		\hline
		System & Star   & $R$\,\tablefootmark{a} &  $T_\mathrm{eff}$ &  $\log L/L_\odot$     & [Fe/H] & $M_\mathrm{K}$ & $E(B-V)$ & Ref.     & $t_\mathrm{parsec}$ & $t_\mathrm{basti}$ & $t_\mathrm{mist}$ & $t_\mathrm{dsep}$ & $t_\mathrm{avg}$\,\tablefootmark{b} & \\
		&               & ($R_\odot$)           &  ($K$)   &  &  (dex)   & (mag) & &  & (Ga) & (Ga) & (Ga) & (Ga) & (Ga) \\
		\hline
		\multirow{2}{*}{AK~For} & A & 0.688(20)  & 4690(100) & $-0.687(45)$  & \multirow{2}{*}{$-0.2(1)$} & 4.38(4) & \multirow{2}{*}{0.000(14)} & \multirow{2}{*}{1,2} & 
		\multirow{2}{*}{1.9(4.2)} & \multirow{2}{*}{8.9(6)} & \multirow{2}{*}{6.1(2.2)} & \multirow{2}{*}{8.6(1.4)} &  \multirow{2}{*}{7.9(1.3)} \\
		& B & 0.610(16) & 4390(150) & $-0.908(64)$ & & 4.75(4)  & & & & & \\
		\hline
		\multirow{2}{*}{HD9312} & A & 2.03(31)  & 5367(166) & 0.488(141)  & \multirow{2}{*}{$0.03(3)$} & 1.06(27) & \multirow{2}{*}{0.002(14)} & \multirow{2}{*}{1,3} & 
		\multirow{2}{*}{5.59(27)} & \multirow{2}{*}{5.46(61)} & \multirow{2}{*}{5.64(37)} & \multirow{2}{*}{5.71(29)} &  \multirow{2}{*}{5.60(9)} \\
		& B & 0.79(23) & 5150(228) & $-0.401(261)$ & & 3.81(28)  & & & & & \\
		\hline
		\multirow{2}{*}{HD41255} & A & 1.01(21)  & 6017(129) & 0.081(183)  & \multirow{2}{*}{$-0.15(6)$} & 2.93(3) & \multirow{2}{*}{0.001(15)} & \multirow{2}{*}{1} & 
		\multirow{2}{*}{0.020(1)} & \multirow{2}{*}{2.45(4)} & \multirow{2}{*}{0.6(1.4)} & \multirow{2}{*}{2.71(21)} &  \multirow{2}{*}{2.58(13)} \\
		& B & 0.89(22) & 6064(98) & $-0.012$(212)   &                                                   & 2.97(3) &    & & & & \\
		\hline
		\multirow{2}{*}{HD70937} & A & 2.87(89)  & 6277(100) & 1.061(271)  & \multirow{2}{*}{0.02(4)} & 1.38(2) & \multirow{2}{*}{0.000(14)} & \multirow{2}{*}{1} & 
		\multirow{2}{*}{1.85(9)} & \multirow{2}{*}{1.68(16)} & \multirow{2}{*}{1.76(5)} & \multirow{2}{*}{1.67(3)} &  \multirow{2}{*}{1.67(7)} \\
		& B & 1.77(49) & 6446(84) & 0.686(241)   &                                                   & 1.94(2) &    & & & & \\
		\hline
		\multirow{2}{*}{HD210763} & A & 2.81(84)  & 6173(38) & 1.015(260)  & \multirow{2}{*}{0.01(4)} & 0.66(3) & \multirow{2}{*}{0.012(16)} & \multirow{2}{*}{1} & 
		\multirow{2}{*}{1.78(5)} & \multirow{2}{*}{1.58(13)} & \multirow{2}{*}{1.63(4)} & \multirow{2}{*}{1.69(4)} &  \multirow{2}{*}{1.67(7)} \\
		& B & 1.84(62) & 6523(87) & 0.743(291)   &                                                   & 1.74(3) &    & & & & \\
		\hline
		\multirow{2}{*}{HD224974} & A & 1.27(19) & 6221(85) & 0.334(132)  & \multirow{2}{*}{0.04(3)} & 2.36(2) & \multirow{2}{*}{0.001(16)} & \multirow{2}{*}{1} & 
		\multirow{2}{*}{0.4(1.5)} & \multirow{2}{*}{3.04(2)} & \multirow{2}{*}{0.5(1.4)} & \multirow{2}{*}{2.91(12)} &  \multirow{2}{*}{2.98(7)} \\
		& B & 1.04(24) & 6171(75) & 0.152(201)   &                                                   & 2.48(3) &    & & & & \\
		\hline
		\multirow{2}{*}{HD188088} & A & 1.59(62) & 4820(200) & 0.087(348)  & \multirow{2}{*}{0.28(5)} & 3.99(25) & \multirow{2}{*}{0.000(15)} & \multirow{2}{*}{1,7,8} & 
		\multirow{2}{*}{0.3(5.3)} & \multirow{2}{*}{0.9(0.4)} & \multirow{2}{*}{1.9(4.3)} & \multirow{2}{*}{3.3(3.3)} &  \multirow{2}{*}{2.0(2.2)} \\
		& B & 1.57(62) & 4820(200) & 0.079(348)   &                                        & 4.10(25) &    & & & & \\            
		\hline
		\multirow{2}{*}{LL~Aqr} & A & 1.321(6) & 6080(45) & 0.332(14)  & \multirow{2}{*}{0.02(5)} & 2.60(2) & \multirow{2}{*}{0.026(16)} & \multirow{2}{*}{1,4} & 
		\multirow{2}{*}{2.95(63)} & \multirow{2}{*}{2.45(69)} & \multirow{2}{*}{2.73(19)} & \multirow{2}{*}{2.93(24)} &  \multirow{2}{*}{2.76(20)} \\
		& B & 1.002(5) & 5703(50) & $-0.019(16)$   &                                 & 3.29(3) &    & & & & \\
		\hline
		\multirow{2}{*}{$o$~Leo} & A & 5.73(34)   & 6107(93) & 1.614(57)  & \multirow{2}{*}{0.11(10)} & -0.39(25) & \multirow{2}{*}{0.001(14)} & \multirow{2}{*}{1,5} & 
		\multirow{2}{*}{1.10(3)} & \multirow{2}{*}{1.03(4)} & \multirow{2}{*}{0.4(5)} & \multirow{2}{*}{1.05(3)} &  \multirow{2}{*}{1.06(3)} \\
		& B & 2.43(35) & 7600(200) & 1.25(12)   &                                          & 1.11(25) &    & & & & \\
		\hline
		\multirow{2}{*}{V963~Cen} & A & 1.446(5)   & 5810(58) & 0.332(18)  & \multirow{2}{*}{0.1(1)}  & 2.441(43) & \multirow{2}{*}{0.018(17)} & \multirow{2}{*}{1,6} & 
		\multirow{2}{*}{7.3(4)} & \multirow{2}{*}{7.41(1)} & \multirow{2}{*}{7.3(3)} & \multirow{2}{*}{7.3(2)} &  \multirow{2}{*}{7.31(6)} \\
		& B & 1.442(7)    & 5820(67) & 0.320(20) &                                          & 2.486(43) &    & & & & \\
		\hline
	\end{tabular}
	\tablefoot{References: 1- This work. 2- \citet{Heminiak_2014_07_7}. 3- \citet{Kiefer_2018_02_6}. 4- \citet{Graczyk_2016_10_0}. 5- \citet{Griffin_2002_02_8}. 6- \citet{Graczyk_2022_10_0}. 
		7- \citet{Luck_2017_01_5}. 8- \citet{Boeche_2016_03_1}.
		\tablefoottext{a}{Values from the literature were re-scaled according to our measured linear semi-major axis.}
		\tablefoottext{b}{Average value from the best fitted models (see text).}
	}
	\label{table__atmospheric_parameter}
\end{sidewaystable*}

\section{Comparison with Gaia parallaxes}

\begin{figure*}[!h]
	\centering
	\resizebox{\hsize}{!}{\includegraphics{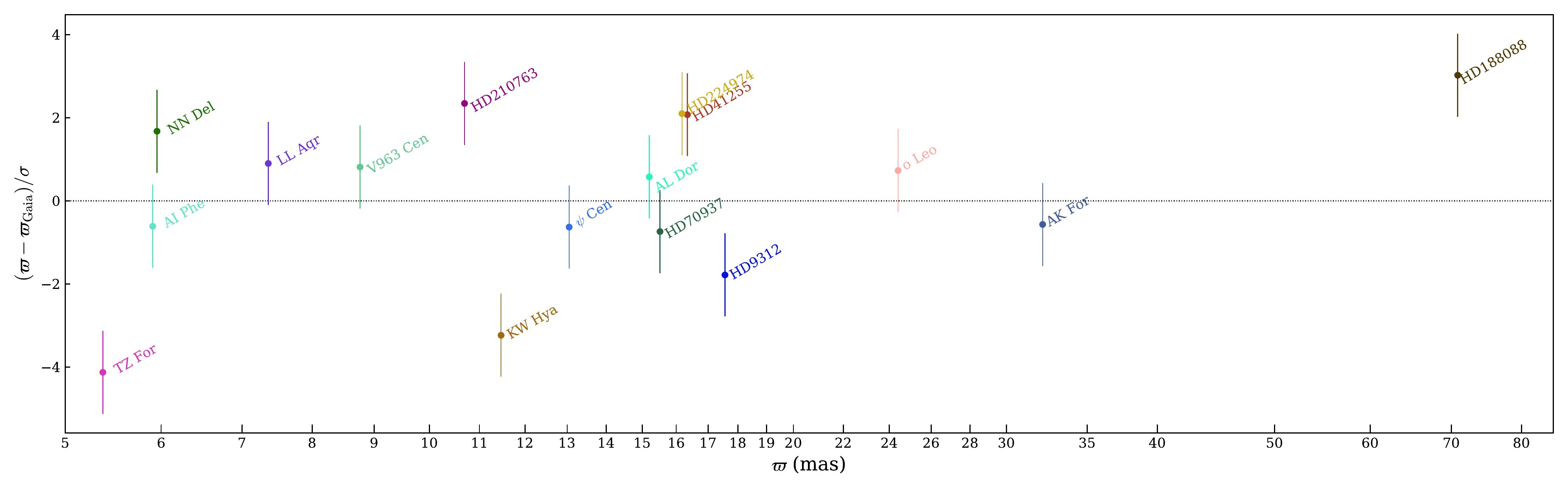}}
	\caption{Comparison of our orbital parallaxes with the Gaia DR3 measurements.}
	\label{figure__gaia_parallaxes}
\end{figure*}

We compared our measured orbital parallaxes from this study and from our previous works \citep{Gallenne_2016_02_0,Gallenne_2018_08_0,Gallenne_2019_10_0} with the Gaia third data release \citep{Gaia-Collaboration_2022_07_3}. We applied a zero-point offset to each Gaia value following the correction recipe from \citet{Lindegren_2021_05_7}. As we can see in Fig.~\ref{figure__gaia_parallaxes}, we found that $\sim 50$\,\% (8/16) of the sample is more than $1\sigma$ away.

To check if Gaia would detect the astrometric signature of the primary star around the centre of mass, that is to say the influence of the secondary star on the primary, we calculated a S/N as follows:
\begin{align}
	a_\mathrm{phot}(\mathrm{mas}) &= a(\mathrm{mas})\ \left( \dfrac{M_2}{M_1 + M_2} - \dfrac{1}{1 + 10^{0.4\Delta m_\mathrm{V}}} \right)  \label{equation__aphot}\\
	\mathrm{S/N} &= \dfrac{a_\mathrm{phot}}{\sigma_\mathrm{fov}}, \label{equation__SN}
\end{align}
where $a_\mathrm{phot}$ is the photocentre semi-major axis related to the angular semi-major axis $a$, the mass of the components $M_1, M_2$, and the magnitude difference $\Delta m_\mathrm{V} = m_2 - m_1$. While $\sigma_\mathrm{fov} = 34.2\,\mu$as is the along-scan accuracy per field of view crossing as defined by \citet[][we adopted the value for stars with $G < 10$\,mag]{Perryman_2014_12_3}. We estimated the magnitude difference between the components using our previously fitted isochrone models, and in the $V$ band as it is similar to the Gaia $G$ band. We list in Table~\ref{table__astrometric_signature} the value of $a_\mathrm{phot}$ and S/N for each star studied here and previously. We also mention the corresponding Gaia Renormalized Unit Weight Error (RUWE), which is the square root of the normalised $\chi^2$ of the astrometric fit to the along-scan observations. This parameter is a good metric to distinguish between a good or bad single-star astrometric solution, whose threshold is usually set to be around 1.4.

First, we notice that most systems have a large S/N and they will be detected by Gaia in the next data release (DR). DR3 provides results for about $800~000$ binary stars, including solutions for astrometric, spectroscopic, and eclipsing binaries \citep{Gaia-Collaboration_2022_07_3,Halbwachs_2022_06_7}. Unfortunately, only the known eclipsing system AK~For has published full orbital solutions, with a corrected parallax improving the agreement with our measurement from $0.7\sigma$ to $0.6\sigma$. However, their given eccentricity of $0.287\pm0.098$ is not at all consistent with the zero eccentricity we measured or previously published. This discrepancy likely comes from the fact that the Gaia solutions only fitted the astrometry, which poorly constrained some orbital parameters of eclipsing systems (as seen with the given eccentricity precision). Combined fits generally provide more robust solutions. DR4 should contain orbital solutions for all available binary stars, including those presented here as they have a large astrometric S/N, and our precise results will likely provide the best reference systems to test and validate the Gaia solutions. The systems HD70937, HD224974, V963 Cen, TZ For, AI Phe, and AL Dor  are also listed in the Gaia DR3 non-single star catalogue, but no astrometric solutions are derived.

\begin{figure}[!h]
	\centering
	\resizebox{\hsize}{!}{\includegraphics{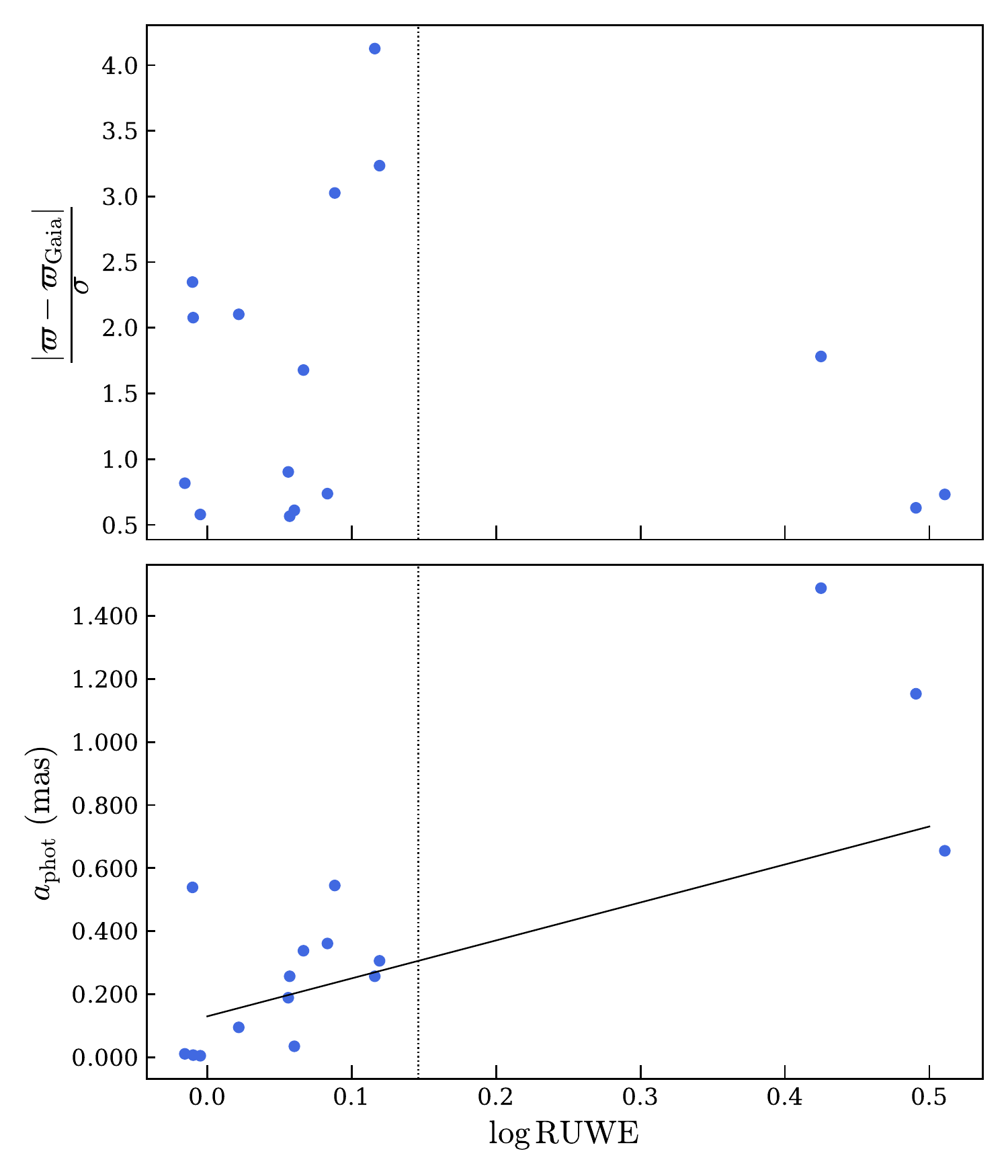}}
	\caption{Correlations between the RUWE metric, the absolute fractional parallax difference (top), and the photocentre semi-major axis of the systems (bottom). The solid line is the linear correlation law estimated by \citet{Stassun_2021_02_1}. The dotted vertical line denotes RUWE = 1.4.}
	\label{figure__gaia_correlations}
\end{figure}

In the top and bottom panel of Fig.~\ref{figure__gaia_correlations}, we plotted the RUWE parameter with respect to the relative parallax difference and the semi-major axis of the photocentre $a_\mathrm{phot}$, respectively. As we previously mentioned, a threshold of RUWE $ \lesssim 1.4$ is usually used to indicate well-behaved Gaia solutions \citep{L.Lindegren_2018_08_2,Lindegren_2021_05_5}, which we show\ with a dotted vertical line. We see that the stars with the largest RUWE ($\psi$~Cen, HD9312, and $o$~leo) have the largest S/N ($> 15$), for which we may think the Gaia parallax is the most biased and less reliable. This is, however, not the case because the agreement with our measurements is $\leq 2\sigma$ (0.6, 1.8, and $0.7\sigma$, respectively). Those stars also have the largest photometric semi-major axis, but they still provide a consistent parallax. The most discrepant Gaia parallaxes are in the 'good' RUWE range $1.0-1.4$. This has already been reported by \citet{Stassun_2021_02_1}, who compare the Gaia parallaxes with benchmark eclipsing binaries and show that the RUWE is highly sensitive to unresolved companions. They also report a correlation with the photocentre semi-major axis, which we also see in the bottom panel of Fig.~\ref{figure__gaia_correlations}. As \citet{Stassun_2021_02_1} and \citet{Kervella_2022_01_0} suggested, a RUWE slightly larger than one may imply the presence of unseen binaries.

\begin{table}[!ht]
	\centering
	\caption{Additional information about the systems and their detection with Gaia.}
	\begin{tabular}{ccccc}
		\hline
		\hline
		Star    & RUWE & $a_\mathrm{phot}$  & S/N & $\dfrac{\varpi -\varpi_\mathrm{Gaia}}{\sigma}$  \\
		&             &   (mas)            &          \\
		\hline
		TZ For & 1.3 & 0.257 & 7.5 & -4.1 \\
		AI Phe & 1.1 & 0.035 & 1.0 & -0.6 \\
		AL Dor & 1.0 & 0.005 & 0.1 & 0.6 \\
		KW Hya & 1.3 & 0.306 & 8.9 & -3.2 \\
		NN Del & 1.2 & 0.338 & 9.9 & 1.7 \\
		$\psi$ Cen & 3.1 & 1.153 & 33.7 & -0.6 \\
		AK For & 1.1 & 0.257 & 7.5 & -0.6 \\
		HD9312 & 2.7 & 1.488 & 43.5 & -1.8 \\
		HD41255 & 1.0 & -0.007 & -0.2 & 2.1 \\
		HD70937 & 1.2 & 0.361 & 10.6 & -0.7 \\
		HD210763 & 1.0 & 0.539 & 15.8 & 2.3 \\
		HD224974 & 1.1 & 0.095 & 2.8 & 2.1 \\
		HD188088 & 1.2 & -0.545 & -15.9 & 3.0 \\
		LL Aqr & 1.1 & 0.189 & 5.5 & 0.9 \\
		$o$ Leo & 3.2 & 0.655 & 19.2 & 0.7 \\
		V963 Cen & 1.0 & 0.011 & 0.3 & 0.8 \\
		\hline
	\end{tabular}
	\tablefoot{\#2: RUWE parameter as estimated by the Gaia team. \#3: estimated photocentre semi-major axis (Equ.~\ref{equation__aphot}). \#4: signal-to-noise ratio to detect the astrometric signature (Equ.~\ref{equation__SN}). \#5: relative difference with our measured orbital parallax.
	}
	\label{table__astrometric_signature}
\end{table}

\section{Conclusions}
\label{section__conclusions}

We have reported new interferometric and spectroscopic observations of double-lined binary systems. We simultaneously fitted the astrometry and RVs to obtain extremely precise and accurate masses and distances for ten systems. We reached uncertainties as low as 0.03\,\% and an average precision of $\sim 0.2$\,\%. A comparison with previous studies and different datasets demonstrated that our measurements are  both precise and accurate. This was possible thanks to the precision and sensitivity of the GRAVITY instrument, which provided exquisite differential astrometry, with a median r.m.s. of $\sim16\,\mu$as.

We confronted our measurements with additional observables to four stellar evolution models and we show that theory is clearly deficient for most of the systems when fitting one common isochrone for the components in a system. We estimated an average age for the system taking into account the uncertainty on the metallicity and the scatter between the ages given by each model. In four cases, the models give a different age for a given system and this may lead to a wrong estimate when using a single evolution model. To reconcile the models, it is likely that a fine-tuning of the models of each star in a given system is necessary, as was done by \citet{Graczyk_2016_10_0}. With such precision as to the masses, stellar interior parameters such as the mixing length and envelop overshooting can now be better constrained and lead to an improved calibration of stellar evolution models.

Our very precise orbital parallaxes also provide a stringent test of Gaia measurements. We found that 50\,\% (8/16 stars, including our previous works) of our sample is $>1\sigma$ away from the Gaia parallax, and within the 'nominal' RUWE range $1 - 1.4$. This can be problematic for stars with unresolved companions which would bias the parallax. We also confirm the correlation between the photocentre semi-major axis and the RUWE parameter reported by \citet{Stassun_2021_02_1}, that is to say the larger the photocentre motion is, the larger the RUWE. This is somewhat expected for large RUWE, meaning the poorly constrained Gaia 5- and 6-parameter astrometric solutions, but not for RUWE $< 1.4$ which is the frequently used cutoff for reliable Gaia astrometry. To reconcile most of the Gaia parallaxes within $1\sigma$ with our measurements, we need to inflate the Gaia errorbars by a factor of two. Several other systems are being observed and will provide a large sample of benchmark stars with high-precision masses and distances.


\begin{acknowledgements}
	We thank F. Widmann and J. Stadler for making public the code to create the GRAVITY aberration maps. A.G. acknowledges the support of the French Agence Nationale de la Recherche (ANR), under grant ANR-15-CE31-0012-01 (project UnlockCepheids). A.G. also acknowledges financial support from the ANID-ALMA fund No. ASTRO20-0059. The research leading to these results has received funding from the European Research Council (ERC) under the European Union's Horizon 2020 research and innovation programme under grant agreement No 695099 (project CepBin), No 951549 (project UniverScale) and from the National Science Center, Poland grants MAESTRO UMO-2017/26/A/ST9/00446 and BEETHOVEN UMO-2018/31/G/ST9/03050. B.P. acknowledges support from the Polish National Science Center grant SONATA BIS 2020/38/E/ST9/00486. We also acknowledge support from the Polish Ministry of Science and Higher Education grant DIR/WK/2018/09. This work has made use of data from the European Space Agency (ESA) mission {\it Gaia} (\url{https://www.cosmos.esa.int/gaia}), processed by the {\it Gaia} Data Processing and Analysis Consortium (DPAC, \url{https://www.cosmos.esa.int/web/gaia/dpac/consortium}). Funding for the DPAC has been provided by national institutions, in particular the institutions participating in the {\it Gaia} Multilateral Agreement.
\end{acknowledgements}


\bibliographystyle{aa}   
\bibliography{/Users/alex/Sciences/Articles/bibliographie}


\begin{appendix} 

	\section{Spectral calibration accuracy from telluric line fits}
	
	\begin{figure*}[!h]
		\centering
		\resizebox{\hsize}{!}{\includegraphics[width = \linewidth]{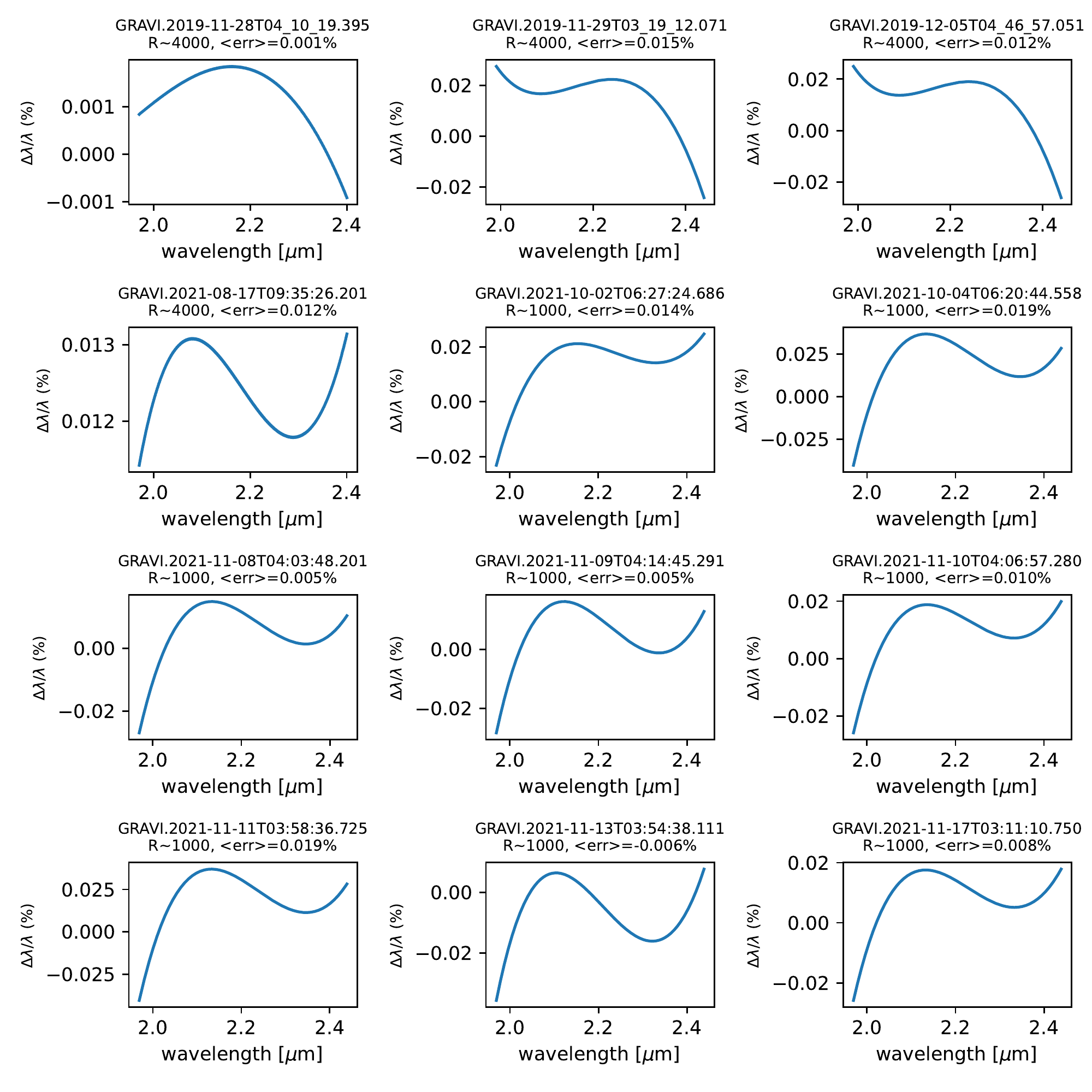}}
		\caption{Accuracy of the spectral calibration by fitting the telluric features in the GRAVITY spectra for each of our epoch measurements for AK~For.}
		\label{appendix__specCal}
	\end{figure*}

	\section{Relative astrometric position of the secondary determined from VLTI/GRAVITY.}
	
	\longtab[1]{
		\begin{landscape}
			\begin{longtable}{ccccccccccc}
				\caption{\label{table__astrometry}Relative astrometric position of the secondary component for all systems.}\\
				\hline\hline
				Date  &  MJD  &  Baselines  & Sp. Res. & Calibrators &  $\Delta \alpha$   &  $\Delta \delta$      & $\sigma_\mathrm{PA}$  & $\sigma_\mathrm{maj}$   &  $\sigma_\mathrm{min}$        &  $f$  \\
				& (day)  &                      &               & \#               &  (mas)                       & (mas)                 & ($^\circ$)      & (mas)         &  (mas)  &  (\%)  \\
				\hline
				\endfirsthead
				\caption{continued.}\\
				\hline\hline
				Date  &  MJD  &  Baselines  & Sp. Res. & Calibrators &  $\Delta \alpha$   &  $\Delta \delta$      & $\sigma_\mathrm{PA}$  & $\sigma_\mathrm{maj}$   &  $\sigma_\mathrm{min}$        &  $f$  \\
				& (day)  &                      &               & \#               &  (mas)                       & (mas)                 & ($^\circ$)      & (mas)         &  (mas)  &  (\%)  \\
				\hline
				\endhead
				\hline
				\endfoot
				\multicolumn{11}{c}{AK~For} \\
				2019-11-28  &  58815.174  &  A0-G1-J2-J3  &  HIGH    & 1 &  0.284 & 1.351 & 170.3 & 0.008 & 0.008  &   $71.91\pm0.08$ \\
				2019-11-29  &  58816.256  &  A0-G1-J2-J3  &  HIGH    & 1 &  $-0.148$ & $-1.234$ & 159.9 & 0.008 & 0.008  & $72.50\pm0.12$ \\
				2019-12-05  &  58822.199  &  A0-G1-J2-J3  &  HIGH    & 1 &  0.111 & 1.172 & 164.1 & 0.008 & 0.008  &   $74.52\pm0.17$ \\
				2021-08-17  &  59443.400  &  A0-G1-J2-J3  &  HIGH    & 2 &  0.219 & 1.435 & 173.1 & 0.008 & 0.008  &   $70.01\pm0.07$ \\
				2021-10-02  &  59489.269  &  A0-G1-J2-K0  &  MEDIUM  & 3 &  $-0.242$ & $-1.566$ & 110.1 & 0.008 & 0.008  & $72.04\pm0.15$ \\
				2021-10-04  &  59491.264  &  A0-G1-J2-K0  &  MEDIUM  & 3 &  0.252 & 1.571 & 125.4 & 0.008 & 0.008  &   $69.38\pm0.18$ \\
				2021-11-08  &  59526.169  &  A0-G1-J2-J3  &  MEDIUM  & 3 &  $-0.183$ & $-0.510$ & 171.5 & 0.008 & 0.008  & 71.70     \\
				2021-11-09  &  59527.177  &  A0-G1-J2-J3  &  MEDIUM  & 3 &  0.266 & 1.649 & 171.0 & 0.008 & 0.008  &   $71.11\pm0.02$ \\
				2021-11-10  &  59528.171  &  A0-G1-J2-J3  &  MEDIUM  & 3 &  0.178 & 0.462 & 161.5 & 0.008 & 0.008  &   71.70     \\
				2021-11-11  &  59529.166  &  A0-G1-J2-J3  &  MEDIUM  & 3 &  $-0.244$ & $-1.645$ & 166.8 & 0.008 & 0.008  & $71.09\pm0.03$ \\
				2021-11-13  &  59531.163  &  A0-G1-J2-J3  &  MEDIUM  & 3 &  0.264 & 1.650 & 170.0 & 0.008 & 0.008  &   $70.51\pm0.03$ \\
				2021-11-17  &  59535.133  &  A0-G1-J2-K0  &  MEDIUM  & 3 &  0.258 & 1.623 & 112.0 & 0.008 & 0.008  &   $73.89\pm0.11$ \\
				\hline
				\multicolumn{11}{c}{HD9312} \\
				2020-12-09  &  59192.090  &  A0-G1-J2-J3  &  HIGH  & 1 &  $-2.535$ & $-2.703$ & 126.0 & 0.050 & 0.050  & $8.59\pm0.01$  \\
				2020-12-14  &  59197.091  &  A0-G1-J2-J3  &  HIGH  & 1 &  0.658 & $-1.114$ & 151.5 & 0.050 & 0.050  &  7.84  \\                   
				2021-10-12  &  59499.179  &  K0-G2-D0-J3  &  HIGH  & 1 &  2.985 & 2.409 & 112.0 & 0.050 & 0.050  & $8.07\pm0.01$  \\
				2021-11-06  &  59524.100  &  K0-G2-D0-J3  &  HIGH  & 1 &  $-0.492$ & $-1.600$ & 78.7 & 0.050 & 0.050  & $8.34\pm0.04$  \\
				2021-11-21  &  59539.096  &  K0-G2-D0-J3  &  HIGH  & 1 &  0.821 & 1.664 & 82.9 & 0.050 & 0.050  & $7.55\pm0.71$  \\   
				2022-08-27  &  59818.296  &  K0-G2-D0-J3  &  HIGH  & 1 &  0.948 & $-0.844$ & 90.3 & 0.050 & 0.050  &   7.84  \\
				2022-08-29  &  59820.287  &  K0-G2-D0-J3  &  HIGH  & 1 &  2.118 & 0.235 & 168.5 & 0.050 & 0.050  & $7.27\pm0.01$  \\
				2022-10-26  &  59878.169  &  K0-G2-D0-J3  &  HIGH  & 1 &  $-4.567$ & $-2.676$ & 116.7 & 0.050 & 0.050  & $7.23\pm0.01$  \\
				2022-11-22  &  59905.087  &  K0-G2-D0-J3  &  HIGH  & 1 &  0.284 & 1.303 & 124.0 & 0.050 & 0.050  &  7.84  \\      
				2022-11-25  &  59908.053  &  K0-G2-D0-J3  &  HIGH  & 1 &  $-1.936$ & $-0.049$ & 168.8 & 0.050 & 0.050  & $7.81\pm0.01$  \\  
				\hline
				\multicolumn{11}{c}{HD41255} \\
				2021-01-17  &  59231.227  &  A0-G1-J2-J3  &  HIGH    & 1 & 4.649 & $-10.969$ & 152.4 & 0.008 & 0.008 & $97.12\pm0.03$ \\
				2021-01-19  &  59233.091  &  A0-G1-J2-J3  &  HIGH    & 1 & 4.040 & $-10.665$ & $-155.1$ & 0.026 & 0.008 & $96.78\pm0.78$ \\
				2022-01-13  &  59592.128  &  K0-B2-D0-J3  &  MEDIUM  & 2 & 3.227 & 4.869 & 135.8 & 0.008 & 0.008 & $95.83\pm0.02$ \\
				2022-02-11  &  59621.079  &  A0-G1-J2-J3  &  MEDIUM  & 2 & 11.306 & $-4.346$ & 91.0 & 0.008 & 0.008 & $95.04\pm0.02$ \\
				2022-03-01  &  59639.104  &  A0-G1-J2-J3  &  MEDIUM  & 2 & 11.961 & $-9.049$ & 130.4 & 0.008 & 0.008 & $96.17\pm0.02$ \\
				2022-03-25  &  59663.007  &  A0-G1-J2-J3  &  MEDIUM  & 2 & 8.463 & $-11.793$ & 114.4 & 0.008 & 0.008 & $97.42\pm0.03$ \\
				2022-04-10  &  59679.987  &  A0-G1-J2-K0  &  MEDIUM  & 2 & 3.388 & $-10.345$ & 97.9 & 0.008 & 0.008 & $97.45\pm0.01$ \\
				2022-10-21  &  59873.333  &  A0-G1-J2-K0  &  MEDIUM  & 2 & $-3.594$ & 6.568 & 94.8 & 0.008 & 0.008 & $94.06\pm0.32$ \\
				2022-11-21  &  59904.282  &  A0-B2-D0-J3  &  MEDIUM  & 2 & 8.660 & 0.041 & 140.7 & 0.008 & 0.008 & $94.87\pm0.01$ \\
				2022-11-27  &  59910.260  &  A0-G2-J2-J3  &  MEDIUM  & 2 & 10.089 & $-1.959$ & 115.3 & 0.008 & 0.008 & $95.02\pm0.08$ \\
				2022-12-15  &  59928.229  &  A0-G1-J2-K0  &  MEDIUM  & 2 & 12.086 & $-7.306$ & 88.3 & 0.008 & 0.008 & $95.90\pm0.04$ \\
				2022-12-19  &  59932.230  &  A0-G1-J2-K0  &  MEDIUM  & 2 & 12.095 & -8.308 & 90.5 & 0.008 & 0.008 & $95.56\pm0.05$ \\
				2022-12-30  &  59943.174  &  A0-G2-J2-J3  &  MEDIUM  & 2 & 11.379 & -10.398 & 98.7 & 0.008 & 0.008 & $95.97\pm0.04$\\
				2023-01-24  &  59968.181  &  A0-G1-J2-J3  &  MEDIUM  & 2 & 6.123 & -11.468 & 143.9 & 0.008 & 0.008 & $94.81\pm0.05$ \\
				\hline
				\multicolumn{11}{c}{HD70937} \\
				2020-12-14  &  59197.328  &  A0-G1-J2-J3  &  HIGH  & 1 & 0.987 & 4.335 & $-163.9$ & 0.012 & 0.012  &  $59.94\pm0.02$ \\
				2021-01-17  &  59231.248  &  A0-G1-J2-J3  &  HIGH  & 1 & 1.980 & 4.983 & 139.6 & 0.012 & 0.012  &  $60.20\pm0.01$ \\
				2021-03-07  &  59280.125  &  A0-G1-J2-J3  &  HIGH  & 1 & 0.781 & 3.943 & $-168.2$ & 0.012 & 0.012  &  $58.73\pm0.03$ \\
				2022-02-28  &  59638.205  &  A0-G1-J2-K0  &  HIGH  & 1 &$ -0.491$ & 0.882 & 86.6 & 0.012 & 0.012  &  59.8 \\
				2022-03-11  &  59649.093  &  A0-B2-D0-J3  &  HIGH  & 2 & 1.947 & 5.070 & 136.3 & 0.012 & 0.012  &  $60.05\pm0.01$ \\
				2022-03-24  &  59662.155  &  A0-G1-J2-J3  &  HIGH  & 2 & $-1.062$ & $-2.178$ & 145.3 & 0.012 & 0.012  &  $60.48\pm0.01$ \\
				2022-03-30  &  59668.077  &  A0-G1-J2-J3  &  HIGH  & 2 & 0.073 & 2.446 & 153.1 & 0.012 & 0.012  &  $60.07\pm0.01$ \\
				2022-04-02  &  59671.128  &  A0-G1-J2-J3  &  HIGH  & 2 & 0.932 & 4.217 & 152.3 & 0.012 & 0.012  &  $58.06\pm0.03$ \\
				2022-04-11  &  59680.051  &  A0-G1-J2-K0  &  HIGH  & 2 & 1.966 & 4.097 & 99.0 & 0.012 & 0.012  &  $60.26\pm0.01$ \\
				2022-05-16  &  59715.000  &  A0-G1-J2-J3  &  HIGH  & 2 & $-0.053$ & $-1.501$ & 148.0 & 0.012 & 0.012  &  $60.41\pm0.03$ \\
				2022-12-15  &  59928.251  &  A0-G1-J2-K0  &  HIGH  & 2 &  1.972 & 5.022 & 104.1 & 0.012 & 0.012 & $60.18\pm0.01$ \\
				2022-12-19  &  59932.290  &  A0-G1-J2-K0  &  HIGH  & 2 & 1.844 & 3.401 & 104.2 & 0.012 & 0.012 & $60.57\pm0.01$ \\
				2022-12-30  &  59943.271  &  A0-G2-J2-J3  &  HIGH  & 2 & -0.897 & -0.561 & 77.0 & 0.012 & 0.012 & $57.06\pm0.09$ \\
				2023-01-27  &  59971.241  &  A0-G1-J2-K0  &  HIGH  & 2 & -0.891 & -0.527 & 73.6 & 0.012 & 0.012 & $59.42\pm0.12$ \\
				\hline
				\multicolumn{11}{c}{HD210763} \\
				2021-08-16  &  59442.248  &  A0-G1-J2-J3  &  HIGH  & 1 & 2.841 & 0.574 & 164.0 & 0.020 & 0.020 & $38.04\pm0.11$ \\
				2021-08-17  &  59443.241  &  A0-G1-J2-J3  &  HIGH  & 1 & 2.362 & 0.284 & 131.7 & 0.020 & 0.020 & $37.31\pm0.01$ \\
				2021-08-21  &  59447.268  &  A0-G1-J2-J3  &  HIGH  & 2 & $-1.254$ & $-0.525$ & $-169.5$ & 0.020 & 0.020 & $37.82\pm0.02$ \\
				2021-09-06  &  59463.216  &  A0-G1-J2-J3  &  HIGH  & 2 & 0.870 & 2.112 & 140.2 & 0.020 & 0.020 & $37.39\pm0.00$ \\
				2021-09-10  &  59467.236  &  A0-G1-J2-J3  &  HIGH  & 2 & 1.935 & 2.314 & $-151.7$ & 0.021 & 0.020 & $36.34\pm0.19$ \\
				2021-11-08  &  59526.088  &  A0-G1-J2-J3  &  HIGH  & 2 & 3.183 & 0.784 & 132.6 & 0.020 & 0.020 & $36.05\pm0.00$ \\
				2021-11-10  &  59528.036  &  A0-G1-J2-J3  &  HIGH  & 2 & 2.341 & 0.256 & 130.7 & 0.020 & 0.020 & $36.15\pm0.00$ \\
				2022-08-06  &  59797.315  &  A0-G1-J2-J3  &  HIGH  & 2 & $-0.516$ & 1.579 & 110.5 & 0.020 & 0.020 & $38.06\pm0.02$ \\
				2022-08-13  &  59804.261  &  A0-G1-J2-K0  &  HIGH  & 1 & 1.411 & 2.250 & 95.5 & 0.020 & 0.020 & $36.32\pm0.01$ \\
				\hline
				\multicolumn{11}{c}{HD224974} \\
				2020-12-05  &  59188.035  &  A0-G1-J2-J3  &  HIGH  & 1 & $-1.522$ & $-1.575$ & 163.4 & 0.008 & 0.008 & $89.70\pm0.02$ \\
				2020-12-09  &  59192.037  &  A0-G1-J2-J3  &  HIGH  & 1 & $-1.056$ & 1.288 & 136.2 & 0.008 & 0.008 & $86.72\pm0.18$ \\
				2021-08-13  &  59439.381  &  A0-G1-J2-J3  &  HIGH  & 1 & 0.540 & 1.746 & 158.1 & 0.008 & 0.008 & $89.10\pm0.01$ \\
				2021-08-16  &  59442.397  &  A0-G1-J2-J3  &  HIGH  & 1 & $-0.449$ & $-1.619$ & 174.1 & 0.016 & 0.009 & $88.00\pm0.88$ \\
				2021-08-17  &  59443.325  &  A0-G1-J2-J3  &  HIGH  & 1 & $-1.247$ & $-1.719$ & 147.3 & 0.008 & 0.008 & $89.57\pm0.02$ \\
				2021-08-20  &  59446.346  &  A0-G1-J2-J3  &  HIGH  & 1 & $-1.731$ & 0.185 & 109.1 & 0.008 & 0.008 & $89.31\pm0.03$ \\
				2021-09-05  &  59462.326  &  A0-G1-J2-J3  &  HIGH  & 1 & 0.859 & $-0.081$ & 98.2 & 0.008 & 0.008 & 90 \\
				2021-09-06  &  59463.294  &  A0-G1-J2-J3  &  HIGH  & 1 & $-0.011$ & $-1.325$ & 94.8 & 0.010 & 0.008 & $90.17\pm0.09$ \\
				2021-09-07  &  59464.312  &  A0-G1-J2-J3  &  HIGH  & 1 & $-0.994$ & $-1.758$ & 161.1 & 0.008 & 0.008 & $88.71\pm0.02$ \\
				2021-09-10  &  59467.263  &  A0-G1-J2-J3  &  HIGH  & 1 & $-1.830$ & $-0.127$ & 119.5 & 0.008 & 0.008 & $88.66\pm0.02$ \\
				2021-09-11  &  59468.251  &  A0-G1-J2-J3  &  HIGH  & 1 & $-1.506$ & 0.636 & 83.6 & 0.008 & 0.008 & $89.52\pm0.03$ \\
				2021-09-15  &  59472.297  &  A0-G1-J2-J3  &  HIGH  & 1 & 0.985 & 0.966 & 149.8 & 0.008 & 0.008 & $89.22\pm0.05$ \\
				\hline
				\multicolumn{11}{c}{HD188088} \\
				2021-08-14  &  59440.242  &  A0-G1-J2-J3  &  HIGH  & 1 & 20.930 & $-3.439$ & 69.1 & 0.031 & 0.018 & $89.35\pm0.36$ \\
				2021-08-16  &  59442.223  &  A0-G1-J2-J3  &  HIGH  & 2 & 19.322 & $-2.484$ & 153.1 & 0.009 & 0.008 & $90.72\pm0.01$ \\
				2021-08-17  &  59443.222  &  A0-G1-J2-J3  &  HIGH  & 2 & 18.358 & $-1.959$ & 150.4 & 0.009 & 0.008 & $89.12\pm0.01$ \\
				2021-09-06  &  59463.163  &  A0-G1-J2-J3  &  HIGH  & 1 & $-8.522$ & 5.264 & 149.3 & 0.008 & 0.008 & $90.56\pm0.00$ \\
				2021-10-06  &  59493.069  &  A0-G1-J2-K0  &  HIGH  & 1 & 15.104 & $-0.432$ & 129.6 & 0.009 & 0.008 & $91.74\pm0.01$ \\
				2022-05-30  &  59729.388  &  K0-G2-D0-J3  &  HIGH  & 1 & 12.345 & 0.721 & 131.0 & 0.008 & 0.008 & $92.10\pm0.01$ \\
				2022-08-07  &  59798.267  &  A0-G1-J2-J3  &  HIGH  & 1 & 14.071 & $-5.959$ & 141.1 & 0.008 & 0.008 & $89.00\pm0.01$ \\
				2022-08-14  &  59805.266  &  A0-G1-J2-K0  &  HIGH  & 1 & 23.436 & $-6.788$ & 139.6 & 0.009 & 0.008 & $91.70\pm0.02$ \\
				2022-08-29  &  59820.137  &  K0-G2-D0-J3  &  HIGH  & 1 & 15.835 & $-0.763$ & 124.8 & 0.009 & 0.008 & $90.99\pm0.01$ \\
				2022-08-30  &  59821.203  &  A0-G1-D0-J3  &  HIGH  & 1 & 14.594 & $-0.217$ & 142.9 & 0.009 & 0.008 & $87.20\pm0.01$ \\
				2022-09-01  &  59823.217  &  A0-G1-J2-J3  &  HIGH  & 1 & 12.150 & 0.804 & 157.1 & 0.008 & 0.008 & $87.85\pm0.02$ \\
				2022-09-02  &  59824.165  &  A0-G1-J2-J3  &  HIGH  & 1 & 10.831 & 1.298 & 145.0 & 0.008 & 0.008 & $88.76\pm0.00$ \\
				2022-09-03  &  59825.209  &  A0-G1-J2-J3  &  HIGH  & 1 & 9.415 & 1.816 & 154.0 & 0.008 & 0.008 & $89.11\pm0.01$ \\
				2022-09-04  &  59826.205  &  A0-G1-J2-K0  &  HIGH  & 1 & 7.997 & 2.307 & 148.9 & 0.008 & 0.008 & $92.46\pm0.01$ \\
				2022-09-28  &  59850.134  &  A0-G1-J2-J3  &  HIGH  & 1 & 22.252 & $-7.065$ & 132.4 & 0.009 & 0.008 & $88.51\pm0.01$ \\
				2022-10-27  &  59879.047  &  K0-G2-D0-J3  &  HIGH  & 1 & $-1.190$ & 4.866 & 128.5 & 0.008 & 0.008 & $92.15\pm0.01$ \\
				\hline
				\multicolumn{11}{c}{LL~Aqr} \\
				2021-09-10  &  59467.177  &  A0-G1-J2-J3  &  MEDIUM  & 1 &  $-0.539$ & 1.285 & 109.2 & 0.012 & 0.012 & $53.02\pm0.14$  \\
				2021-10-04  &  59491.179  &  A0-G1-J2-K0  &  MEDIUM  & 1 &  $-0.689$ & 1.647 & 112.5 & 0.013 & 0.012 &  53.3  \\
				2021-11-08  &  59526.061  &  A0-G1-J2-J3  &  HIGH    & 1 &  $-0.395$ & 0.865 & 171.3 & 0.012 & 0.012 & 53.3  \\
				2021-11-11  &  59529.008  &  A0-G1-J2-J3  &  MEDIUM  & 1 &  $-0.643$ & 1.515 & 122.7 & 0.012 & 0.012 & $52.91\pm0.12$  \\
				2021-11-12  &  59530.048  &  A0-G1-J2-J3  &  MEDIUM  & 1 &  $-0.681$ & 1.622 & 110.7 & 0.012 & 0.012 & $52.93\pm0.06$  \\
				2022-08-05  &  59796.310  &  A0-G1-J2-J3  &  MEDIUM  & 1 &  $-0.515$ & 1.270 & 118.1 & 0.012 & 0.012 & 53.3  \\
				2022-08-31  &  59822.283  &  A0-G1-J2-J3  &  MEDIUM  & 1 &  0.317 & $-0.835$ & 160.5 & 0.012 & 0.012 & 53.3  \\
				2022-09-29  &  59851.138  &  A0-G1-J2-J3  &  MEDIUM  & 1 &  $-0.574$ & 1.398 & 121.9 & 0.012 & 0.012 & $54.45\pm0.08$  \\
				2022-09-30  &  59852.134  &  A0-G1-J2-K0  &  MEDIUM  & 2 &  $-0.644$ & 1.573 & 110.5 & 0.013 & 0.012 & 53.3  \\
				\hline
				\multicolumn{11}{c}{$o$~Leo} \\
				2022-02-28  &  59638.185  &  A0-G1-J2-K0  &  HIGH  & 1 & $-0.917$ & 3.552 & 102.4 & 0.020 & 0.020 & $25.14\pm0.01$\\
				2022-03-02  &  59640.236  &  A0-G1-J2-J3  &  HIGH  & 1 & 1.215 & 4.266 & 162.2 & 0.020 & 0.020 & $25.51\pm0.01$\\
				2022-03-25  &  59663.102  &  A0-G1-J2-J3  &  HIGH  & 1 & $-2.104$ & $-3.248$ & 100.0 & 0.020 & 0.020 & $25.03\pm0.01$\\
				2022-03-26  &  59664.127  &  A0-G1-J2-J3  &  HIGH  & 1 & $-2.48$5 & $-1.660$ & 82.0 & 0.020 & 0.020 & $24.90\pm0.01$\\
				2022-03-30  &  59668.098  &  A0-G1-J2-J3  &  HIGH  & 1 & 0.060 & 4.287 & 117.4 & 0.020 & 0.020 & $24.90\pm0.01$\\
				2022-03-31  &  59669.118  &  A0-G1-J2-J3  &  HIGH  & 1 & 1.105 & 4.324 & 157.7 & 0.020 & 0.020 & $24.97\pm0.01$\\
				2022-04-01  &  59670.130  &  A0-G1-J2-J3  &  HIGH  & 1 & 1.967 & 3.516 & 130.9 & 0.020 & 0.020 & $24.75\pm0.01$\\
				2022-04-02  &  59671.150  &  A0-G1-J2-J3  &  HIGH  & 1 & 2.450 & 2.037 & 160.6 & 0.020 & 0.020 & $25.12\pm0.01$\\
				2023-01-27  &  59971.261  &  A0-G1-J2-K0  &  HIGH  & 1 & -1.286 & 3.109 & 110.6 & 0.020 & 0.020 &  $24.98\pm0.01$ \\
				\hline
				\multicolumn{11}{c}{V963~Cen} \\
				2020-02-28  &  58907.345  &  A0-G1-J2-J3  &  HIGH    & 1 & -0.529 & 1.636 & -165.3 & 0.008 & 0.008 & $94.98\pm0.30$ \\
				2020-03-01  &  58909.372  &  A0-G1-J2-J3  &  HIGH    & 1 & -0.495 & 1.667 & 46.4 & 0.008 & 0.008 & $95.15\pm0.66$ \\
				2022-03-01  &  59639.362  &  A0-G1-J2-J3  &  MEDIUM  & 2 & -0.463 & 1.438 & -147.6 & 0.008 & 0.008 & $98.32\pm0.41$ \\
				2022-03-31  &  59669.132  &  A0-G1-J2-J3  &  MEDIUM  & 2 & -0.430 & 1.255 & 173.0 & 0.008 & 0.008 & $95.83\pm0.26$ \\
				2022-04-01  &  59670.221  &  A0-G1-J2-J3  &  MEDIUM  & 2 & -0.475 & 1.501 & -177.3 & 0.008 & 0.008 & $94.86\pm0.19$ \\
				2022-04-02  &  59671.232  &  A0-G1-J2-J3  &  MEDIUM  & 2 & -0.507 & 1.639 & 125.9 & 0.022 & 0.011 & $95.72\pm4.48$ \\
				2022-04-03  &  59672.261  &  A0-G1-J2-J3  &  MEDIUM  & 2 & -0.493 & 1.657 & -157.9 & 0.008 & 0.008 & $96.95\pm0.20$ \\
				2022-08-04  &  59795.972  &  A0-G1-J2-J3  &  MEDIUM  & 2 & -0.402 & 1.432 & -148.7 & 0.008 & 0.008 & $96.07\pm0.50$ \\
				\hline
			\end{longtable}
			\tablefoot{Values without errors were kept fixed. The calibrator number is related to the number in Table~\ref{table__calibrators}.}
		\end{landscape}
	}
	
	\section{Radial velocities determined from our VLT/UVES observations.}

	\longtab[1]{
		\begin{landscape}
			\begin{longtable}{cccccc|cccccc}
				\caption{\label{table__rv} Measured radial velocities for all stars.}\\
				\hline\hline
				MJD             &       $V_1$   &       $\sigma_\mathrm{V_1}$   &  $V_2$  &       $\sigma_\mathrm{V_2}$ & Inst. & MJD &   $V_1$   &       $\sigma_\mathrm{V_1}$   &  $V_2$  &       $\sigma_\mathrm{V_2}$  & Inst.  \\
				(days)                  &       (\kms)  &       (\kms)  & (\kms) & (\kms) & (days) &      (\kms)  &       (\kms)  & (\kms) & (\kms)        \\
				\hline
				\endfirsthead
				\caption{continued.}\\
				\hline\hline
				MJD             &       $V_1$   &       $\sigma_\mathrm{V_1}$   &  $V_2$  &       $\sigma_\mathrm{V_2}$ & Inst. & MJD &   $V_1$   &       $\sigma_\mathrm{V_1}$   &  $V_2$  &       $\sigma_\mathrm{V_2}$  & Inst.  \\
				(days)                  &       (\kms)  &       (\kms)  & (\kms) & (\kms) & (days) &      (\kms)  &       (\kms)  & (\kms) & (\kms)        \\
				\hline
				\endhead
				\hline
				\endfoot
				\multicolumn{12}{c}{AK~For}  \\
				59200.03232 & $-9.300$ & 0.073 & 15.186 & 0.161 & HARPS & 59439.37426 & $-53.918$ & 0.073 & 63.748 & 0.161 & HARPS \\
				59208.07606 & $-18.206$ & 0.073 & 24.853 & 0.161 & HARPS & 59440.30004 & $-45.709$ & 0.073 & 54.868 & 0.161 & HARPS \\
				59209.12947 & $-62.826$ & 0.073 & 73.709 & 0.161 & HARPS & 59511.32594 & $-66.804$ & 0.073 & 77.885 & 0.161 & HARPS \\
				59235.08570 & 63.851 & 0.073 & $-64.842$ & 0.161 & HARPS & 59512.21072 & $-22.065$ & 0.073 & 29.282 & 0.161 & HARPS \\
				59438.42522 & 40.481 & 0.073 & $-39.206$ & 0.161 & HARPS & 59513.20822 & 68.181 & 0.073 & $-69.933$ & 0.161 & HARPS \\
				\hline
				\multicolumn{12}{c}{HD9312}  \\
				59510.17125 & 25.186 & 0.031 & $-31.073$ & 0.246 & UVES & 59557.04329 & 20.175 & 0.031 & $-23.990$ & 0.246 & UVES \\
				59512.28142 & 29.769 & 0.031 & $-36.796$ & 0.246 & UVES & 59566.15560 & $-25.473$ & 0.031 & 35.778 & 0.246 & UVES \\
				59530.04681 & $-27.628$ & 0.031 & 38.493 & 0.246 & UVES & 59893.18459 & $-16.416$ & 0.031 & 23.632 & 0.246 & UVES \\
				59532.08202 & $-35.974$ & 0.031 & 49.478 & 0.246 & UVES & 59899.07437 & $-38.838$ & 0.031 & 52.891 & 0.246 & UVES \\
				59541.11543 & $-2.857$ & 0.031 & 6.384 & 0.246 & UVES & 59901.16776 & $-34.881$ & 0.031 & 48.349 & 0.246 & UVES \\
				59554.17390 & 27.824 & 0.031 & $-34.340$ & 0.246 & UVES & 59910.04219 & 18.555 & 0.031 & $-22.129$ & 0.246 & UVES \\
				\hline
				\multicolumn{12}{c}{HD41255} \\
				59533.34910 & -10.938 & 0.081 & 6.971 & 0.058 & UVES & 59640.12538 & -16.583 & 0.081 & 12.310 & 0.058 & UVES \\
				59551.30753 & 4.833 & 0.081 & -8.954 & 0.058 & UVES & 59645.14140 & -17.099 & 0.081 & 12.928 & 0.058 & UVES \\
				59556.33712 & 12.031 & 0.081 & -16.239 & 0.058 & UVES & 59866.36630 & 28.709 & 0.081 & -32.377 & 0.058 & UVES \\
				59563.23017 & 22.456 & 0.081 & -26.616 & 0.058 & UVES & 59878.35591 & 21.496 & 0.081 & -25.397 & 0.058 & UVES \\
				59580.15487 & 23.166 & 0.081 & -27.334 & 0.058 & UVES & 59893.34072 & 3.894 & 0.081 & -8.059 & 0.058 & UVES \\
				59591.08240 & 9.910 & 0.081 & -13.962 & 0.058 & UVES & 59942.34011 & -17.323 & 0.081 & 12.825 & 0.058 & UVES \\
				59596.10157 & 4.461 & 0.081 & -8.602 & 0.058 & UVES & 59957.30449 & -17.210 & 0.081 & 12.728 & 0.058 & UVES \\
				59612.17374 & -7.631 & 0.081 & 3.263 & 0.058 & UVES & 59208.18826 & -17.503 & 0.040 & 13.172 & 0.060 & HARPS \\
				59618.08028 & -10.483 & 0.081 & 6.214 & 0.058 & UVES & 59234.21303 & -12.303 & 0.040 & 8.033 & 0.060 & HARPS \\
				59623.07546 & -12.492 & 0.081 & 8.170 & 0.058 & UVES & 59511.37579 & -17.197 & 0.040 & 12.861 & 0.060 & HARPS \\
				\hline
				\multicolumn{12}{c}{HD70937} \\
				59513.34325 &   $-4.546$ &  0.066  & $-58.451$  & 0.077 & UVES  & 59610.15877 &  $-42.563$ &  0.066  & $-16.827$  & 0.077 & UVES \\
				59547.34384 &  $-68.817$ &  0.066  &  12.137  & 0.077 & UVES  & 59615.10141 &   $-8.371$ &  0.066  & $-54.692$  & 0.077 & UVES \\
				59563.23580 &    1.072 &  0.066  & $-64.833$  & 0.077 & UVES  & 59618.08526 &   $-0.255$ &  0.066  & $-63.438$  & 0.077 & UVES \\
				59566.29308 &    1.371 &  0.066  & $-65.303$  & 0.077 & UVES  & 59621.10894 &    1.858 &  0.066  & $-66.022$  & 0.077 & UVES \\
				59580.16282 &  $-71.002$ &  0.066  &  14.446  & 0.077 & UVES  & 59208.18449 & $-12.135$ & 0.060 & $-50.176$ & 0.090 & HARPS\\
				59589.32604 &   $-1.974$ &  0.066  & $-61.597$  & 0.077 & UVES  & 59234.22756 & $-3.590$ & 0.060 & $-59.305$ & 0.090 & HARPS \\
				59592.12271 &    1.913 &  0.066  & $-65.670$  & 0.077 & UVES  & 59235.22436 & $-7.554$ & 0.060 & $-55.047$ & 0.090 & HARPS \\
				59595.07390 &    0.190 &  0.066  & $-64.099$  & 0.077 & UVES  & 59373.95123 & $-4.571$ & 0.060 & $-58.236$ & 0.090 & HARPS \\
				59606.13544 &  $-97.210$ &  0.066  &  43.213  & 0.077 & UVES  & 59509.32495 & 2.195 & 0.060 & $-65.703$ & 0.090 & HARPS  \\
				59609.36583 &  $-52.096$ &  0.066  &  $-6.396$  & 0.077 & UVES  & 59510.37819 & 1.769 & 0.060 & $-65.239$ & 0.090 & HARPS  \\
				\hline
				\multicolumn{12}{c}{HD210763} \\
				59510.16368 & 6.805 & 0.072 & 24.530 & 0.172 & UVES & 59865.16489 & $-22.477$ & 0.072 & 58.547 & 0.172 & UVES \\
				59533.15095 & 77.493 & 0.072 & $-58.502$ & 0.172 & UVES & 59870.14780 & 44.534 & 0.072 & $-19.768$ & 0.172 & UVES \\
				59536.08947 & 59.289 & 0.072 & $-36.978$ & 0.172 & UVES & 59891.04787 & 7.924 & 0.072 & 23.192 & 0.172 & UVES \\
				59538.11783 & 48.205 & 0.072 & $-23.703$ & 0.172 & UVES & 59896.12739 & $-2.009$ & 0.072 & 34.779 & 0.172 & UVES \\
				59541.10544 & 35.941 & 0.072 & $-9.283$ & 0.172 & UVES & 59899.06219 & $-7.632$ & 0.072 & 41.407 & 0.172 & UVES \\
				59546.07020 & 21.353 & 0.072 & 7.949 & 0.172 & UVES & 59906.03501 & $-20.703$ & 0.072 & 56.594 & 0.172 & UVES \\
				59555.03566 & 1.999 & 0.072 & 30.262 & 0.172 & UVES & 59910.02409 & $-16.946$ & 0.072 & 52.280 & 0.172 & UVES \\
				59855.02342 & $-4.649$ & 0.072 & 37.526 & 0.172 & UVES & 59916.03523 & 69.377 & 0.072 & $-48.499$ & 0.172 & UVES \\
				59861.13714 & $-16.343$ & 0.072 & 51.187 & 0.172 & UVES & & & & &  \\
				\hline
				\multicolumn{12}{c}{HD224974} \\
				59508.15556 & $-62.238$ & 0.082 & 20.346 & 0.090 & UVES & 59896.13236 & 8.276 & 0.082 & $-51.298$ & 0.090 & UVES \\
				59509.22156 & $-47.072$ & 0.082 & 5.124 & 0.090 & UVES & 59899.06807 & 13.702 & 0.082 & $-56.886$ & 0.090 & UVES \\
				59510.28563 & $-30.107$ & 0.082 & $-12.233$ & 0.090 & UVES & 59901.03850 & $-71.848$ & 0.082 & 30.255 & 0.090 & UVES \\
				59512.27610 & 5.411 & 0.082 & $-48.392$ & 0.090 & UVES & 59910.02926 & $-5.435$ & 0.082 & $-37.349$ & 0.090 & UVES \\
				59535.06680 & 32.357 & 0.082 & $-75.868$ & 0.090 & UVES & 59911.04934 & $-61.462$ & 0.082 & 19.925 & 0.090 & UVES \\
				59536.09456 & 35.192 & 0.082 & $-79.004$ & 0.090 & UVES & 59925.06899 & $-44.436$ & 0.082 & 2.704 & 0.090 & UVES \\
				59537.11937 & $-14.930$ & 0.082 & $-27.758$ & 0.090 & UVES & 57362.18855 & 22.442 & 0.058 & $-65.573$ & 0.070 & HARPS \\
				59538.16960 & $-65.775$ & 0.082 & 24.243 & 0.090 & UVES & 59199.02701 & $-62.502$ & 0.058 & 21.026 & 0.070 & HARPS \\
				59541.11069 & $-48.431$ & 0.082 & 6.456 & 0.090 & UVES & 59202.01912 & $-16.083$ & 0.058 & $-26.735$ & 0.070 & HARPS \\
				59546.07537 & 36.675 & 0.082 & $-80.250$ & 0.090 & UVES & 59208.03110 & $-70.055$ & 0.058 & 28.753 & 0.070 & HARPS \\
				59554.15904 & $-8.518$ & 0.082 & $-33.955$ & 0.090 & UVES & 59209.06678 & $-69.171$ & 0.058 & 27.787 & 0.070 & HARPS \\
				59556.15975 & 28.780 & 0.082 & $-72.135$ & 0.090 & UVES & 59235.01630 & 2.792 & 0.058 & $-45.526$ & 0.070 & HARPS \\
				59563.10454 & $-37.906$ & 0.082 & $-4.236$ & 0.090 & UVES & 59374.40763 & 18.663 & 0.058 & $-61.711$ & 0.070 & HARPS \\
				59575.12233 & $-14.953$ & 0.082 & $-27.599$ & 0.090 & UVES & 59375.41487 & 35.429 & 0.058 & $-78.820$ & 0.070 & HARPS \\
				59578.07905 & 36.999 & 0.082 & $-80.400$ & 0.090 & UVES & 59438.35386 & 18.580 & 0.058 & $-61.634$ & 0.070 & HARPS \\
				59579.08825 & 25.140 & 0.082 & $-68.225$ & 0.090 & UVES & 59439.30190 & 34.679 & 0.058 & $-78.020$ & 0.070 & HARPS \\
				59865.16985 & 27.169 & 0.082 & $-70.958$ & 0.090 & UVES & 59440.20093 & 34.897 & 0.058 &$ -78.208$ & 0.070 & HARPS \\
				59870.16359 & $-66.655$ & 0.082 & 24.895 & 0.090 & UVES & 59511.18575 & $-14.741$ & 0.058 & $-27.894$ & 0.070 & HARPS \\
				59885.03753 & $-0.016$ & 0.082 & $-42.971$ & 0.090 & UVES & 59512.18261 & 3.669 & 0.058 & $-46.422$ & 0.070 & HARPS \\
				59891.05273 & $-70.426$ & 0.082 & 28.939 & 0.090 & UVES & 59513.19091 & 22.802 & 0.058 & $-66.019$ & 0.070 & HARPS \\
				59893.18047 & $-43.082$ & 0.082 & 1.240 & 0.090 & UVES & & & & & & \\
				\hline
				\multicolumn{12}{c}{HD188088} \\
				59508.14059 & $-41.300$ & 0.112 & 31.766 & 0.103 & UVES & 55722.24620 & $-9.665$ & 0.032 & $-0.357$ & 0.042 & HARPS \\
				59530.03570 & 11.775 & 0.112 & $-22.098$ & 0.103 & UVES & 55722.24832 & $-9.595$ & 0.032 & $-0.438$ & 0.042 & HARPS \\
				59532.07695 & 9.127 & 0.112 & $-19.153$ & 0.103 & UVES & 55811.03862 & $-49.606$ & 0.032 & 40.378 & 0.042 & HARPS \\
				59535.01529 & 5.161 & 0.112 & $-14.974$ & 0.103 & UVES & 55811.04000 & $-49.615$ & 0.032 & 40.386 & 0.042 & HARPS \\
				59538.02463 & 0.635 & 0.112 & $-10.774$ & 0.103 & UVES & 55812.01082 & $-56.913$ & 0.032 & 47.788 & 0.042 & HARPS \\
				59648.38497 & $-40.070$ & 0.112 & 30.769 & 0.103 & UVES & 55812.01226 & $-56.926$ & 0.032 & 47.802 & 0.042 & HARPS \\
				59667.39384 & 16.039 & 0.112 & $-26.409$ & 0.103 & UVES & 55812.07947 & $-57.463$ & 0.032 & 48.347 & 0.042 & HARPS \\
				59670.38139 & 11.879 & 0.112 & $-22.291$ & 0.103 & UVES & 55812.08088 & $-57.477$ & 0.032 & 48.374 & 0.042 & HARPS \\
				59859.14120 & 9.702 & 0.112 & $-20.365$ & 0.103 & UVES & 55812.08227 & $-57.492$ & 0.032 & 48.373 & 0.042 & HARPS \\
				59861.13277 & 7.025 & 0.112 & $-17.581$ & 0.103 & UVES & 55813.10141 & $-65.597$ & 0.032 & 56.613 & 0.042 & HARPS \\
				59865.10971 & 1.451 & 0.112 & $-11.822$ & 0.103 & UVES & 55813.10266 & $-65.604$ & 0.032 & 56.615 & 0.042 & HARPS \\
				59875.11233 & $-15.618$ & 0.112 & 6.102 & 0.103 & UVES & 55813.10391 & $-65.616$ & 0.032 & 56.632 & 0.042 & HARPS \\
				59881.05136 & $-33.649$ & 0.112 & 23.973 & 0.103 & UVES & 56137.05270 & $-39.461$ & 0.032 & 30.105 & 0.042 & HARPS \\
				59887.02527 & $-68.195$ & 0.112 & 59.308 & 0.103 & UVES & 56137.10775 & $-39.744$ & 0.032 & 30.390 & 0.042 & HARPS \\
				59891.06231 & 24.551 & 0.112 & $-35.258$ & 0.103 & UVES & 56137.24634 & $-40.471$ & 0.032 & 31.115 & 0.042 & HARPS \\
				59899.05115 & 19.541 & 0.112 & $-29.811$ & 0.103 & UVES & 56138.03680 & $-44.994$ & 0.032 & 35.624 & 0.042 & HARPS \\
				53279.13535 & $-30.323$ & 0.032 & 20.626 & 0.042 & HARPS & 56138.09850 & $-45.370$ & 0.032 & 35.993 & 0.042 & HARPS \\
				53280.03581 & $-33.872$ & 0.032 & 24.230 & 0.042 & HARPS & 56138.19538 & $-45.959$ & 0.032 & 36.610 & 0.042 & HARPS \\
				53332.01740 & $-66.817$ & 0.032 & 57.744 & 0.042 & HARPS & 56178.15045 & $-19.343$ & 0.032 & 9.510 & 0.042 & HARPS \\
				53891.33812 & $-48.004$ & 0.032 & 38.668 & 0.042 & HARPS & 56179.03994 & $-21.683$ & 0.032 & 11.878 & 0.042 & HARPS \\
				55721.13777 & $-55.138$ & 0.032 & 46.060 & 0.042 & HARPS & 56179.14919 & $-21.992$ & 0.032 & 12.181 & 0.042 & HARPS \\
				55721.26571 & $-50.750$ & 0.032 & 41.594 & 0.042 & HARPS & 56885.05123 & $-34.516$ & 0.032 & 24.963 & 0.042 & HARPS \\
				55721.26725 & $-50.693$ & 0.032 & 41.538 & 0.042 & HARPS & 56887.07058 & $-44.843$ & 0.032 & 35.440 & 0.042 & HARPS \\
				55721.34824 & $-47.606$ & 0.032 & 38.370 & 0.042 & HARPS & 57119.42519 & $-35.828$ & 0.032 & 26.336 & 0.042 & HARPS \\
				55721.35006 & $-47.533$ & 0.032 & 38.295 & 0.042 & HARPS & 57119.42980 & $-35.842$ & 0.032 & 26.350 & 0.042 & HARPS \\
				55721.43193 & $-44.265$ & 0.032 & 34.954 & 0.042 & HARPS & 57120.42356 & $-40.667$ & 0.032 & 31.326 & 0.042 & HARPS \\
				55722.14422 & $-13.435$ & 0.032 & 3.658 & 0.042 & HARPS & 57281.11653 & 14.179 & 0.032 & $-24.550$ & 0.042 & HARPS \\
				55722.14632 & $-13.350$ & 0.032 & 3.574 & 0.042 & HARPS & & & & & & \\
				\hline
				\multicolumn{12}{c}{$o$~Leo} \\
				59509.36177 & -28.463 & 0.085 & 88.090 & 0.085 & UVES & 59959.35809 & -27.395 & 0.085 & 86.393 & 0.085 & UVES \\
				59533.34135 & 58.279 & 0.085 & -9.871 & 0.085 & UVES & 55259.95943 & -16.980 & 0.054 & 75.010 & 0.082 & SOPHIE \\
				59536.33618 & -7.796 & 0.085 & 64.638 & 0.085 & UVES & 55261.95506 & -27.322 & 0.054 & 86.810 & 0.082 & SOPHIE \\
				59547.33935 & 66.906 & 0.085 & -19.924 & 0.085 & UVES & 55266.99879 & 66.284 & 0.054 & -18.776 & 0.082 & SOPHIE \\
				59566.28758 & -22.310 & 0.085 & 80.770 & 0.085 & UVES & 55268.02101 & 78.391 & 0.054 & -32.472 & 0.082 & SOPHIE \\
				59589.32020 & 80.659 & 0.085 & -35.189 & 0.085 & UVES & 55270.95938 & 57.327 & 0.054 & -8.642 & 0.082 & SOPHIE \\
				59592.12908 & 41.217 & 0.085 & 9.082 & 0.085 & UVES & 55271.96791 & 35.285 & 0.054 & 16.053 & 0.082 & SOPHIE \\
				59595.12107 & -20.433 & 0.085 & 78.642 & 0.085 & UVES & 55281.86844 & 71.859 & 0.054 & -24.846 & 0.082 & SOPHIE \\
				59596.12677 & -28.252 & 0.085 & 87.513 & 0.085 & UVES & 55283.91363 & 78.545 & 0.054 & -32.661 & 0.082 & SOPHIE \\
				59903.33916 & -0.123 & 0.085 & 55.737 & 0.085 & UVES & 55285.87765 & 48.729 & 0.054 & 1.006 & 0.082 & SOPHIE \\
				\hline
			\end{longtable}
		\end{landscape}
	}

				%
				%

			\section{Parameters of the  interferometric calibrators used for the GRAVITY observations}
			
			\begin{table*}[!ht]
				\centering
				\caption{Interferometric calibrators used for our GRAVITY observations taken from the \texttt{SearchCal} software.}
				\begin{tabular}{cccccc} 
					\hline
					\hline
					\# & HD         &       Sp.~type        &       $V$  & $K$  &  $\theta_\mathrm{UD}$  \\
					&       &                                       &       (mag)   &       (mag)   &       (mas)                            \\
					\hline
					\multicolumn{6}{c}{\object{AK~For}} \\
					1       &       \object{HD22575}  &  K0/1III  &  7.54  & 6.93  & $0.198\pm0.005$  \\
					2       &       \object{HD19873}  &  K0III/IVCNII/III  &  9.17  & 6.84  & $0.203\pm0.005$        \\
					3       &       \object{HD21967}  &  K1III  &  7.37  & 6.77  & $0.212\pm0.005$     \\
					\hline
					\multicolumn{6}{c}{\object{HD9312}} \\
					1       &  \object{HD8910}  &  K0III  &  7.98  & 5.39  & $0.409\pm0.010$ \\
					\hline
					\multicolumn{6}{c}{\object{HD41255}} \\
					1       &       \object{HD41648}  & K0III       &  9.15   & 6.90  & $0.196\pm0.005$   \\
					2       &       \object{HD41917}  & K1(III)       &  9.47   & 6.94  & $0.195\pm0.005$   \\
					\hline
					\multicolumn{6}{c}{\object{HD70937}} \\
					1       &       \object{HD69687}  & K1III       &  8.58   & 5.67  & $0.358\pm0.008$   \\
					2       &       \object{HD71263}  & K1III       &  8.48   & 5.86  & $0.328\pm0.008$   \\
					\hline
					\multicolumn{6}{c}{\object{HD210763}} \\
					1       &       \object{HD212458}  & K0III       &  8.48   & 5.80  & $0.332\pm0.008$   \\
					2       &       \object{HD211650}  & K2III       &  9.03   & 5.98  & $0.322\pm0.008$   \\
					\hline
					\multicolumn{6}{c}{\object{HD224974}} \\
					1       &       \object{HD318}  & K0III       &  8.55   & 6.09  & $0.288\pm0.007$   \\
					\hline
					\multicolumn{6}{c}{\object{HD188088}} \\
					1       &       \object{HD187516}  & K0III       &  7.52   & 4.93  & $0.534\pm0.013$   \\
					2       &       \object{HD188276}  & K2/3III    &  7.94  & 4.97  & $0.506\pm0.011$   \\
					\hline
					\multicolumn{6}{c}{\object{LL~Aqr}} \\
					1       &       \object{HD214548}  & K0       &  10.4   & 7.88  & $0.127\pm0.004$   \\
					2       &       \object{HD213691}  & G5V       &  9.47   & 7.95  & $0.113\pm0.003$   \\
					\hline
					\multicolumn{6}{c}{\object{$o$~Leo}} \\
					1       &       \object{HD83511}  & M0       &  8.60   & 3.89  & $1.047\pm0.074$   \\
					\hline
					\multicolumn{6}{c}{\object{V963~Cen}} \\
					1       &       \object{HD119887}  & K1/2III       &  10.2   & 7.51 & $0.154\pm0.004$   \\
					2       &       \object{HD114082}  & F3V       &  8.20   & 7.16  & $0.146\pm0.004$   \\
					\hline
				\end{tabular}
				\label{table__calibrators}
			\end{table*}
			
			%
			
			\section{Isochrones fit}
			
			\begin{figure*}[!h]
				\centering
				\resizebox{\hsize}{!}{\includegraphics{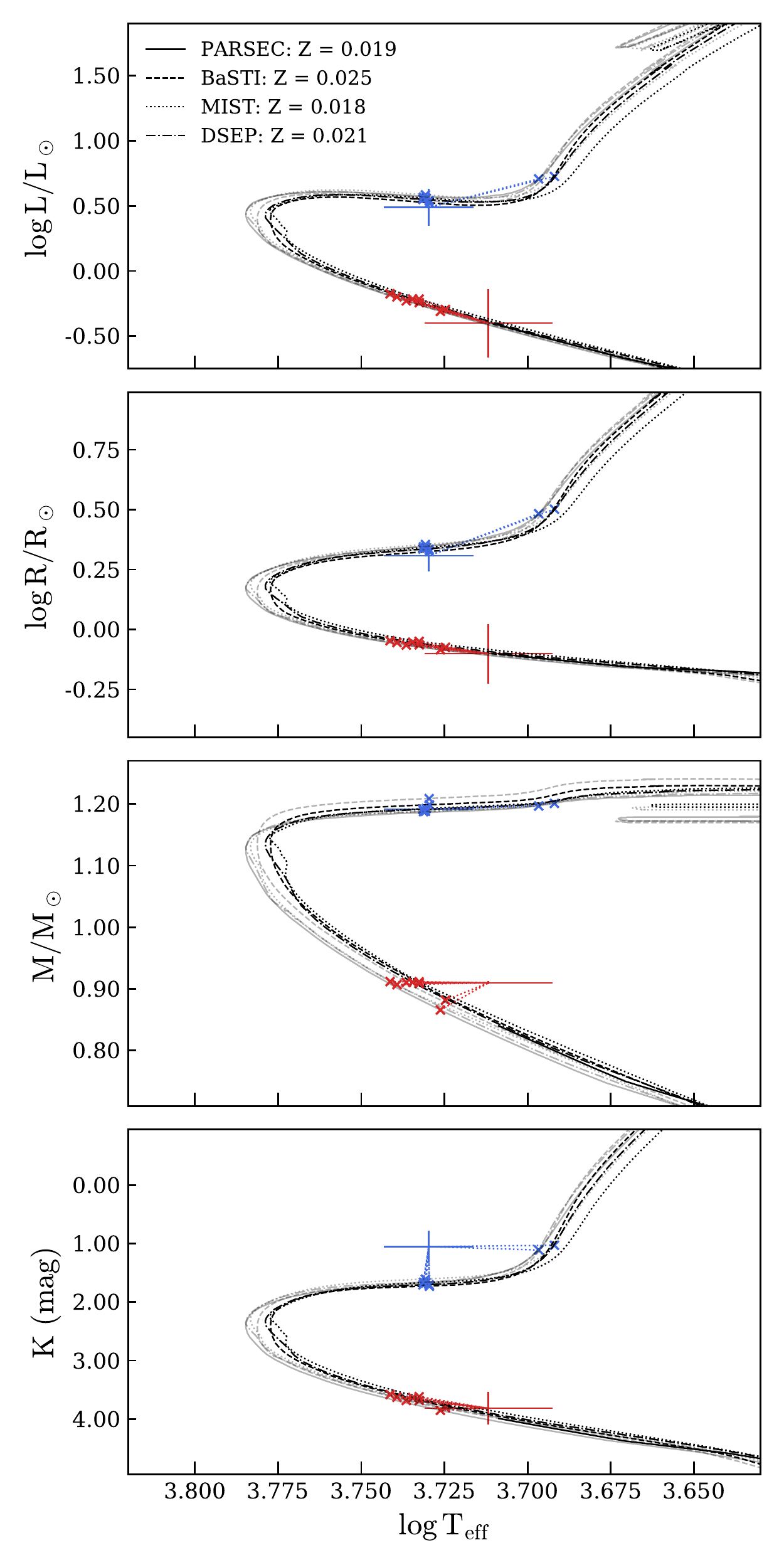}\includegraphics{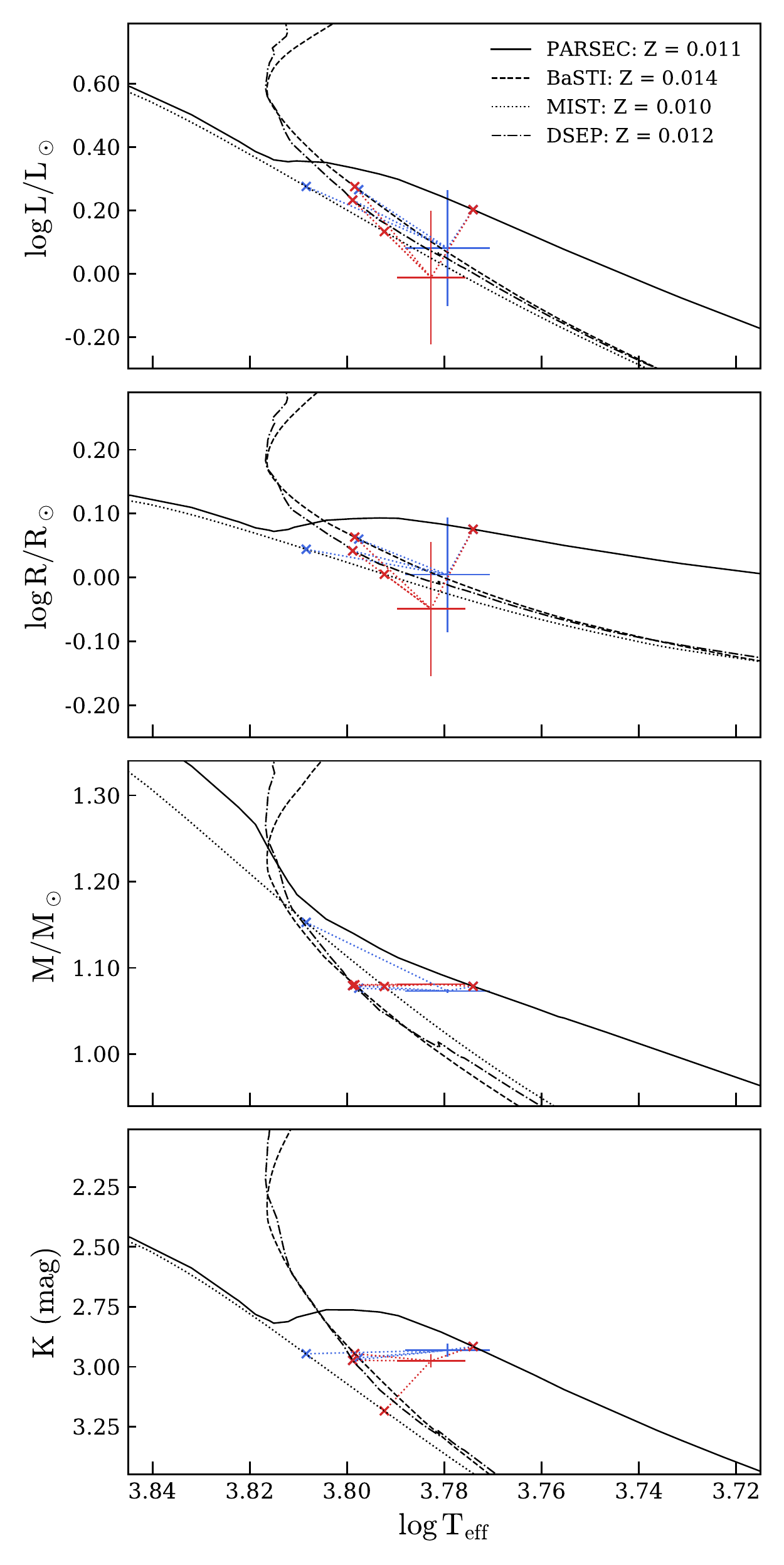}}
				\caption{Fitted \parsec, \basti, \mist, and \dsep\ isochrones for the HD9312 (left) and HD41255 (right) systems. Luminosities of both components and the radius of the secondary are displayed but not fitted (see text). For HD9312, the black colour is for isochrones with a chosen metallicity of $0.1$\,dex, while the grey colour is for the metallicity from \citet{Kiefer_2018_02_6}.}
				\label{figure__isochrones_hd9312_hd41255}
			\end{figure*}
			
			
			\begin{figure*}[!h]
				\centering
				\resizebox{\hsize}{!}{\includegraphics{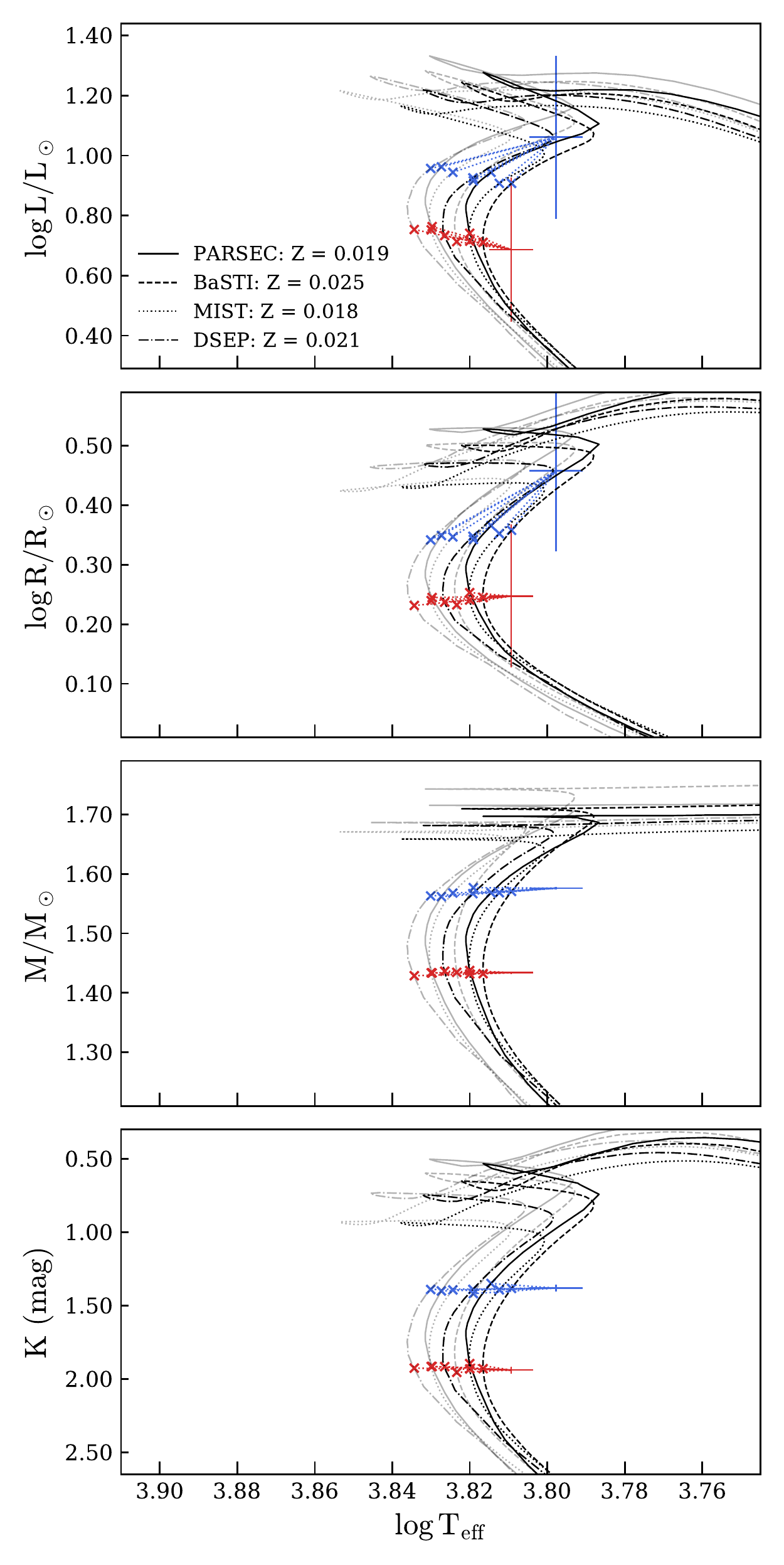}\includegraphics{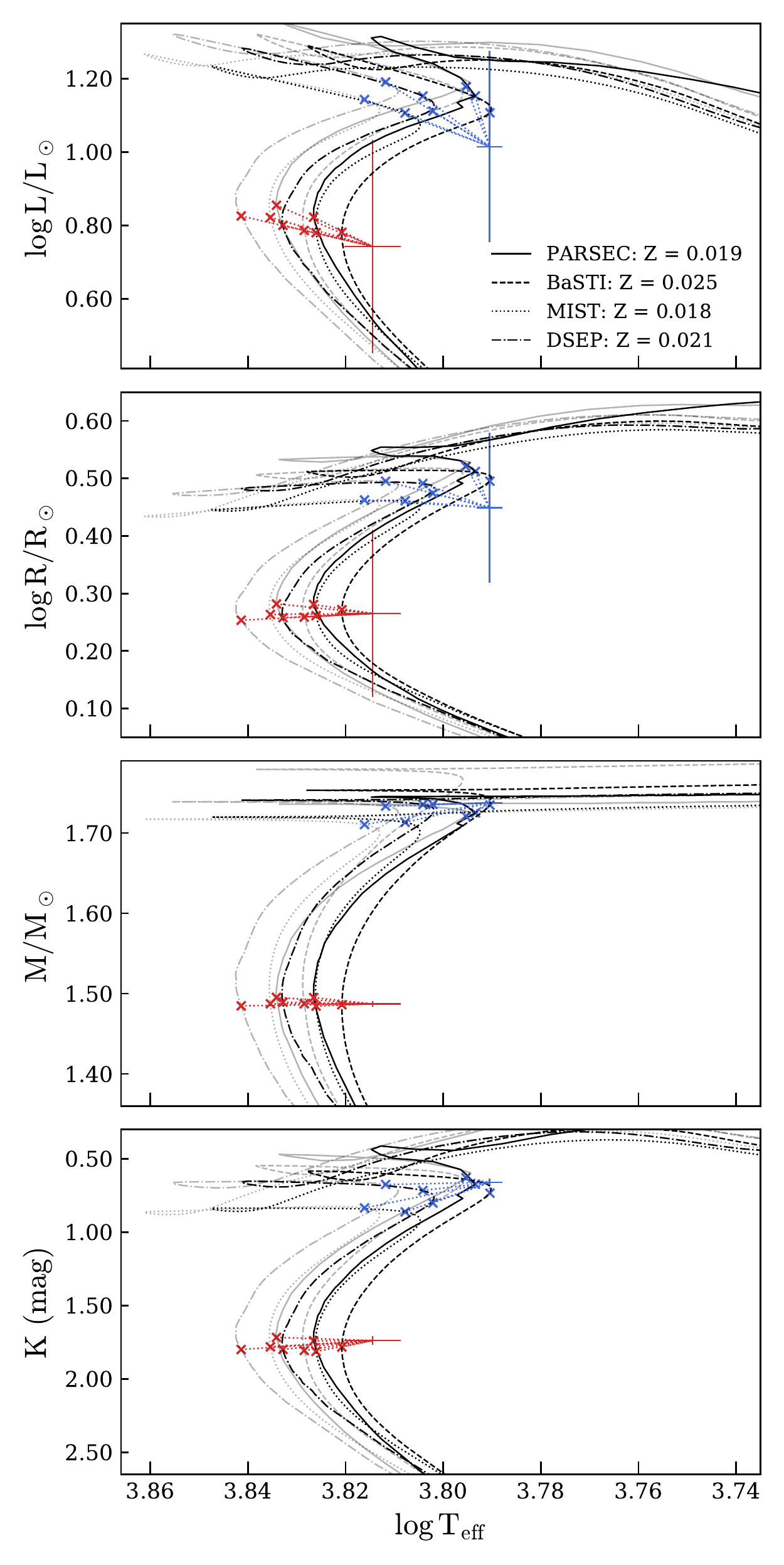}}
				\caption{Fitted \parsec, \basti, \mist, and \dsep\ isochrones for the HD70937 (left) and HD210763 (right) systems. The black colour is for isochrones with a chosen metallicity of $0.1$\,dex, while the grey colour is for our measured metallicity.}
				\label{figure__isochrones_hd70937_hd210763}
			\end{figure*}
			
			
			\begin{figure*}[!h]
				\centering
				\resizebox{\hsize}{!}{\includegraphics{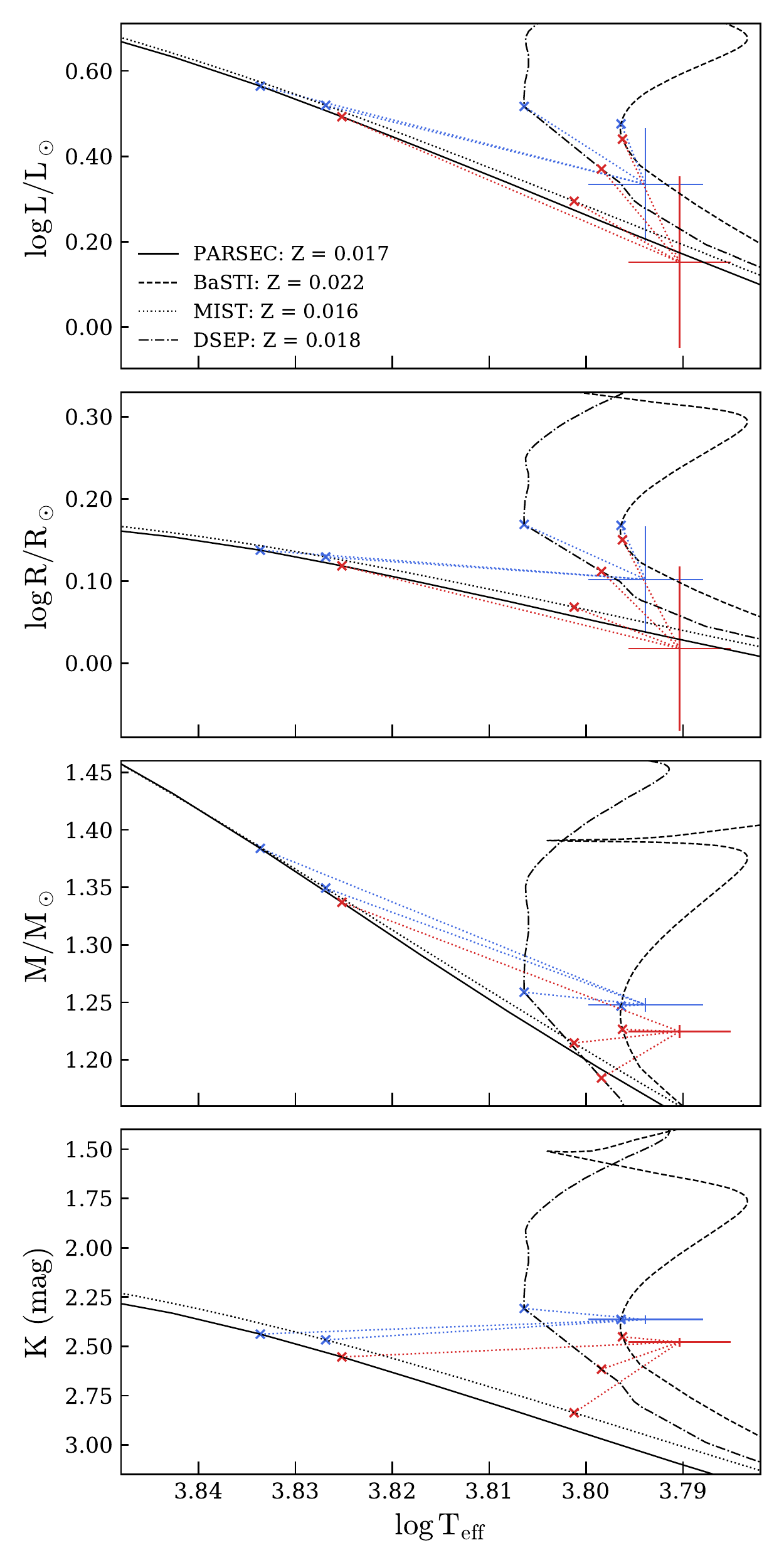}\includegraphics{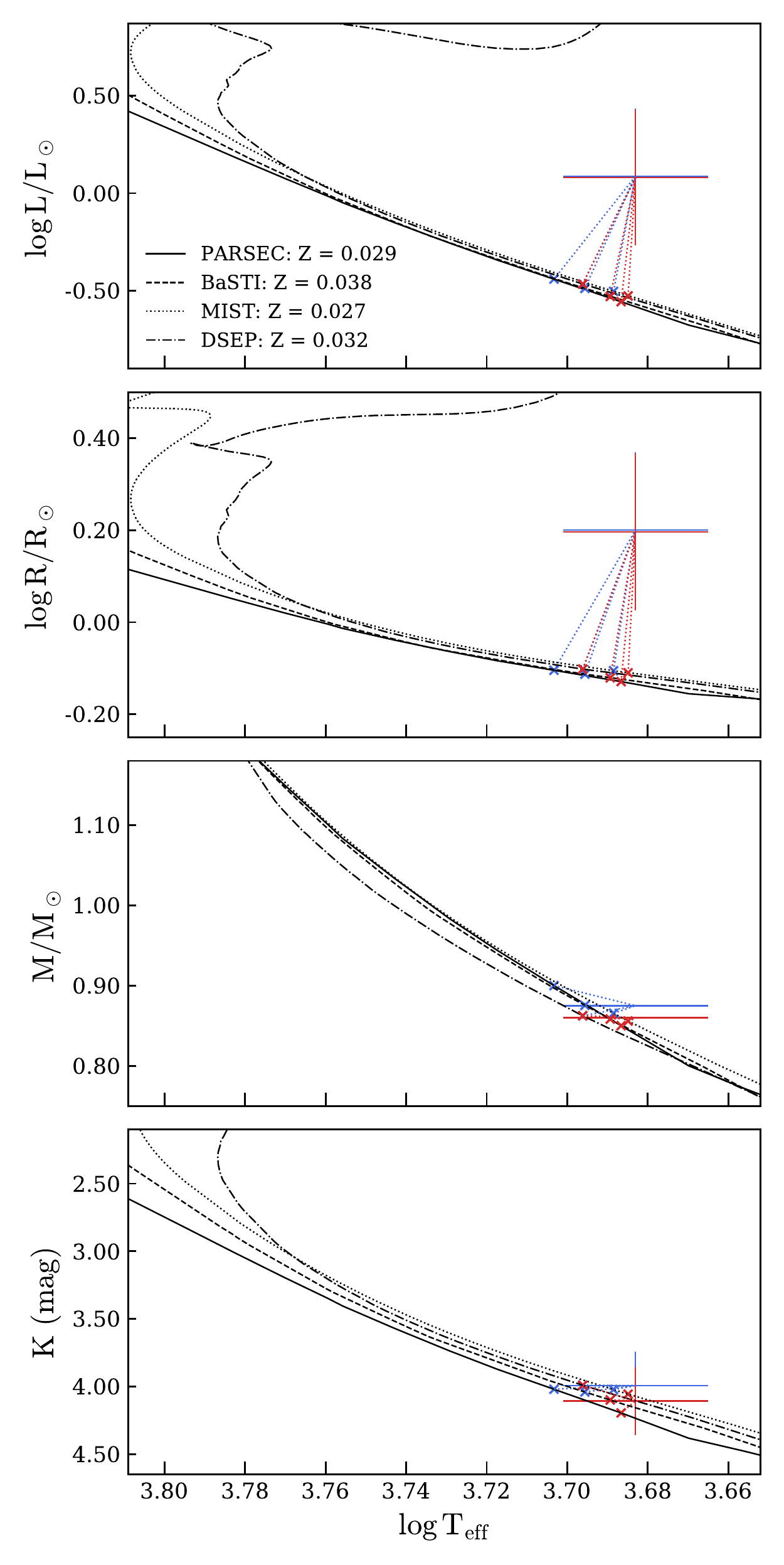}}
				\caption{Fitted \parsec, \basti, \mist, and \dsep\ isochrones for the HD224974 (left) and HD188088 (right) systems.}
				\label{figure__isochrones_hd224974_hr7578}
			\end{figure*}
			
			
			\begin{figure*}[!h]
				\centering
				\resizebox{\hsize}{!}{\includegraphics{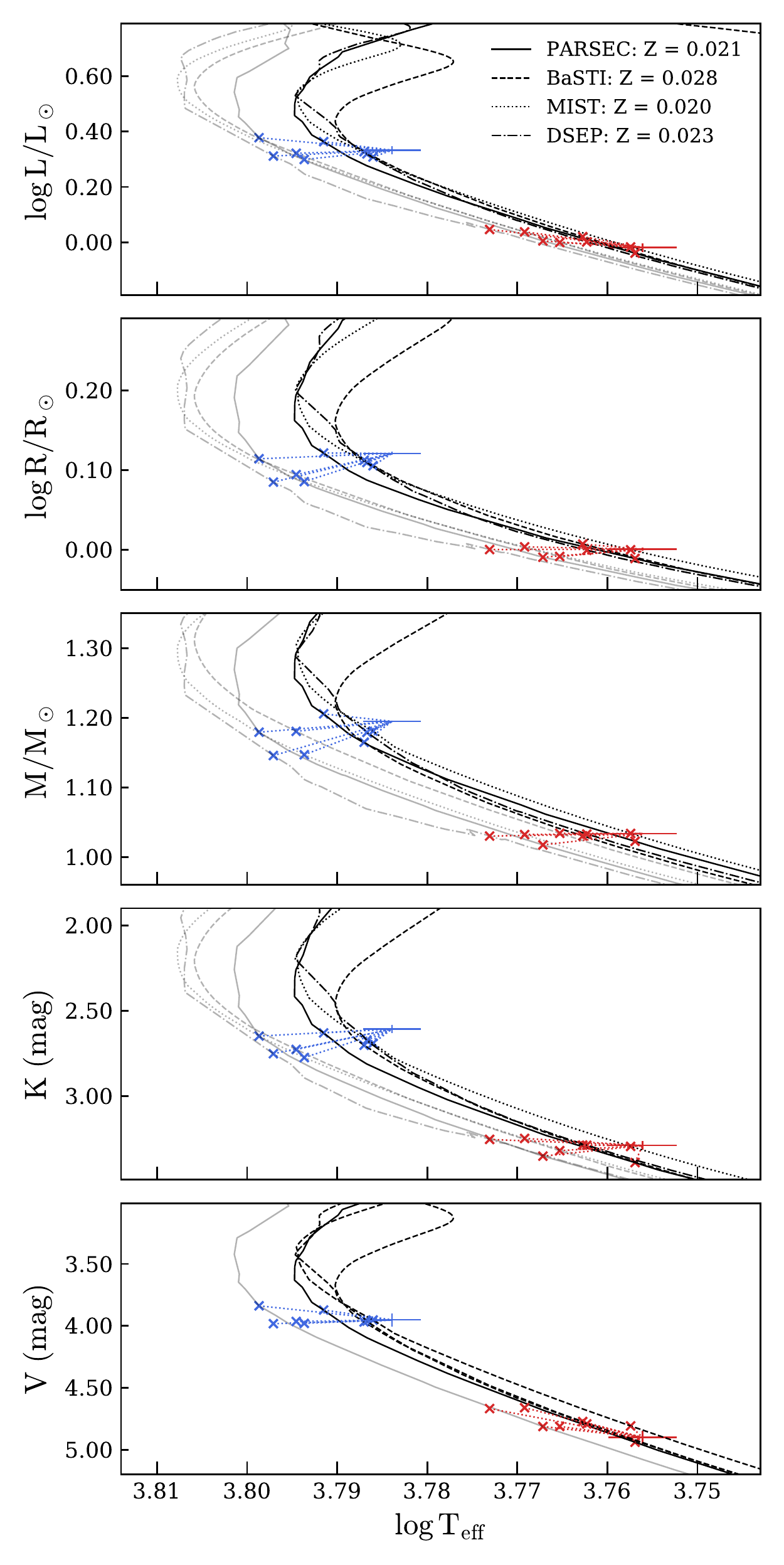}\includegraphics{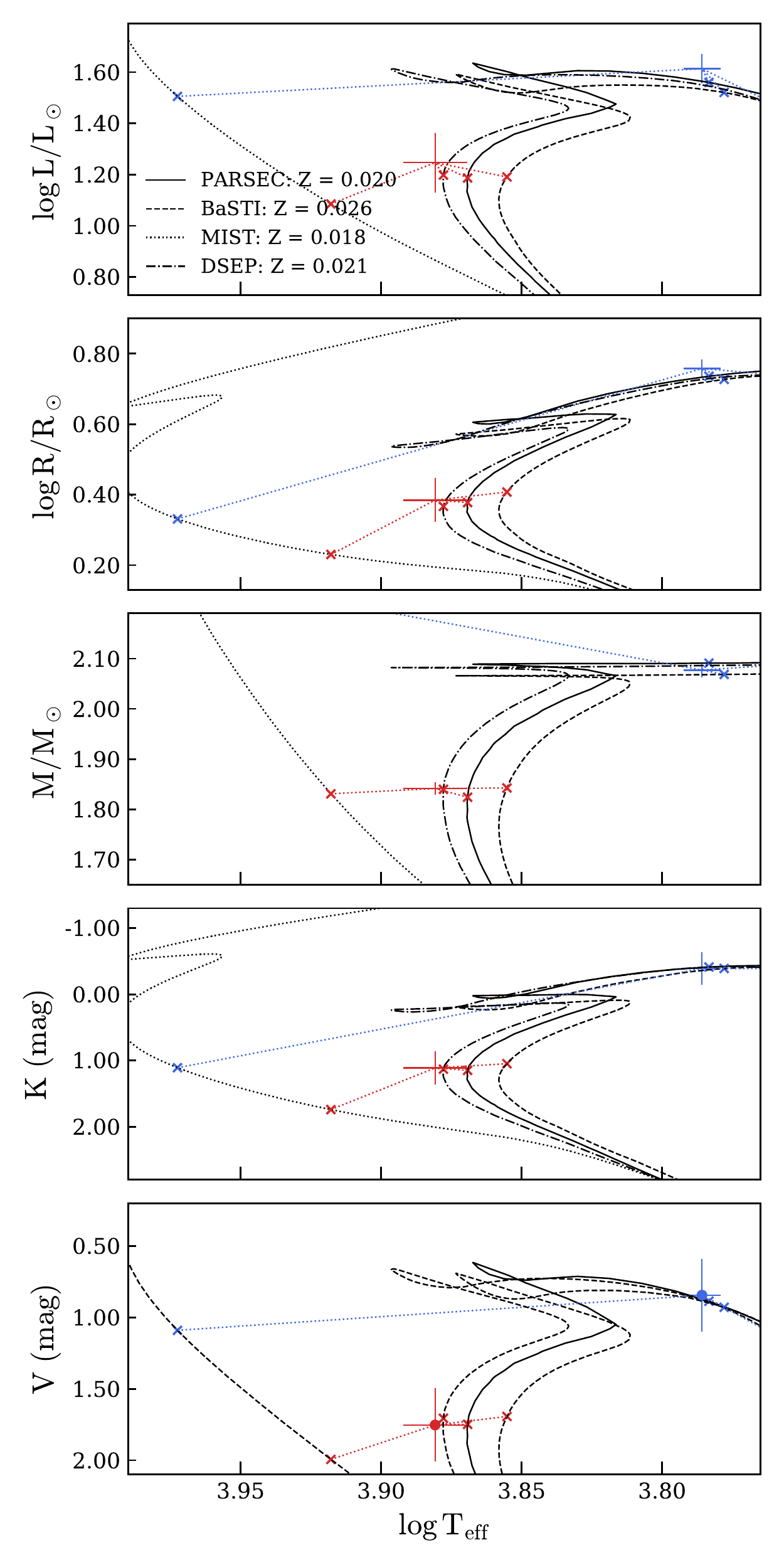}}
				\caption{Fitted \parsec, \basti, \mist, and \dsep\ isochrones for the LL Aqr (left) and $o$~Leo (right) systems. For LL~Aqr, the black colour is for isochrones with a chosen metallicity of $0.15$\,dex. while the grey colour is for the metallicity from \citet{Graczyk_2022_10_0}.}
				\label{figure__isochrones_llaqr_oleo}
			\end{figure*}
			
			
			\begin{figure}[!h]
				\centering
				\resizebox{\hsize}{!}{\includegraphics[width = \linewidth]{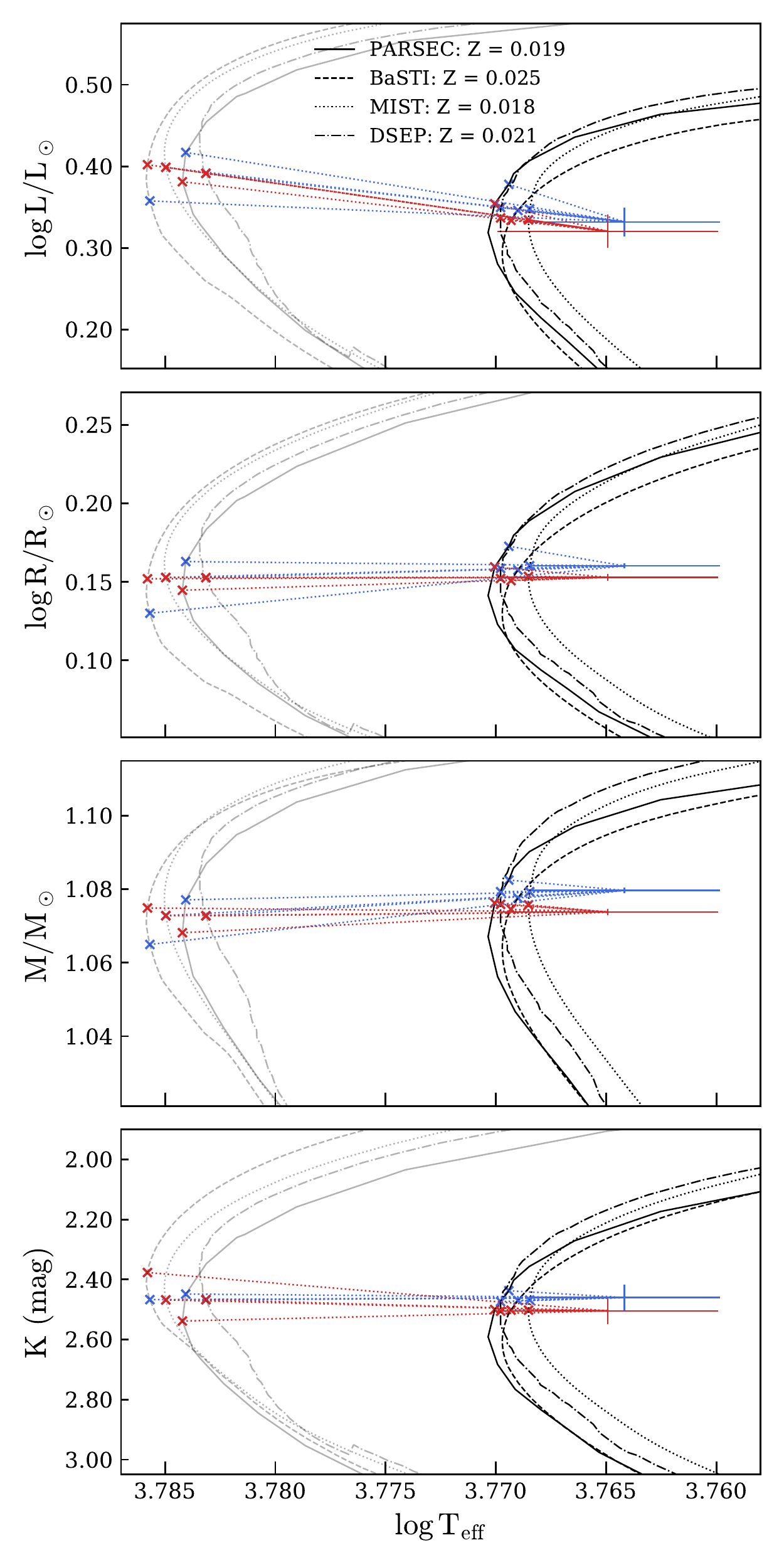}}
				\caption{Fitted \parsec, \basti, \mist, and \dsep\ isochrones for the V963~Cen system. The black colour is for isochrones with a chosen metallicity of $0.1\pm0.1$\,dex, while the grey colour is for the metallicity from \citet{Graczyk_2022_10_0}.}
				\label{figure__isochrones_v963cen}
			\end{figure}
			
		\end{appendix}
		
	\end{document}